\documentclass[12pt,letterpaper]{article}
\usepackage{epsfig,pslatex}

\begin{document}

\begin{titlepage}

\vspace*{1cm}

\begin{center}
{\bf \Large  ATLAS sensitivity to top quark

and $W$~boson polarization in $t\bar{t}$ events}
\vspace{2.cm}
\end{center}

\begin{center}

{\bf F. Hubaut, E. Monnier, P. Pralavorio} \\
\vspace{0.1cm}
{Centre de Physique des Particules de Marseille, CNRS/IN2P3 - Univ. M\'editerran\'ee, Marseille~-~France} \\
\vspace{0.5cm}

{\bf K. Smolek} \\
\vspace{0.1cm}
{Institute of Experimental and Applied Physics, 
Czech Technical University, Prague~-~Czech~Republic} \\
\vspace{0.5cm}

{\bf V. Simak} \\
\vspace{0.1cm}
{Faculty of Nuclear Sciences and Physical Engineering, Czech Technical University,
and Institute of Physics of the Czech Academy of Sciences, Prague~-~Czech Republic} \\
\vspace{0.5cm}

\end{center}

\vspace{1.5cm}

\begin{center}
{\bf Abstract}
\end{center}
Stringent tests on top quark production and decay mechanisms
are provided by the measurement of the top quark and $W$~boson polarization.
This paper presents a detailed study of these two measurements with the ATLAS detector, in the 
semileptonic 
($t\overline{t} \rightarrow W W b \overline{b}\rightarrow l \nu j_1 j_2 b \overline{b}$)
and dileptonic 
($t\overline{t} \rightarrow W W b \overline{b}\rightarrow l \nu l \nu b \overline{b}$)
$t\bar{t}$ channels. It is based on leading-order 
Monte Carlo generators and on a fast simulation of the detector. A particular attention is paid to 
the systematic uncertainties, which dominate the statistical errors after one LHC year at low luminosity (10~fb$^{-1}$), and to 
the background estimate. Combining results from both channel studies, the longitudinal component of 
the $W$ polarization ($F_0$) can be measured with a 2\% accuracy and the right-handed component 
($F_R$, which is zero in the Standard Model) with a 1$\%$ precision with 10~fb$^{-1}$.
Even though the top quarks in $t\bar{t}$ pairs are not polarized, a large asymmetry
is expected within the Standard Model in the like-spin versus unlike-spin pair production. 
A 4\% precision on this asymmetry measurement is possible with 
10~fb$^{-1}$, after combining results from both channel studies.
These promising results are converted in a sensitivity to new physics,
such as $tWb$ anomalous couplings, top decay to charged Higgs boson, or new $s$-channels
(heavy resonance, gravitons) in $t\bar{t}$ production.

\vspace{3.3cm}
\end{titlepage}

\begin{titlepage}
{\sf \tableofcontents}
\end{titlepage}


\section{Introduction}
\label{sec:intro}

Because of its high mass, intriguingly close to the electroweak symmetry breaking scale,
the top quark raises interesting questions and 
its sector is an ideal place to look for new physics~\cite{CERN_WORKSHOP, THEORY_41}.
Consequently, the search for non Standard Model interactions both in top quark production
and decay is one of the main motivations for top quark physics. 
A consequence of the very high top mass is that this quark decays before it can form 
hadronic bound states~\cite{THEORY_1}. 
This unique feature among quarks allows direct top spin studies, since spin properties 
are not washed out by hadronization 
and since the typical top spin-flip time is much larger than the top lifetime. 
Therefore, top spin polarization~\cite{THEORY_11} and correlation~\cite{THEORY_2} are precisely 
predicted by the Standard Model~(SM) and
reflect fundamental interactions involved in the top quark production and decay.
By testing only the top decay,
the $W$ boson polarization measurement complements top spin studies, helping to disentangle the origin of new physics, if observed.
Namely, the $t \rightarrow W^+ b$ decay mode is responsible for 99.9\% of top quark decays in the SM. 
Therefore, the $W$ polarization in the top decay is unambiguously predicted by the SM and its measurement 
provides a direct test of the $tWb$ vertex understanding and more particularly of its V-A structure~\cite{Kane,Larios}.\\ 

As a consequence, $W$ polarization in top decay and top spin observables are sensitive 
probes of new physics in top production and decay.
At the production level, a non-exhaustive list involves 
either anomalous $gt\bar{t}$ couplings~\cite{CPVIOL,chromomag},
which naturally arise in dynamical electroweak symmetry breaking models~\cite{THEORY_41} such as technicolor~\cite{TECHNICOLOR} or topcolor~\cite{TOPCOLOR},
or new interactions, as for example
a strong coupling of the top quark with a heavy spin~0 resonance, such as a heavy (pseudo)scalar Higgs boson~\cite{HIGGS_RESO} 
as predicted e.g. by SUSY models (~$gg \rightarrow H \rightarrow t\bar{t}$~), 
or the presence of extra dimensions~\cite{EXTRA_DIM}.
At the decay level, deviations from the Standard Model can for example arise from $tWb$ anomalous couplings, such as a V+A contribution 
in the vertex structure~\cite{THEORY_3}, or from a decay to charged Higgs boson~\cite{THEORY_43}.\\

Precise measurements of $W$ and top polarization require a higher statistics than currently available from Tevatron data.
The Large Hadron Collider (LHC) will be a top factory, producing more than 8 millions of $t\bar{t}$ events per year during its low luminosity
running phase (10$^{33}$ cm$^{-2}$s$^{-1}$), corresponding to an integrated luminosity of 10~fb$^{-1}$.
The production will occur through the $gg \rightarrow t\bar{t}$ (90$\%$) and 
$q\bar{q} \rightarrow t\bar{t}$ (10$\%$) hard processes. 
Depending on the $W$ decay modes, the $t\bar{t}$ events can be classified into three channels: 
the semileptonic channel ($t\overline{t} \rightarrow W W b \overline{b}\rightarrow l \nu j_1 j_2 b \overline{b}$), 
the dileptonic channel   ($t\overline{t} \rightarrow W W b \overline{b}\rightarrow l \nu l \nu b \overline{b}$)
and the all hadronic channel ($t\overline{t} \rightarrow W W b \overline{b}\rightarrow j_1 j_2 j_3 j_4 b \overline{b}$).
The latter will be difficult to extract from the huge QCD background and has not been considered in this work.
The electroweak single top production processes, which amount to approximately one third of the $t\bar{t}$ 
cross-section, can also be used to measure the polarization effects, but with a lower precision~\cite{ATLAS_SINGLE_TOP}.
They are not investigated in the following.\\

The goal of this paper is to evaluate the precision to which $W$~boson and top quark
polarization can be measured with the ATLAS detector, by combining results from dileptonic
and semileptonic $t\bar{t}$ channels. 
It is organized as follows. 
Section~\ref{sec:wtop_spin} discusses the $W$ boson and top quark polarization in $t\bar{t}$ events, 
and gives the related physics observables. 
Section~\ref{sec:tools} presents the event simulation, reconstruction and selection, as well as a detailed background estimate. 
Section~\ref{sec:wpola} gives the expected ATLAS sensitivity to the $W$ polarization, including
a complete study of the systematic uncertainties. 
From these results, the sensitivity to the magnitude
of $tWb$ anomalous couplings that parametrize new physics is also extracted. 
Using the same selected events, section~\ref{sec:cd} presents the expected ATLAS sensitivity to the top polarization, and to the related physics
beyond the SM.
Section~\ref{sec:conclu} is dedicated to conclusions.


\section{$W$ boson and top quark polarization in $t\bar{t}$ events}
\label{sec:wtop_spin}

This section presents the observables used to measure the polarization 
of the $W$~boson (section~\ref{sec:wpola_observables}) 
and of the top quark (section~\ref{sec:cd_observables}). 

\subsection{$W$ polarization observables}
\label{sec:wpola_observables}

\begin{figure}[htbp]
\begin{center}
\rotatebox{0}{\includegraphics*[width=.85\linewidth]{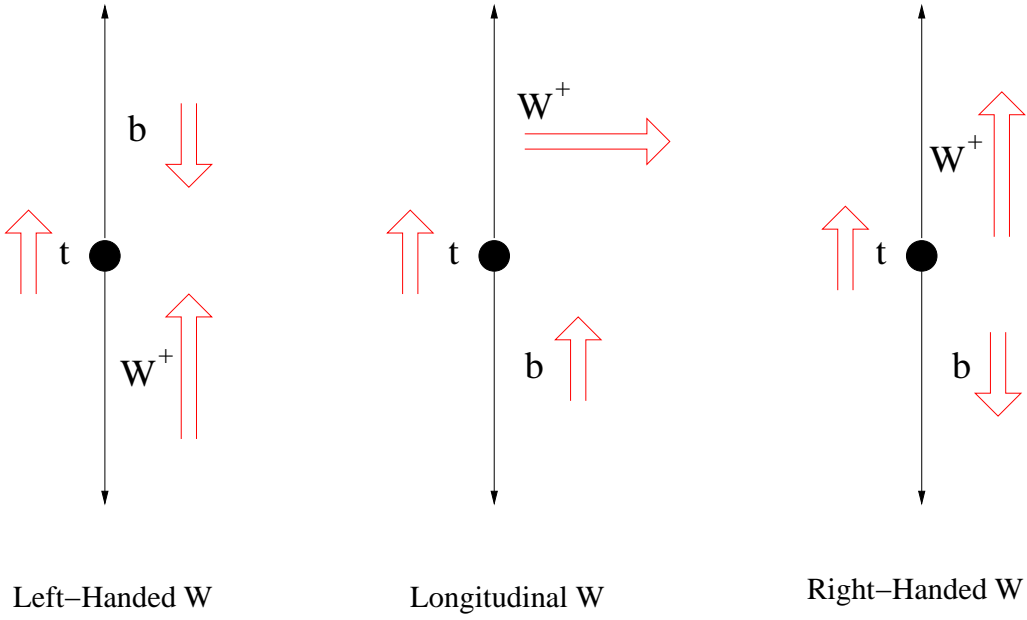}}
\end{center}
\caption{\it Sketches of angular momentum conservation in $t \rightarrow W^+ b$ decay in the top 
rest frame. Simple (open) arrows denote particle direction of motion (spin).
As a massless $b$-quark must be left-handed, the rightmost plot is forbidden in the SM at tree 
level.}
\label{fig:sketch}
\end{figure}

The real $W^+$ in the $t \rightarrow W^+ b$ decay can be produced with a longitudinal, left-handed 
or right-handed helicity as sketched in Figure~\ref{fig:sketch}.
The corresponding probabilities are $F_0$, $F_L$ and $F_R$, respectively, whose SM expectations at tree level in the zero
$b$-mass approximation are:

\begin{equation}
\left\{ 
\begin{array}{l}
\vspace*{.5cm}
F_0 = \frac{\mathrm{M}_t^2}{\mathrm{M}_t^2 +2 \mathrm{M}_W^2} = 0.703 + 0.002 \times (\mathrm{M}_t - 175) \\
\vspace*{.5cm}
F_L = \frac{2 \mathrm{M}_W^2}{\mathrm{M}_t^2 +2 \mathrm{M}_W^2} = 0.297 - 0.002 \times (\mathrm{M}_t - 175) \\ 
F_R=0.000
\end{array}
\right.
\label{eq:fractions}
\end{equation}
where M$_t$ and M$_W$  are the top and $W$ masses in~GeV.  
Left and right components are inverted for $W^-$ bosons.
By definition, we have the restriction $F_0+F_L+F_R=1$.
Since massless particles must be left-handed in the SM, right-handed $W^+$ bosons 
do not exist in the zero $b$-mass approximation, due to angular momentum conservation sketched in Figure~\ref{fig:sketch}.
Including QCD and electroweak radiative corrections, finite width corrections and non-zero $b$-quark 
mass induces small variations: $F_0=0.695$, $F_L=0.304$ and $F_R=0.001$ for 
M$_t = 175$~GeV~\cite{Groote}. Because the top quark is very heavy, $F_0$ is large and the top decay is 
the only significant source of longitudinal $W$ bosons\footnote{QCD production, the only 
other source of real $W$ bosons, produces nearly all $W$ transversely polarized.}.
Deviations of $F_0$ from its SM value would bring into question 
the validity of the Higgs mechanism of the spontaneous symmetry breaking, responsible for the 
longitudinal degree of freedom of the massive gauge bosons.
Any deviation of $F_R$ from zero could point to a non-SM V+A admixture to the standard 
left-handed weak current, as for example predicted by $SU(2)_L \times SU(2)_R \times U(1)$ 
extensions of the SM~\cite{su2right}.\\

\begin{figure}[htbp]
\begin{center}
\rotatebox{0}{\includegraphics*[width=.85\linewidth]{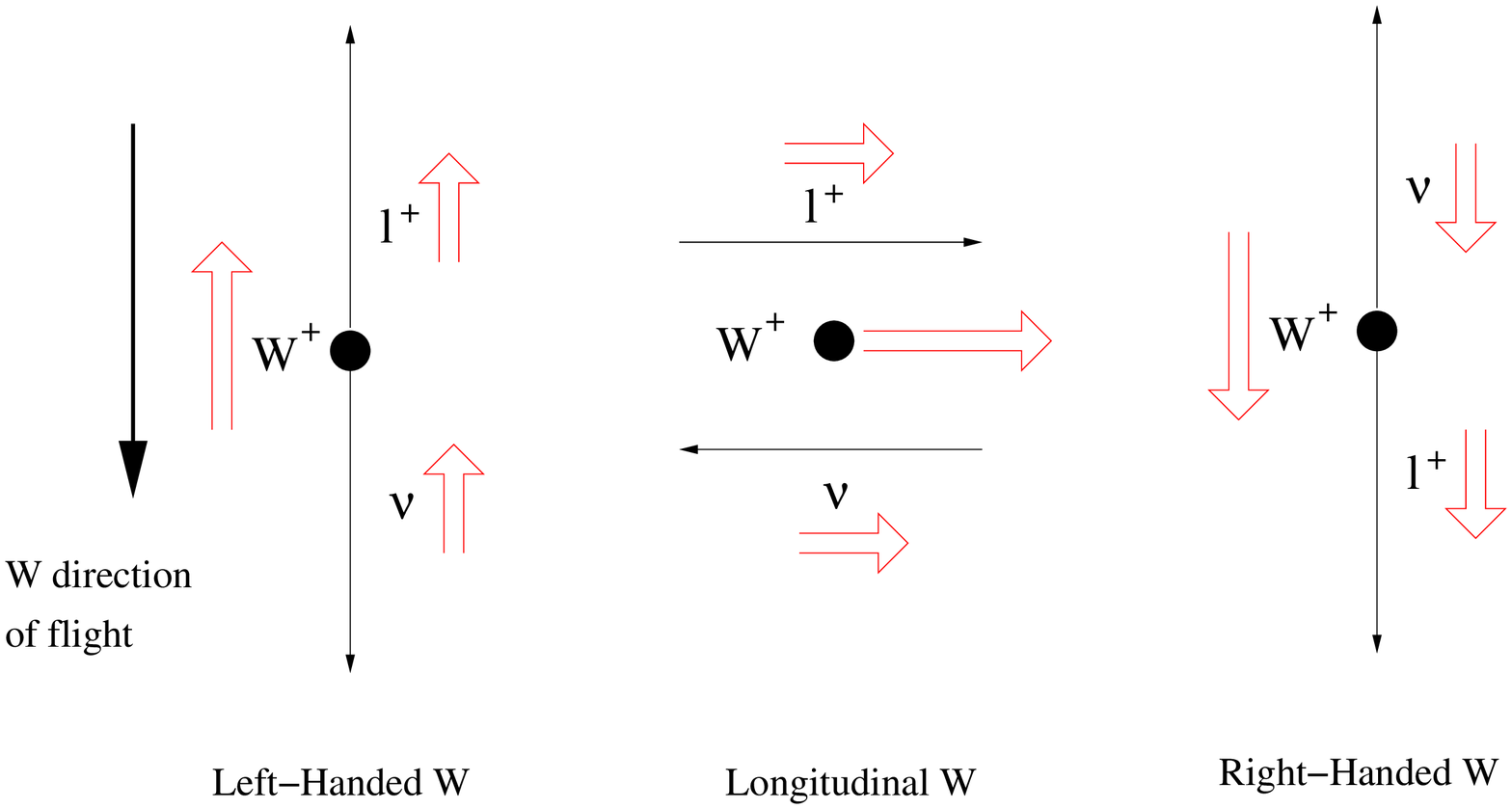}}
\end{center}
\caption{\it Sketches of the different $W^+$ polarization modes in $t \rightarrow W^+ b$ decay 
and resulting lepton directions.
Simple (open) arrows denote particle direction of motion (spin).
For $W^-$, left and right-handed components are inverted.}
\label{fig:illus_wpola}
\end{figure}

The best way to access particle spin information is to 
measure the angular distribution of its decay products, thereby called spin analyzers.
As an example, illustrated in Figure~\ref{fig:illus_wpola}, the charged lepton from the decay of 
longitudinally polarized $W^+$ tends to be emitted transversally to the $W^+$ 
direction, due to angular momentum conservation. Similarly, the charged lepton from a left-handed 
(right-handed) $W^+$ is preferentially emitted in the opposite (same) $W^+$ direction, 
leading to a softer (harder) p$_T$ spectrum with respect to the leptons from longitudinal $W^+$.
The resulting angular lepton distributions are therefore very distinct for each $W$ helicity state.\\

\begin{figure}[htbp]
\begin{center}
\rotatebox{0}{\includegraphics*[width=.6\linewidth]{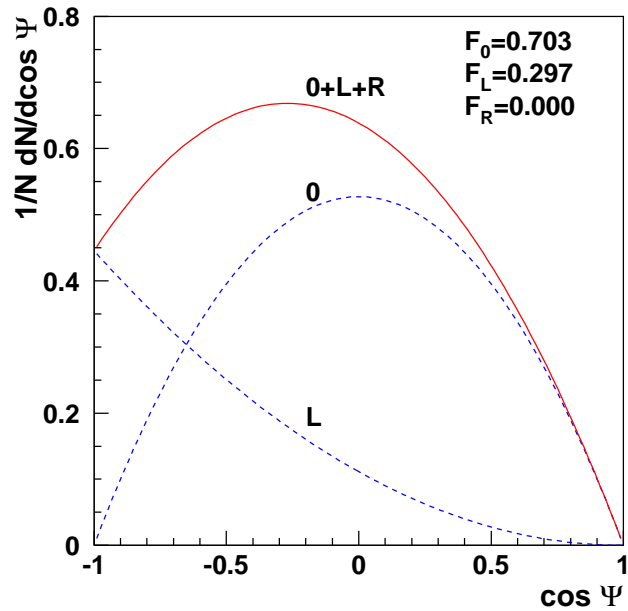}}
\end{center}
\vspace*{-.5cm}
\caption{\it Angular distribution of Equation~(\ref{eq:cosphi}) in the SM. 
The predicted contributions from longitudinal (0) and left-handed (L) 
helicity states are shown separately with dashed lines. 
The right-handed contribution is null in the SM. The sum (0+L+R) is depicted with a full line.}
\label{fig:parton}
\end{figure}

As it is necessary to know the weak isospin of the $W$ spin analyzer, the charged lepton is the best choice
since $u$-like jets can not be distinguished experimentally from $d$-like jets.
Consequently, the $W$ polarization is better measured in dileptonic and semileptonic $t\bar{t}$ channels 
through the distribution of the $\Psi$ angle
between the charged lepton direction in the $W$ rest frame and the
$W$ direction in the top quark rest frame.
The $\Psi$ angular distribution is given by the following expression~\cite{Kane}:
\vspace*{.2cm}
\begin{equation}
\frac{1}{N}\frac{dN}{d\cos\Psi}=\frac{3}{2}
\left\lbrack 
F_0 \left(\frac{\sin \Psi}{\sqrt{2}} \right)^2 +
F_L \left(\frac{1-\cos \Psi}{2} \right)^2 +
F_R \left(\frac{1+\cos \Psi}{2} \right)^2
\right\rbrack 
\label{eq:cosphi}
\end{equation}
Its SM expectation is shown in Figure~\ref{fig:parton}. 
It reflects the superposition of the three terms of Equation~(\ref{eq:cosphi}), corresponding 
to the longitudinal $(\sin \Psi)^2$, the left-handed $(1-\cos \Psi)^2$ and 
the right-handed $(1+\cos \Psi)^2$ $W$ helicity states.
Each term is weighted by the fraction $F_0$, $F_L$ or $F_R$ given in 
Equation~(\ref{eq:fractions}).\\
\vspace{3cm}

Since the $W$ and top rest frames are used in the $\Psi$ angle measurement, it requires a complete event topology reconstruction.
This is rather easy in the semileptonic $t\overline{t}$ channel, with only one neutrino in the 
final state and a high signal over background ratio (see section~\ref{sec:evt_rec_semilep}).
In the dileptonic channel\footnote{This is also the case at the
Tevatron~\cite{CDF_WPOLA_FR,CDF_WPOLA_F0,D0_WPOLA,TEV_RUNII} and for
single top analysis at LHC~\cite{ATLAS_SINGLE_TOP}.}, the event reconstruction is more challenging (see section~\ref{sec:evt_rec_dilep}). 
Therefore, the $\Psi$ angle is reconstructed in terms of the invariant mass of the lepton and the $b$-quark, 
M$_{lb}$~\cite{Kane}:
\begin{equation}
\cos\Psi\sim\frac{2 \mathrm{M}_{lb}^2}{\mathrm{M}_t^2-\mathrm{M}_W^2} -1
\label{eq:mlb}
\end{equation}
which is valid in the zero $b$-mass limit, and where M$_t$ and M$_W$ are set to 175~GeV and 80.41~GeV, respectively.
In this approach, the dependence on the $b$-jet energy scale and on the top mass uncertainty is high. 
On the contrary, these two dependences cancel at first order when measuring directly $\cos\Psi$.
A study performed in the semileptonic channel
shows a two times lower systematics on $F_0$, $F_L$ and $F_R$ using the $\cos\Psi$ observable compared to that obtained with M$_{lb}^2$.

\subsection{Top polarization observables}
\label{sec:cd_observables}

Similarly as for the $W$, the top polarization can be analyzed trough the angular
distribution of its daughters. In this case, the spin analyzer, denoted by $i$, can be either
a direct daughter ($W$, $b$) or a $W$ decay product ($l$, $\nu$, $j_1$ or $j_2$). The 
relevant angular distribution is~\cite{THEORY_3}:
\vspace*{.5cm}
\begin{equation}
\frac{1}{N}\frac{dN}{d\cos\theta_i}= \frac{1}{2} (1+S\alpha_i \cos\theta_i)
\label{eq:top_decay_polarized}
\end{equation}
where $S$ is the modulus of the top polarization and $\theta_i$ is the 
angle between the direction of particle $i$ in the top quark rest frame and
the direction of the top polarization. $\alpha_i$ is the spin analyzing power 
of this particle. It is the degree to which its direction is correlated to the 
spin of the parent top quark. 
It has been computed at the next-to-leading order (NLO) since long for the lepton~($l$)~\cite{THEORY_31} 
and more recently for the $b$ quark, the $W$ boson and the quarks from the $W$~decay~\cite{THEORY_32}.
The theoretical values are given both at LO and NLO in Table~\ref{tab:spin_analysing_power}
for a spin up top quark (signs are reversed for a spin down or for an anti-quark).
Even if $W$ and $b$ are direct daughters of the top, their analyzing power is low
due to the intrinsic polarization of the $W$ which interferes destructively 
with the top one. Consequently, at LO, the difference $F_0-F_L$ provides a measurement of $\alpha_W$~\cite{guillian}.
Charged leptons and down-type quarks, which are almost 100\% correlated 
with the top spin direction, are optimal spin analyzers. 
But on the contrary to leptons, $d$ and $s$-jets cannot be 
distinguished experimentally from $u$ and $c$-jets. 
Therefore, the analyzing power of light jets is the average value $\alpha_{\mathrm{jet}} \sim(1-0.31)/2 = 0.35$. 
This can be improved by choosing the least energetic jet ($lej$) in the top rest frame, 
which is of $d$ type in 61\% of the case, resulting to $\alpha_i\sim0.5$~\cite{THEORY_32}.\\

\begin{table}[htbp]
\begin{center}
\begin{tabular}{|l||c|c||c|c|c|c|}
\hline
 Particle      &  $b$-jet & $W^+$ & $l^+$ & $j_1$=$\bar{d}$-jet, $\bar{s}$-jet & $j_2$=$u$-jet, $\bar{c}$-jet & $lej$  \\
\hline
$\alpha_i$ (LO)  &   -0.41  & 0.41  &     1 & 	 1           &   -0.31			                 &  0.51  \\
\hline
$\alpha_i$ (NLO) &   -0.39  & 0.39  & 0.998 &  0.93	      & -0.31					 &  0.47  \\
\hline
\end{tabular}
\caption{\it SM spin analyzing power at LO and NLO of top quark daughters: $b$-jet, $W^+$, 
and $W$ decay products : lepton ($l^+$), $j_1$, $j_2$ or the least energetic non b-jet in 
the top rest frame, called $lej$~\cite{THEORY_32}. The top quark is spin up. 
Signs are reversed for a spin down or for an anti-quark.} 
\label{tab:spin_analysing_power}
\end{center}
\end{table}

\begin{figure}[htb]
\unitlength 1mm 
\noindent
  \begin{minipage}[htbp]{0.53\linewidth}
    \begin{center}
    \epsfig{file=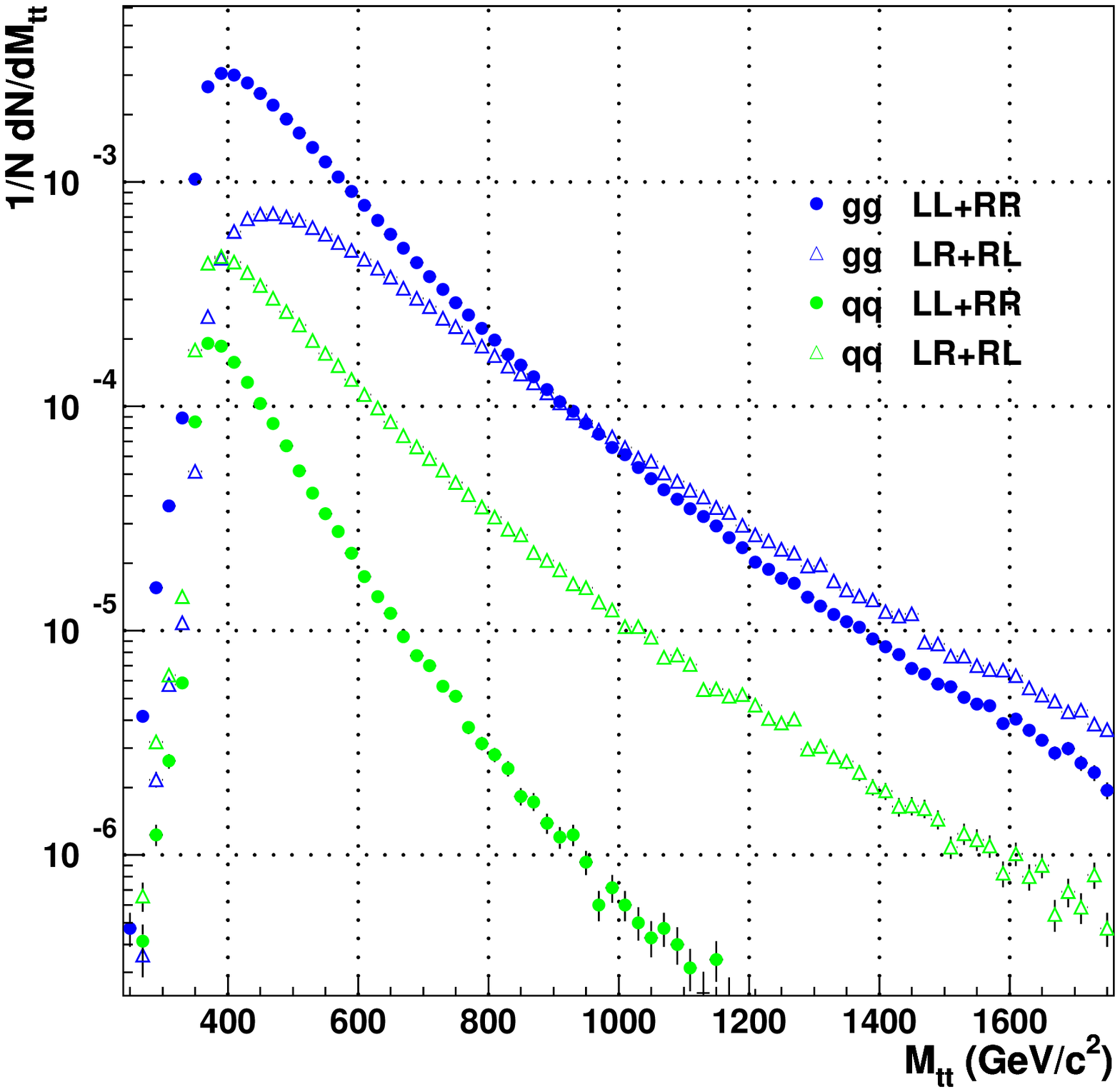, width=\linewidth} 
    \caption{\it Invariant mass distribution of the $t\bar{t}$ system with like (LL+RR) and unlike 
(LR+RL) helicities for the two possible production mechanisms ($gg$ and $q\bar{q}$)~\protect{\cite{TT_SEMILEP}}.}
\label{fig:tt_cross_sect}
    \end{center}
  \end{minipage}
  \hfill
  \begin{minipage}[htb]{0.42\linewidth}
    \begin{center}
    \epsfig{file=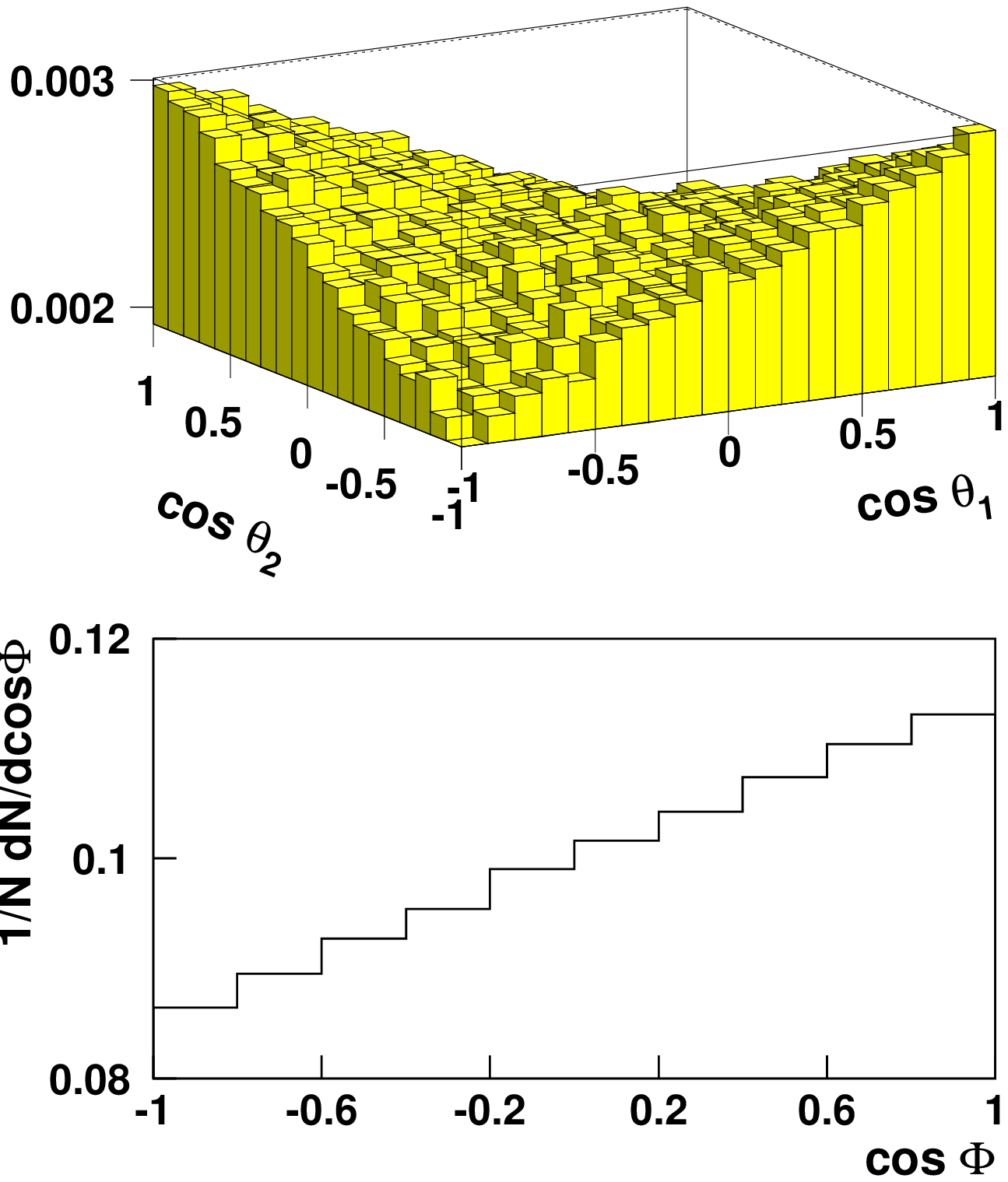, width=\linewidth} 
    \caption{\it Up, double differential angular distribution of Equation~(\ref{eq:param_C}). 
    Down, opening angle distribution of Equation~(\ref{eq:param_D}).
Both are shown at parton level with M$_{t\bar{t}}<550$~GeV.}
\label{fig:tt_cross_sect2}
    \end{center}
  \end{minipage}
\end{figure}

Equation~(\ref{eq:top_decay_polarized}) can be directly used for
top quarks produced lonely via the weak interaction, which are polarized ($S\sim1$).
This is not the case of the top quarks produced in $t\bar{t}$ pairs, which are not polarized~\cite{THEORY_42}. 
However, the top and the anti-top spins are correlated, which
can be easily understood.
Close to the $t\bar{t}$ production threshold,
the $t\bar t$ system produced by $q\bar{q}$ annihilation is in a $^3S_1$ state,
while it is in a $^1S_0$ state with the $gg$ fusion process. 
Therefore, in the first case, the top quarks
tend to have their spins aligned while in the second case
their spins tend to be opposite to each other. Away from threshold,
this simple picture is modified due to the presence of angular
momentum. In the absence of polarization, a direct measurement of the
correlation at the level of the top quarks 
is obtained from observables of the form:
\begin{equation}
  ( {\bf \hat a}\cdot {\bf S}_t) ( {\bf \hat b}\cdot {\bf S}_{\bar
  t}),\,\, ({\bf S}_t\cdot {\bf S}_{\bar t})
  \label{PartoniSpinObservables}
\end{equation}
where ${\bf S}_t$, ${\bf S}_{\bar t}$ are the spin operators of the top and
anti-top, and $\bf \hat a$, $\bf \hat b$ are arbitrary directions ($|{\bf \hat a}| = |{\bf \hat b}| = 1$).
A more familiar representation of the observables shown in
Equation~(\ref{PartoniSpinObservables}) can be obtained from the relation:
\begin{equation}
  A = 4 \langle ( {\bf \hat a}\cdot {\bf S}_t) ( {\bf \hat b}\cdot {\bf S}_{\bar
    t}) \rangle
  = \frac{\sigma(t_{\uparrow} \bar{t}_{\uparrow})+\sigma(t_{\downarrow} 
\bar{t}_{\downarrow})
-\sigma(t_{\uparrow} \bar{t}_{\downarrow})-\sigma(t_{\downarrow} 
\bar{t}_{\uparrow})}
{\sigma(t_{\uparrow} \bar{t}_{\uparrow})+\sigma(t_{\downarrow} 
\bar{t}_{\downarrow})
+\sigma(t_{\uparrow} \bar{t}_{\downarrow})+\sigma(t_{\downarrow} 
\bar{t}_{\uparrow})}
  \label{eq:SpinAsymmetry}
\end{equation}
where $\sigma(t_{\uparrow/\downarrow} \bar{t}_{\uparrow/\downarrow})$ 
denotes the
cross section for the production of a top quark pair with spins up or
down with respect to a
quantization axis defined by $\bf \hat a$ in case of the top quark and
$\bf \hat b$ in case of the anti-top quark.
Note that rewriting the observable $({\bf S}_t\cdot {\bf S}_{\bar t})$
using
\begin{equation}
  A_D = ({\bf S}_t\cdot {\bf S}_{\bar t}) = \sum_i { S}_{ti} {S}_{\bar t i} =
  \sum_i ( {\bf \hat e}^{(i)}\cdot {\bf S}_t) ( {\bf \hat e}^{(i)}\cdot 
{\bf S}_{\bar
    t})
\end{equation}
where ${\bf \hat e}^{(i)}_k = \delta_{ik}$,
the observable $({\bf S}_t\cdot {\bf S}_{\bar t})$ can also be cast
into the form shown in Equation~(\ref{eq:SpinAsymmetry}).
Computation in the Standard Model gives favorably like-spin pairs ($A>0$) when
using the `helicity' basis for the spin basis at LHC~\cite{THEORY_43}. 
In this basis\footnote{Another basis 
was recently found to be more optimal, but more complicated to reconstruct~\cite{OPTI_BASIS}.}, the top (anti-top) spin quantization axis 
corresponds to the top (anti-top) direction of flight
in the $t\bar{t}$ center of mass system, and the notation $\uparrow$, $\downarrow$
is replaced by L (Left) and R (Right).
Figure~\ref{fig:tt_cross_sect} shows the invariant mass distribution 
of the $t\bar{t}$ system with like and unlike helicities for the two possible
production mechanisms. 
As already explained, $gg$ and $q\bar{q}$ processes contribute to the asymmetry with opposite signs
($A>0$ for $gg$ and $A<0$ for $q\bar{q}$).
The theoretical Standard Model value integrated over the whole LHC spectrum at 
LO is $A=0.319$ and $A_D$=-0.217. At NLO these values become~\cite{THEORY_5}:
\begin{equation}
\left\{ 
\begin{array}{l}
\vspace*{.5cm}
A = 0.326^{+0.003}_{-0.002}(\mu)^{+0.013}_{+0.001}(PDF)\\
A_D =-0.237^{+0.005}_{-0.007}(\mu)^{+0.000}_{-0.006}(PDF)
\end{array}
\right.
\label{eq:asym_values}
\end{equation}
Systematic uncertainties come from factorization and renormalization 
scales ($\mu=\mu_F=\mu_R$) and from Parton Distribution Function (PDF). 
As $A$ and $A_D$ are defined as
ratios between two cross sections, PDF, $\mu$ and $\alpha_S$ dependences 
cancel to a large extent. Moreover, NLO QCD corrections are small and thus 
theoretical uncertainties are well under control.\\

At LHC, it is possible to increase the asymmetry by applying an upper cut
on the $t\bar{t}$ invariant mass M$_{t\bar{t}}$.
As shown in Figure~\ref{fig:tt_cross_sect}, the asymmetry is maximal at low invariant masses
for the $gg$ contribution, which is by far dominant at the LHC, and
is equal to 0 around \mbox{M$_{t\bar{t}}=900$~GeV}. 
Therefore, selecting low energetic top quarks with M$_{t\bar{t}}<550$~GeV
rejects only 30\% of the events while $A$ and $A_D$ are enhanced by about 30\% at LO:
\vspace*{.2cm}
\begin{equation}
\left\{ 
\begin{array}{l}
\vspace*{.5cm}
A = 0.422\\
A_D =-0.290
\end{array}
\right.
\label{eq:asym_values_cut}
\end{equation}

\newpage
Similarly to Equation~(\ref{eq:top_decay_polarized}) for top polarization, angular
distributions can be used to probe the $t\bar{t}$ spin correlation, as:

\begin{itemize}
\item the double differential angular distribution of top and anti-top quark 
decay pro\-ducts~\cite{THEORY_44}:
\begin{equation}
\frac{1}{N}\frac{d^2N}{d\cos\theta_1 d\cos\theta_2}= 
\frac{1}{4} (1 + B_1 \cos\theta_1 + B_2 \cos\theta_2 - C \cos\theta_1 \cos\theta_2)
\label{eq:param_C}
\end{equation}
where $\theta_1$ ($\theta_2$) is the angle between the direction of the $t$ ($\bar{t}$)
spin analyzer in the $t$ ($\bar{t}$) rest frame and the $t$ ($\bar{t}$)
direction in the $t\bar{t}$ center of mass system. These complicated angles are the direct
consequence of the choice of the 'helicity' basis for measuring the asymmetry. 
As top and anti-top quarks are not polarized in this basis,
$B_1 =  B_2 = 0$. 
Figure~\ref{fig:tt_cross_sect2} (top panel) illustrates this double angular 
distribution for the SM in the semileptonic channel.
\item the opening angle distribution~\cite{THEORY_5}:
\begin{equation}
\frac{1}{N}\frac{dN}{d\cos\Phi}= \frac{1}{2} (1 - D \cos\Phi)
\label{eq:param_D}
\end{equation}
where $\Phi$ is the angle between the direction of flight of the two spin 
analyzers, defined in the $t$ and $\bar{t}$ rest frames respectively.
Figure~\ref{fig:tt_cross_sect2} (down panel) shows this opening angle 
distribution for the SM in the semileptonic channel.
\end{itemize}

In Equations (\ref{eq:param_C}) and (\ref{eq:param_D}), $C$ and $D$ are the spin correlation observables. 
Before any phase-space cut, they can be easily measured using the following
unbiased estimators~\cite{TT_DILEP2}:
\vspace*{.2cm}
\begin{equation}
\left\{ 
\begin{array}{l}
\vspace*{0.5cm}
C   =  -9 <\cos \theta_1 \cos \theta_2> \\
D   =  -3 <\cos \Phi>
\end{array}
\right.
\label{eq:estimators}
\end{equation}
The production asymmetries $A$ and $A_D$ are then directly deduced by 
simply unfolding the decay contribution through the spin analyzing power of the daugther particles $\alpha_1$ and $\alpha_2$:
\begin{equation}
\left\{ 
\begin{array}{l}
\vspace*{0.5cm}
A   = \frac{C}{| \alpha_1 \alpha_2 |} \\
A_D = \frac{D}{| \alpha_1 \alpha_2 |} 
\end{array}
\right.
\label{eq:asym}
\end{equation}
Since leptons are the most powerful spin 
analyzers ($\alpha=1$), the dileptonic case is a priori the most promising. 
On the contrary, the all hadronic case is the most unfavorable, with a low spin analyzing power 
and a huge background.
Several choices exist in the semileptonic channel for the spin analyzers on the 
hadronic side ($W$, $b$ and $lej$) and experimentally the $lej$ is the best choice~\cite{CPPM_SPINCO}. 
In any case the semileptonic channel is more challenging compared to the dileptonic one because
its spin analyzers are less powerful than the leptons. 
But on the contrary to the dileptonic channel, the number of events is 6 times larger
and the event reconstruction is much easier (only one neutrino in the final state).

\section{Event simulation, selection and reconstruction}
\label{sec:tools}

This section describes the software tools used to generate and simulate signal and background
events using a modeling of the ATLAS detector. 
Then the event selection and reconstruction is explained in both $t\bar{t}$ semileptonic and dileptonic channels.

\subsection{Signal and background definition}
\label{sec:sig}

General figures of $t\bar{t}$ pairs decay are: $Br(t\rightarrow Wb)\sim 1$, $Br(W \rightarrow 
l\nu_l)\sim 1/3$ with $l=e,\mu,\tau$ in equal probabilities and $Br(W\rightarrow 
q_1 q_2) \sim 2/3$ with $q_1(q_2)= u(d), c(s)$ in equal probabilities.  
With a NLO cross-section around 850 pb~\cite{TT_XSECTION}, 3.8 (0.9) millions of $t\bar{t}$ 
semileptonic (dileptonic) events will be produced 
with an integrated luminosity of 10~fb$^{-1}$, corresponding to one LHC year at low luminosity. 
Among them, 2.5 (0.4) millions are signal events, defined as: 
\begin{equation}
\left\{ 
\begin{array}{l}
\vspace*{.5cm}
t\bar{t} \rightarrow W b W \bar{b} \rightarrow l \nu b j_1 j_2 \bar{b} \\
t\bar{t} \rightarrow W b W \bar{b} \rightarrow \bar{l} \nu b l \bar{\nu} \bar{b}
\end{array}
\right.
\label{eq:ttbar_sig}
\nonumber
\end{equation}
with $l=e,\mu$.
The 1.3 (0.4) million events with at least one $l=\tau$
in the semileptonic (dileptonic) channel are considered as background.
Non-$t\bar{t}$ background is composed of QCD background, from which mainly 
$b\bar{b}$ production is relevant for our study, and of electroweak backgrounds, which are
$W$+jets, $Z(\rightarrow ll)$+jets, $Wb\bar{b}$, 2 vector bosons ($ZZ$, $ZW$, $WW$) and single top production. 

\subsection{Event generation}
\label{sec:mc_gene}

The Monte Carlo leading-order generator TopReX~4.05~\cite{TOPREX} is used for the $t\bar{t}$
event generation. It includes the Standard Model LO $t\bar{t}$ spin 
correlation\footnote{NLO spin correlation 
simulations are expected to be included in the future in MCatNLO~\cite{MCATNLO} generator.}.
A top mass of 175~GeV is assumed and the structure function
CTEQ5L~\cite{CTEQ5L} is used. The $Q^2$-scale ($p_T(t)^2 + \mathrm{M}_t^2$) used for 
$\alpha_S$ is the same as for the structure function. 
The proportion of $gg$ and $q\bar{q}$ processes, which directly impacts the 
spin correlation (see Figure~\ref{fig:tt_cross_sect}), is $86\%/14\%$.
Partons are fragmented and hadronized using PYTHIA~6.2~\cite{PYTHIA}, including
initial and final state radiations, as well as multiple interactions, in agreement with CDF data extrapolated to LHC~\cite{MIN_BIAS}.
The $b$-fragmentation is performed using the Peterson parametrization with $\epsilon_b=-0.006$. 
TAUOLA and PHOTOS~\cite{TAUOLA} are used to process the $\tau$-decay and radiative corrections.
All results correspond to one LHC year at low luminosity.
For the systematics study, samples corresponding to three times (ten times) more statistics are generated for each 
source of uncertainties in the semileptonic (dileptonic) channel. \\

For what concerns the non-$t\bar{t}$ background generation in the semileptonic
channel, PYTHIA is used, except for $W$+4~jets and $Wb\bar{b}$ which are treated with 
AlpGen~\cite{ALPGEN} and AcerMC~\cite{ACERMC} generators, respectively. 
About $3\cdot10^{10}$ $W(\rightarrow l\nu)$+4~jets weighted events are generated with cuts 
on the four extra light jets: p$_T>$10~GeV, $|\eta|<2.5$ and 
$\Delta R($jet-jet$)>$0.4\footnote{$\Delta R=\sqrt{\Delta \phi^2 + \Delta \eta^2}$}. 
Despite this huge effort, it only represents 1/63$^{rd}$ 
of the statistics for one LHC year (380\,000 events). 
For $b\bar{b}$ background, given the very high cross-section ($\sim$500~$\mu$b) of the process, the cut 
$\sqrt{\hat{s}}>$120~GeV is applied at the parton level.
750 million of events have been generated, corresponding to 1/8$^{th}$ of the statistics for one year.
Anyway, as this QCD background is very difficult to estimate, it should be extracted from the data.
Except for $W$+4~jets and $b\bar{b}$,
the statistics corresponding to one LHC year at low luminosity
is simulated for each background without any cut at the parton level. In the dileptonic case, 
all backgrounds are simulated using PYTHIA~6.2.\\

\subsection{Detector modeling}
\label{sec:sim_det}

A simplified modeling of the ATLAS detector, ATLFAST 2.6.0~\cite{ATLFAST}, is used. 
This essentially accounts for resolution smearing of objects 
accepted within the detector geometry, according to the 
expected performances~\cite{ATLAS_PHYS_TDR}. Only settings of particular importance are recalled here:
\begin {itemize}
\item Isolation criteria only for lepton (electron and muon) consists in:
$i)$ asking $E_T<10$~GeV in a pointing cone of 0.2 around the lepton and
$ii)$ requiring the nearest calorimeter cluster at $\Delta R > 0.4$.
\item Jets are reconstructed with a cone algorithm, with a size $\Delta R = 0.4$. 
They are calibrated to obtain a correct jet energy scale~\cite{ATLAS_PHYS_TDR}.
\item A 60\% $b$-tagging efficiency is assumed, as well as a $c$-jet
rejection of 10. For the other jets, the rejection is 100. This is a rather pessimistic
assumption compared to the latest simulation results~\cite{B-TAGGING}.
\end{itemize}
No trigger inefficiencies and no detailed acceptance (as crack 
between barrel and endcap) are included in this analysis. 
A 90\% lepton and a 95\% jet reconstruction efficiency are assumed.

\subsection{Event selection and reconstruction}
\label{sec:evt_rec}

The heavy top mass makes the event topology at LHC outstanding:
$t$ and $\bar{t}$ are preferentially produced in the central region ($|\eta|<2.5$) 
and back to back in the transverse plane. 
Therefore, $t$ and $\bar{t}$ are naturally well separated, which will ease the event reconstruction.

\subsubsection{Semileptonic $t\bar{t}$ events}
\label{sec:evt_rec_semilep}

Semileptonic signal events are characterized by one (and only one) isolated lepton, 
at least 4~jets of which 2~are $b$-jets, and missing energy. They are selected by requiring an isolated 
lepton from first and second level trigger. The offline kinematic cut on the lepton is 
directly given by the trigger threshold which is set to p$_T>20$~GeV at $10^{33}$cm$^{-2}$s$^{-1}$.
Moreover, at least four jets with p$_T>30$~GeV are required, among which at least 
two are $b$-tagged. This p$_T$ cut is a good compromise between a low combinatorial background 
and a good statistics. The 20~GeV cut on the missing transverse energy (p$_T^{miss}$) is standard for events with 
one neutrino and rejects almost no signal events. All these kinematic cuts are summarized 
in Table~\ref{tab:sel_cuts}. Their resulting efficiency on signal events is~6.5\%. 
The average p$_T$ of the lepton and least energetic jet in the top rest frame ($lej$) are around 50~GeV.\\
 
After kinematic cuts, the full event topology is reconstructed.
On the top hadronic decay side, the two non $b$-jets with M$_{jj}$ closest and within 20~GeV to the known M$_W$~\cite{PDG}
are selected. 
Then, the $b$-jet with M$_{jjb}$ closest to the known M$_t$~\cite{PDG} is chosen to reconstruct the 
hadronic top and removed from the $b$-jet list. 
On the leptonic side, the missing transverse momentum is used to evaluate the neutrino p$_T$.
Its longitudinal component, p$_z$, is determined by constraining M$_{l\nu}$ to M$_W$:
\vspace*{.5cm}
\begin{equation}
\Big[ E^l+\sqrt{(p^\nu_T)^2+(p^{\nu}_z)^2} \Big] ^2 - (p^l_x+p^\nu_x)^2 - 
(p^l_y+p^\nu_y)^2 - (p^l_z+p^\nu_z)^2  = \mathrm{M}_W ^2
\end{equation}
For 25\% of the events, there is no solution since p$_T^{miss}$ overestimates p$_T^\nu$.
In these cases, p$_T^\nu$ is decreased step by step by 1\% until a solution is reached~\cite{PHD_PR}.
Then, for the 5\% of events with more than two $b$-jets, the one closest to the lepton in $\Delta R$ is chosen to reconstruct the 
leptonic top.\\

\begin{figure}[htbp]
\begin{center}
\rotatebox{0}{\includegraphics[height=10cm,width=14.cm]{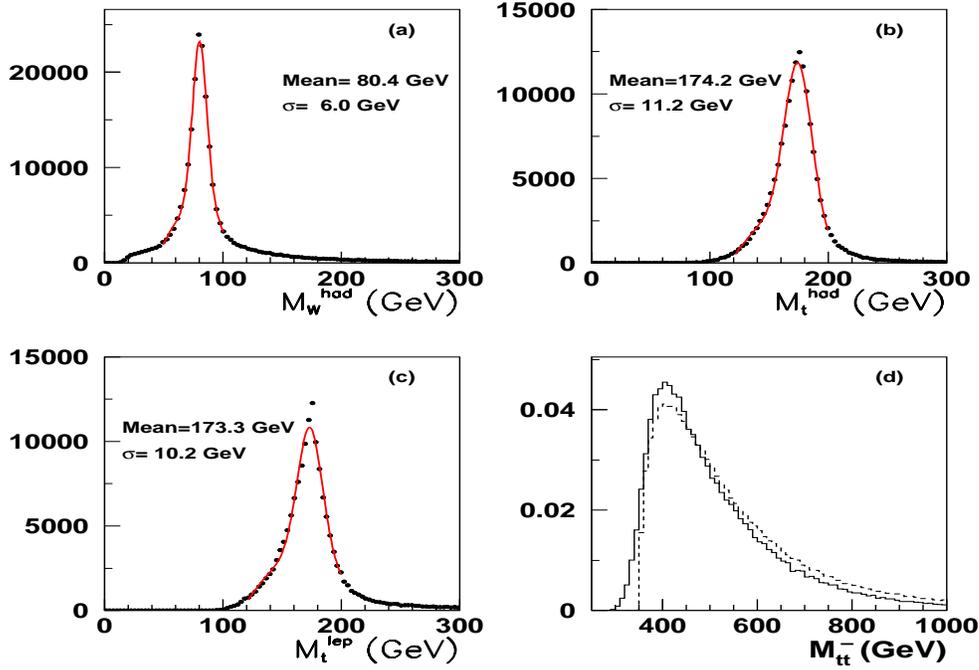}}
\end{center}
\vspace*{-.5cm}
\caption{\it (a-c): Reconstructed masses of $W$ from hadronic side, top from hadronic and leptonic sides
in the semileptonic channel. Lines correspond to Gaussian+third order polynomial fits,
from which Mean and $\sigma$ are extracted. (d): Normalized reconstructed mass distributions of 
the $t\bar{t}$ system are shown both in 
the semileptonic (full) and dileptonic (dashed) channels.}
\label{fig:recmass}
\end{figure}

Figure~\ref{fig:recmass}~(a-c) shows reconstructed $W$ and top masses.
Results are comparable with those of the top mass study~\cite{TT_MASS}.
Quality cuts are then applied on top and anti-top 
reconstructed masses ($|$M$_t^{had}-$M$_t| < $ 35 GeV and $|$M$_t^{lep}-$M$_t| < $ 35 GeV) to reject badly reconstructed 
events. At this stage, 3.3\% of the signal events are kept, 
corresponding to 85000 signal events for one LHC year at low luminosity.
Table~\ref{tab:sel_cuts} lists all the selection cuts.

\begin{table}[htbp]
\begin{center}
\begin{tabular}{|l||c|c|c|}
\hline
   Selection type   &  Variables               &  Cuts      \\
\hline
\hline
                    &  =1 isolated lepton      &  p$_T > $ 20 GeV, $|\eta|<2.5$ \\
  Kinematic and     &  $\ge$ 4 jets            &  p$_T > $ 30 GeV, $|\eta|<2.5$ \\
  acceptance        &  $b$-tagged jets         &  $\ge$ 2       \\
                    &  Missing energy ($\nu$)  &  p$_T^{miss} > $ 20 GeV        \\
\hline
                    &  $|$M$_W^{had}$-M$_W|$ &  $ < $ 20 GeV  \\
 Reconstruction     &  $|$M$_t^{had}$-M$_t|$ &  $ < $ 35 GeV  \\
 quality            &  $|$M$_t^{lep}$-M$_t|$ &  $ < $ 35 GeV  \\
\hline
\end{tabular}
\vspace*{.2cm}
\caption{\it Selection cuts in the semileptonic $t\bar{t}$ channel.} 
\label{tab:sel_cuts}
\end{center}
\end{table}

After selection criteria are applied, the  background is composed for more than 80\% of 
$t\bar{t} \rightarrow \tau + X$ events, as shown in Table~\ref{tab:SB_eff}.
The amplitude and shape of this $t\bar{t}$ background should be easily under control. 
The remaining non-$t\bar{t}$ background is dominated by $W(\rightarrow l\nu)$+4~jets, $b\bar{b}$ and single top events. 
In the first two cases, since only
a few tens of events subsist after all cuts, Poisson statistics is used to give
an estimate of the expected number of events. 
Given its very low contribution to the overall background, 
the non-$t\bar{t}$ background will be neglected in the rest of the analysis.
In total, about 7000 background events are expected for one LHC year at low luminosity,
giving a signal over background ratio of 12.\\

\begin{table}[htbp]
\begin{center}
\begin{tabular}{|l||c||c|c|}
\hline
                    & Initial number of 	& Number of    \\
                    & Events ($\times 10^6$)&  selected events     \\
\hline
\hline
 {\bf Signal} ($t\bar{t}$ semileptonic)      &  2.5              	&  85000      \\
\hline
\hline 
$\mathbf{t\bar{t}}$ {\bf  background} &  		   &  	      \\
 $t\bar{t} \rightarrow \tau+X$        &  1.3       &  6200	  \\
 $t\bar{t} \rightarrow$ all had       &  3.7       &  70 	  \\
\hline
\hline
{\bf Non-}$\mathbf{t\bar{t}}$ {\bf background}  &	      &             \\
 $W(\rightarrow l\nu)$+4~jets                   &  24     &  [400,1000] \\
 $b\bar{b}$ ($\sqrt{\hat{s}} > $120~GeV)        &  6000   &  200        \\
 $Z(\rightarrow ll)$+jets                       &  49     &  12	        \\
 $ZZ,WW,ZW$        				&  1.1    &  5	        \\
 $W(\rightarrow l\nu)b\bar{b}$        		&  0.7    &  3	        \\
 single top       				&  1.0    &  350 	    \\
\hline
\end{tabular}
\vspace*{.5cm}
\caption{\it Number of events in the semileptonic $t\bar{t}$ channel (signal and background)
for one LHC year at low luminosity, 10~fb$^{-1}$, before and after selection cuts.} 
\label{tab:SB_eff}
\end{center}
\end{table}

For the top spin study, to enhance the correlation, a further cut, M$_{t\bar{t}} <$~550 GeV, is applied
on the $t\bar{t}$ reconstructed mass (section~\ref{sec:cd_observables}), 
whose distribution is shown in Figure~\ref{fig:recmass}~(d). 
The total efficiency becomes 2.3\%, corresponding to 60000 signal events for one LHC year at low luminosity.

\subsubsection{Dileptonic $t\bar{t}$ events}
\label{sec:evt_rec_dilep}

Dileptonic events are characterized by two (and only two) opposite charged isolated leptons, 
at least two jets of which two are $b$-jets, and missing energy. 
They are selected by requiring two leptons from first and second level trigger. 
The offline p$_T$ cut on opposite sign leptons is conservatively set to 20~GeV, well above the trigger 
thresholds which are lower or equal to 15~GeV for two leptons. 
Moreover, two $b$-tagged jets with p$_T>20$~GeV are required. 
The 40~GeV cut on p$_T^{miss}$ is standard for events with two neutrinos.
Table~\ref{tab:sel_cuts_dilep} shows the selection cuts in the dileptonic channel.
Their total efficiency on signal events is 6.5\%. \\

\begin{table}[htbp]
\begin{center}
\begin{tabular}{|l||c|c|c|}
\hline
   Selection type   &  Variables             &  Cuts      \\
\hline
\hline
                    &  =2 isolated leptons      &  p$_T > $ 20 GeV, $|\eta|<2.5$ \\
  Kinematic and     &  $\ge$ 2 jets            &  p$_T > $ 20 GeV, $|\eta|<2.5$ \\
  acceptance        &  $b$-tagged jets         &  $=$ 2       \\
                    &  Missing energy ($\nu$)  &  p$_T^{miss} > $ 40 GeV        \\
\hline
\end{tabular}
\vspace*{.2cm}
\caption{\it Selection cuts in the dileptonic $t\bar{t}$ channel.} 
\label{tab:sel_cuts_dilep}
\end{center}
\end{table}

After kinematic cuts, the event topology is reconstructed using the algorithm developed 
in~\cite{DILEP_REC}. The aim of the reconstruction is to obtain the unknown momenta of neutrino 
and anti-neutrino and the association between the two $b$-jets and the $b$ and $\bar{b}$ quarks. To solve 
the set of six non-linear equations coming from the momenta 
and energy conservation, the known M$_t$ and M$_W$ are assumed. The set of equations can have up to  
four solutions for each combination of the association b-jets to $b$ and $\bar{b}$ quarks. The choice 
of the solution is based on the computation of weights from known distribution of transversal 
momenta of $t$, $\bar{t}$ and $\nu$, $\bar{\nu}$. The reconstruction efficiency of this algorithm is $80\%$ 
with the correct solution found in $65\%$ of the cases. Most of the dilution comes from the 
wrong $b$ assignation. After cuts and reconstruction, 5.3\% of 
the signal events are kept, corresponding to 21000 signal events for one LHC year at low luminosity. 
The background is then composed for 90\% of
$t\bar{t} \rightarrow \tau + X$ events, as shown in Table~\ref{tab:SB_eff_dilep}.
In total, less than 4000 background events are expected for one LHC year at low luminosity,
giving a signal over background ratio of 6, two times lower than in the semileptonic channel.\\

As in the semileptonic case, a cut on the $t\bar{t}$ reconstructed mass, M$_{t\bar{t}} <$~550 GeV, 
whose distribution is shown in Figure~\ref{fig:recmass}~(d), dashed lines, is applied to enhance 
the spin correlation. In this case, the total efficiency becomes 3.5\%, 
corresponding to 15000 signal events for one LHC year at low luminosity.\\ 

\vspace{1cm}
\begin{table}[htbp]
\begin{center}
\begin{tabular}{|l||c||c|c|c|}
\hline
	  & Number of 	           & Number of   \\
	  & Events ($\times 10^6$)&  selected events     \\
\hline
\hline
 {\bf Signal} ($t\bar{t}$ dileptonic) &  0.4  &  21000  \\
\hline
\hline 
$\mathbf{t\bar{t}}$ {\bf  background}  &	&  	  \\
 $t\bar{t} \rightarrow \tau+l$         &  0.5   &  3700	  \\
 $t\bar{t} \rightarrow l+ jet$         &  3.8   &   40 	  \\
\hline
\hline
{\bf Non-}$\mathbf{t\bar{t}}$ {\bf background}  &	    &	        \\
 $b\bar{b}$ ($\hat{p_T} > $20~GeV)		&  30000    &  $<$200   \\
 $Z$+jets, $W$+jets, $ZZ,WW,ZW$,$Wb\bar{b}$     &  4500     &  $<$100	\\
 single top 					&  1.0      &   7	\\
\hline
\end{tabular}
\vspace*{.5cm}
\caption{\it Number of events in the dileptonic $t\bar{t}$ channel (signal and background)
for one LHC year at low luminosity, 10~fb$^{-1}$, before and after selection cuts.} 
\label{tab:SB_eff_dilep}
\end{center}
\end{table}

\section{Sensitivity to $W$~boson polarization in $t\bar{t}$ events}
\label{sec:wpola}
In this section the method to extract the $W$ polarization observables is explained 
(section~\ref{sec:wpola_meas_method}), the complete systematics study is presented
(section~\ref{sec:wpola_syst}) and results combining dileptonic and semileptonic channels
are given (section~\ref{sec:wpola_results}). Finally, using these results, the sensitivity to $tWb$ 
anomalous couplings is discussed (\ref{sec:wpola_wtb}).

\subsection{Measurement method}
\label{sec:wpola_meas_method}
\begin{figure}[htbp]
\begin{center}
\rotatebox{0}{\includegraphics*[width=.48\linewidth,bbllx=0pt,bblly=20pt,bburx=400pt,bbury=375pt]{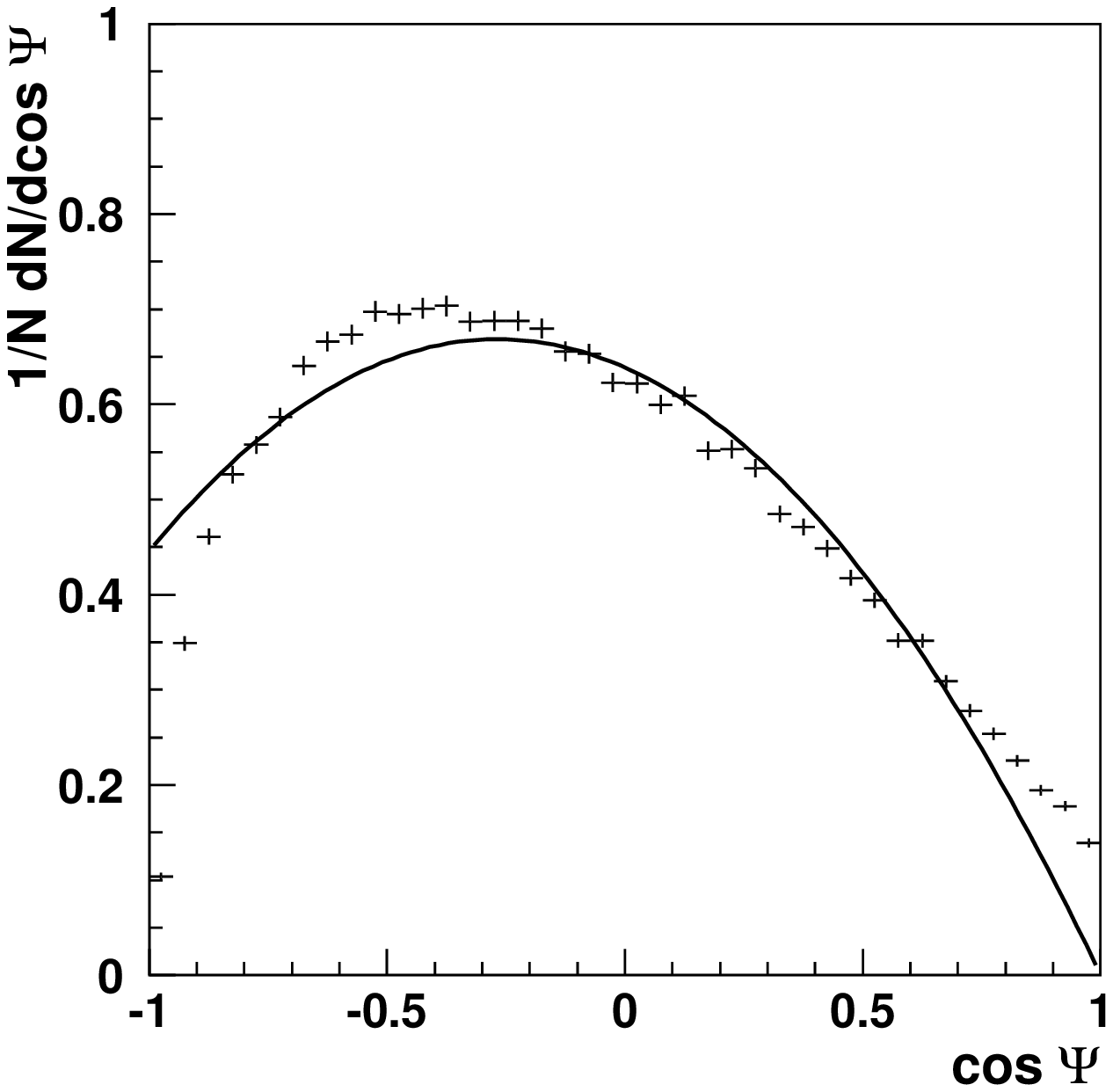}
\includegraphics*[width=.48\linewidth,height=6.8cm,bbllx=0pt,bblly=10pt,bburx=540pt,bbury=415pt]{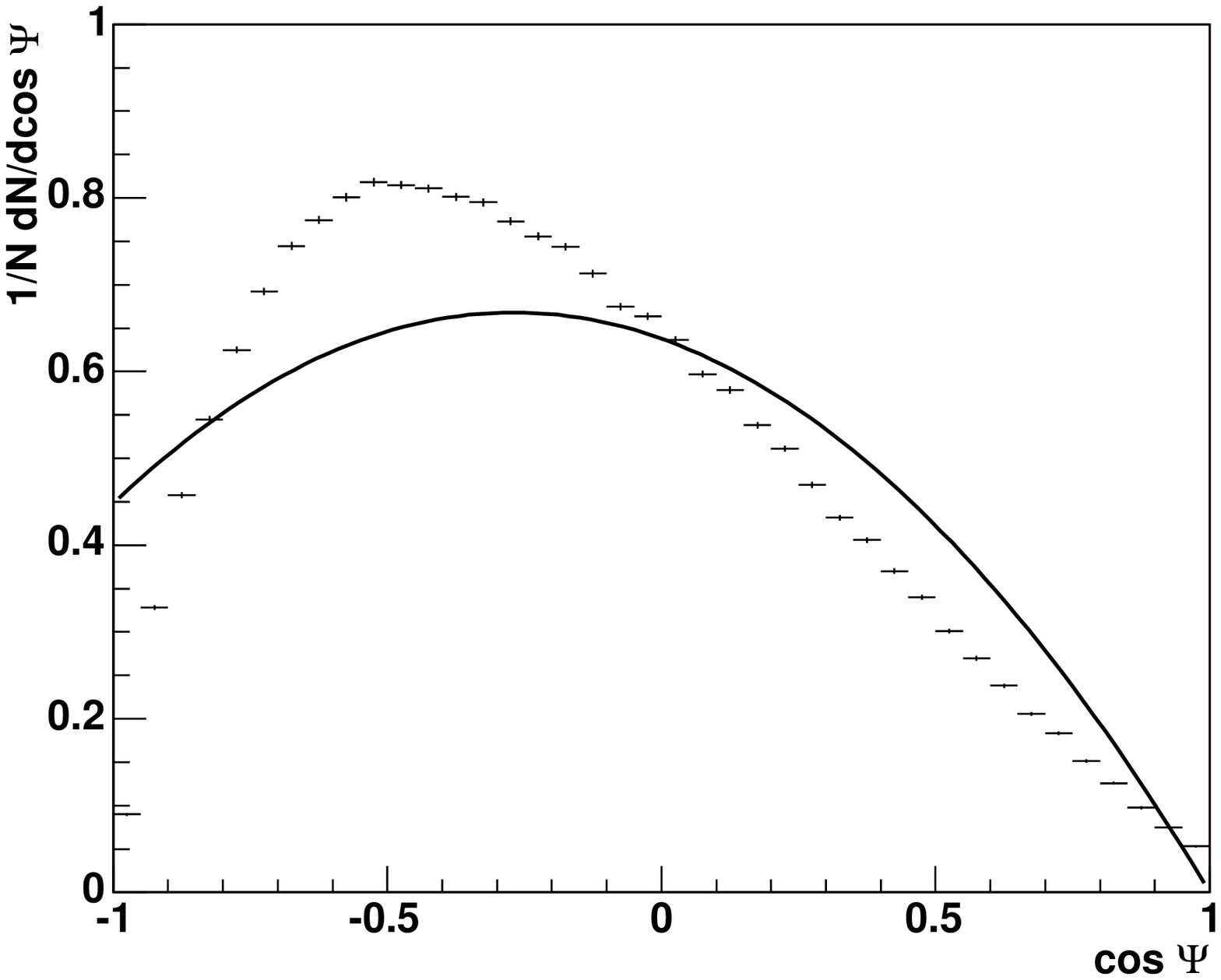}}
\end{center}
\caption{\it Normalized $\cos\Psi$ distribution
    after selection cuts (histogram) for semileptonic (left) and 
    dileptonic (right) $t\bar{t}$ events. For comparison, the SM parton level distribution
    (Figure~\ref{fig:parton}) is superimposed (full line). }
\label{fig:distort}
\end{figure}

\begin{figure}[htbp]
\begin{center}
\rotatebox{0}{\includegraphics*[width=.48\linewidth,bbllx=0pt,bblly=20pt,bburx=400pt,bbury=375pt]{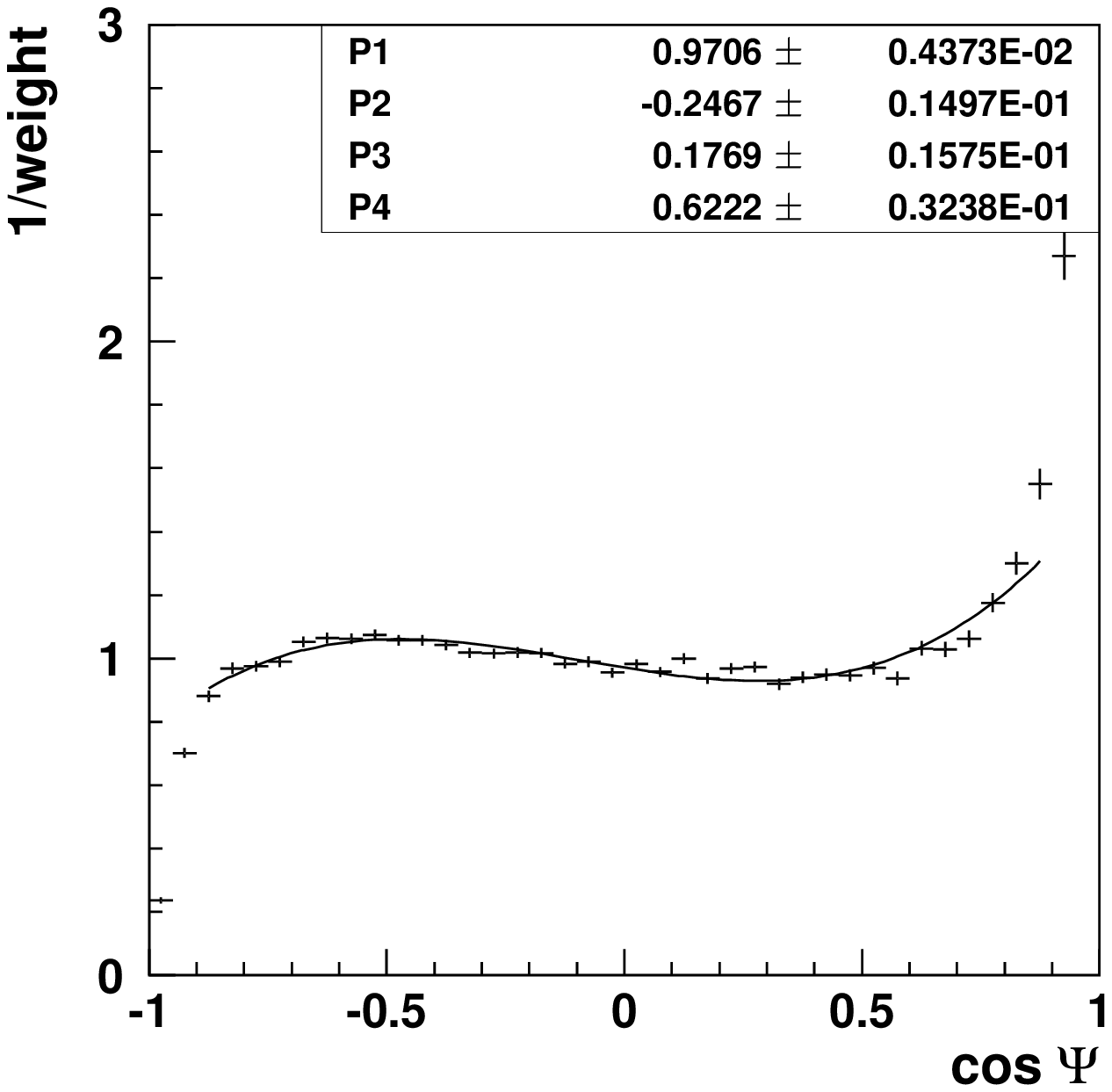}
\includegraphics*[width=.48\linewidth,height=6.8cm,bbllx=0pt,bblly=10pt,bburx=540pt,bbury=415pt]{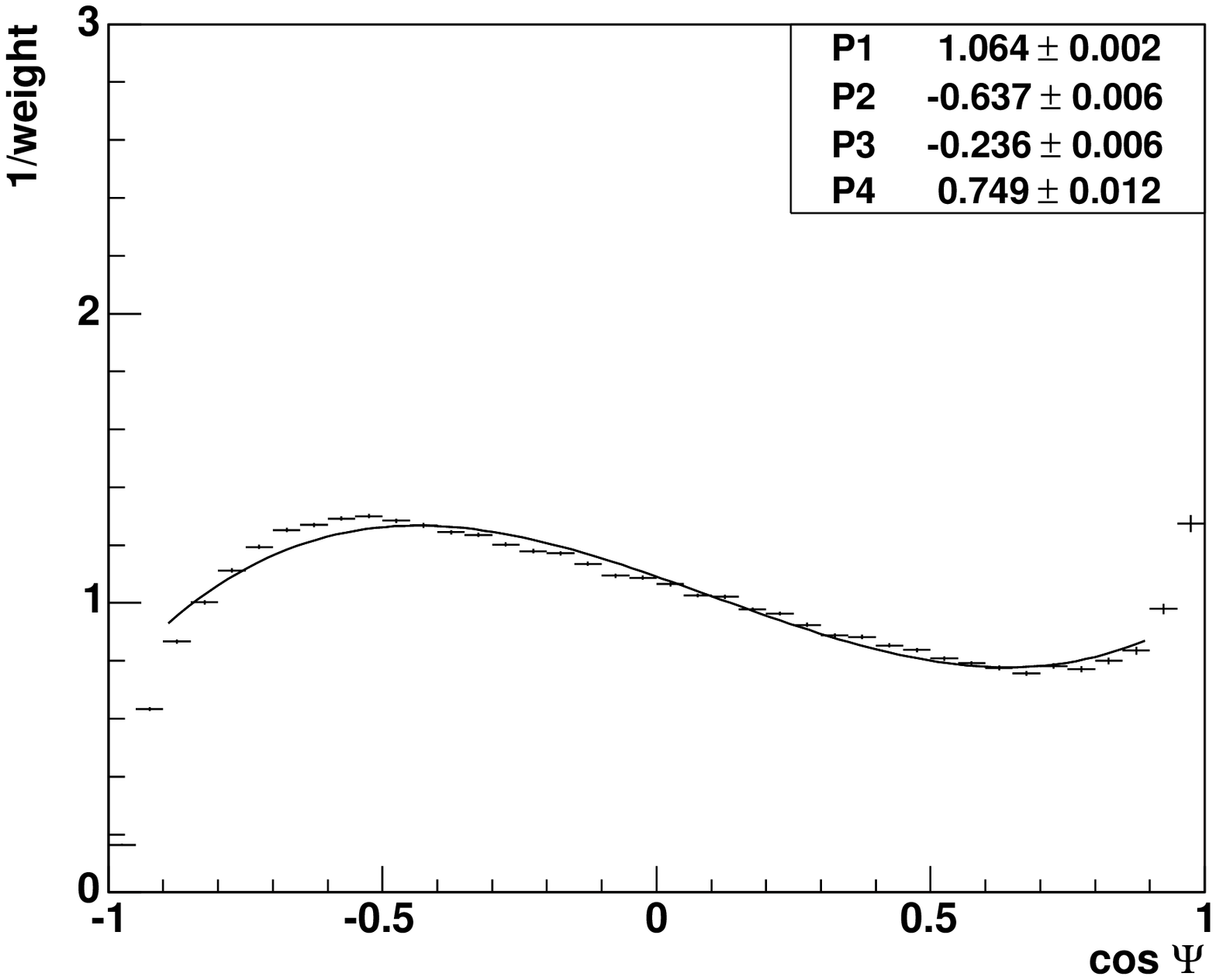}}
\end{center}
\caption{\it Ratio between the two normalized distributions of $\cos\Psi$ 
    shown in Figure~\ref{fig:distort}, i.e. after selection cuts and at parton level, 
    for semileptonic (left) and dileptonic (right) $t\bar{t}$ events.
    The full line is the result of a third order polynomial fit in the range $-0.9 < \cos\Psi < 0.9$.}
\label{fig:correction}
\end{figure}

As charged leptons from longitudinal $W$ have a harder p$_T$ spectrum than those from left-handed~$W$, 
they are more likely to pass the trigger threshold and offline selection requirements. 
More generally, the reconstructed $\cos\Psi$ angular distribution is distorted by all the acceptance and reconstruction
effects compared to the parton level one, as shown in Figure~\ref{fig:distort}.
This leads to a bias in the measurement, which is more pronounced as the cut on the lepton p$_T$ increases. 
To correct for it, a weight is applied on an event by event basis, allowing
to recover, as much as possible, the original shape.
The weighting function is obtained from the ratio between the two normalized distributions of 
Figure~\ref{fig:distort} (i.e. after selection cuts and at parton level) for semileptonic and dileptonic
events.
In both channels, this ratio, shown in Figure~\ref{fig:correction}, is fitted by a 
third order polynomial function to extract a smooth correction. 
The fit is restricted to the region $-0.9 < \cos\Psi < 0.9$, which is 
the most extended region where the correction is varying slowly.
The correction functions, computed on an independent data sample, are then applied event by event on
the analysis samples.\\

\begin{figure}[htbp]
\begin{center}
\rotatebox{0}{\includegraphics*[width=.48\linewidth,bbllx=0pt,bblly=20pt,bburx=400pt,bbury=375pt]{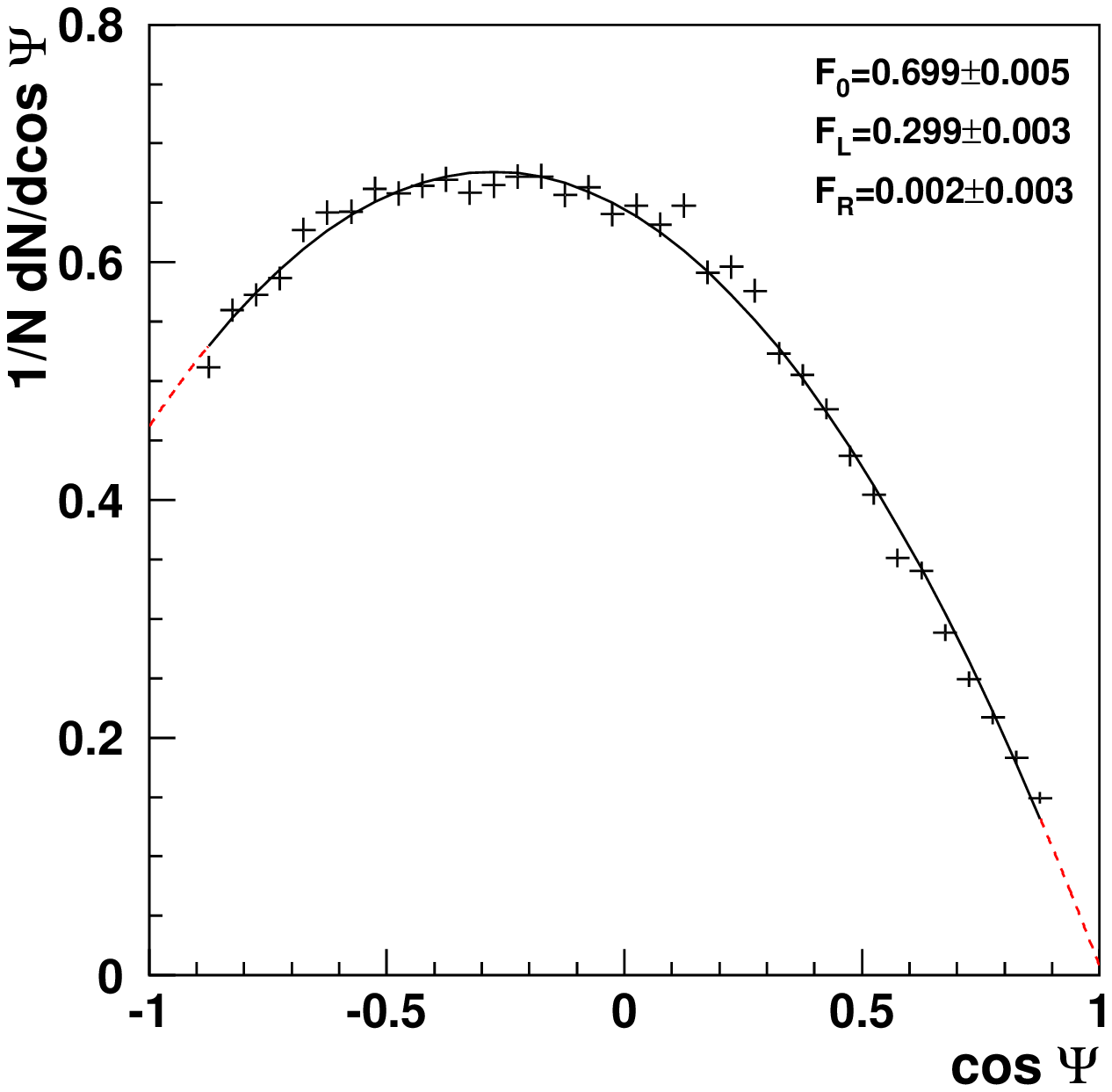}
\includegraphics*[width=.48\linewidth,height=6.8cm,bbllx=0pt,bblly=10pt,bburx=540pt,bbury=415pt]{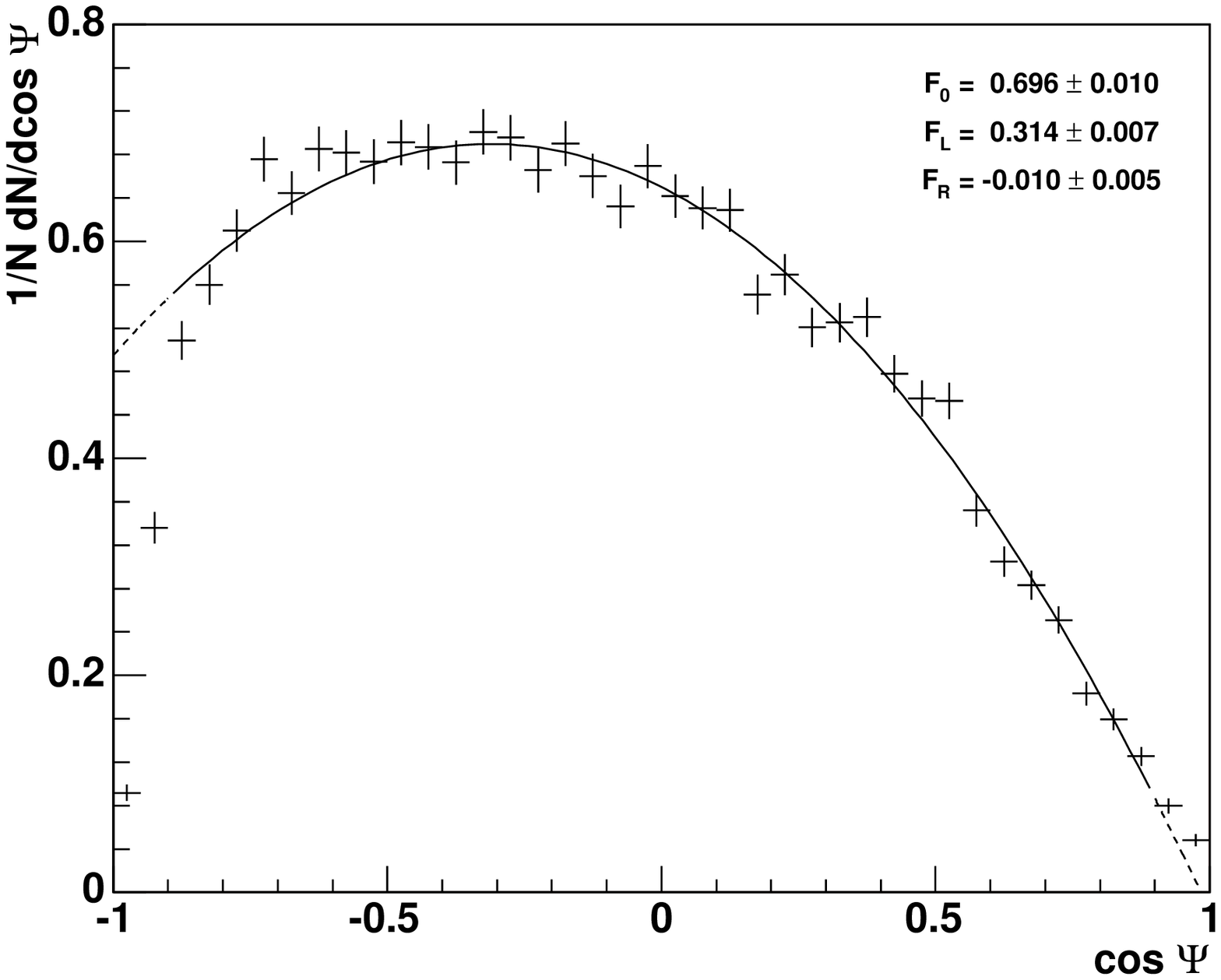}}
\end{center}
\vspace*{-.5cm}
\caption{\it Normalized $\cos\Psi$ distribution after selection cuts and correction 
    for semileptonic (left) and dileptonic (right) $t\bar{t}$ events after one year at low luminosity, 10~fb$^{-1}$. 
    The full line is the result of a fit with function of Equation~(\ref{eq:cosphi}) in the range 
    $-0.9 < \cos\Psi < 0.9$, with $F_0+F_L+F_R=1$. The dashed 
    line represents the continuation of the function outside the fit region.}
\label{fig:recons}
\end{figure}

The final distribution after event selection and correction is shown in 
Figure~\ref{fig:recons} for semileptonic and dileptonic events, corresponding
to one year at low luminosity, 10~fb$^{-1}$. 
The $W$ polarization is extracted from a fit in the restricted region with the Equation~(\ref{eq:cosphi}) function
and the constraint $F_0+F_L+F_R=1$.
The results for $F_0$, $F_L$ and $F_R$ are compatible with their SM expectations.
The statistical errors are 0.005 for $F_0$ and 0.003 for $F_L$ and $F_R$ in the semileptonic case 
and increase to 0.010, 0.007 and 0.005 in the dileptonic case.
The correlation between the parameters for the couples ($F_0$,$F_L$), ($F_0$,$F_R$) and ($F_L$,$F_R$) 
are -0.9, -0.8 and 0.4.\\

The systematic uncertainty induced by the weighting method has been estimated 
by varying the number of bins (from 40 to 25), 
the fit limits (from [-0.9;0.9] to [-0.8;0.8]) and the polynomial 
order (from P3 to P5 and P7). All variations are below the statistical error.
Therefore, the total uncertainty of the method is estimated to be smaller than the statistical error. \\

\begin{figure}[htbp]
\unitlength 1mm 
\noindent
  \begin{minipage}[htbp]{0.48\linewidth}
    \begin{center}
    \epsfig{file=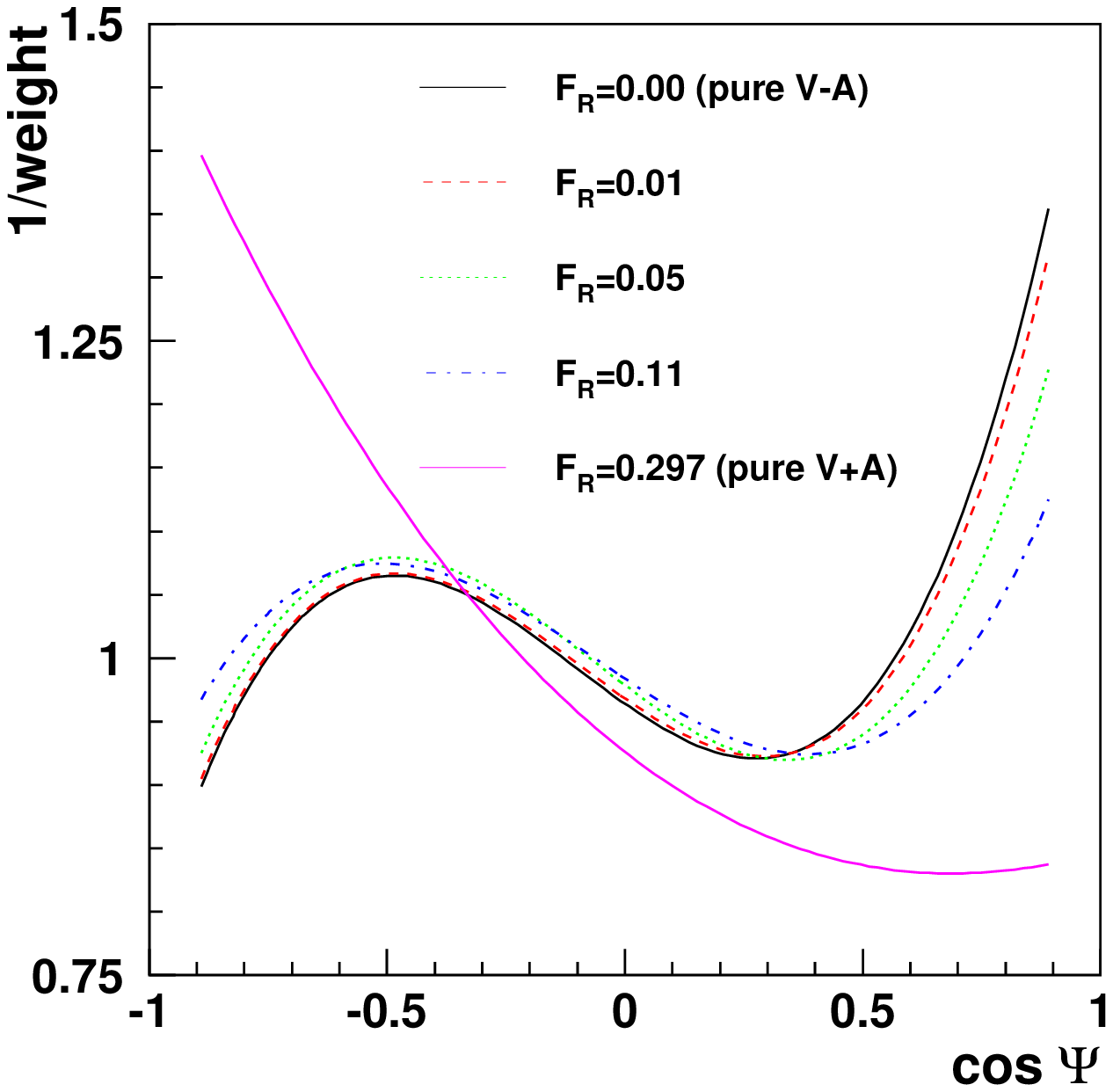, width=\linewidth,height=7cm,bbllx=10pt,bblly=20pt,bburx=390pt,bbury=375pt} 
    \caption{\it Correction functions for different $F_R$ input values in the semileptonic channel. 
    The case $F_R=0$ corresponds to the SM, and the case $F_R=0.297$ corresponds to a pure V+A model.
    The SM function (pure V-A) is the one of Figure~\ref{fig:correction}.}
    \label{fig:bias}
    \end{center}
  \end{minipage}
  \hfill
  \begin{minipage}[htbp]{0.48\linewidth}
    \begin{center}
    \epsfig{file=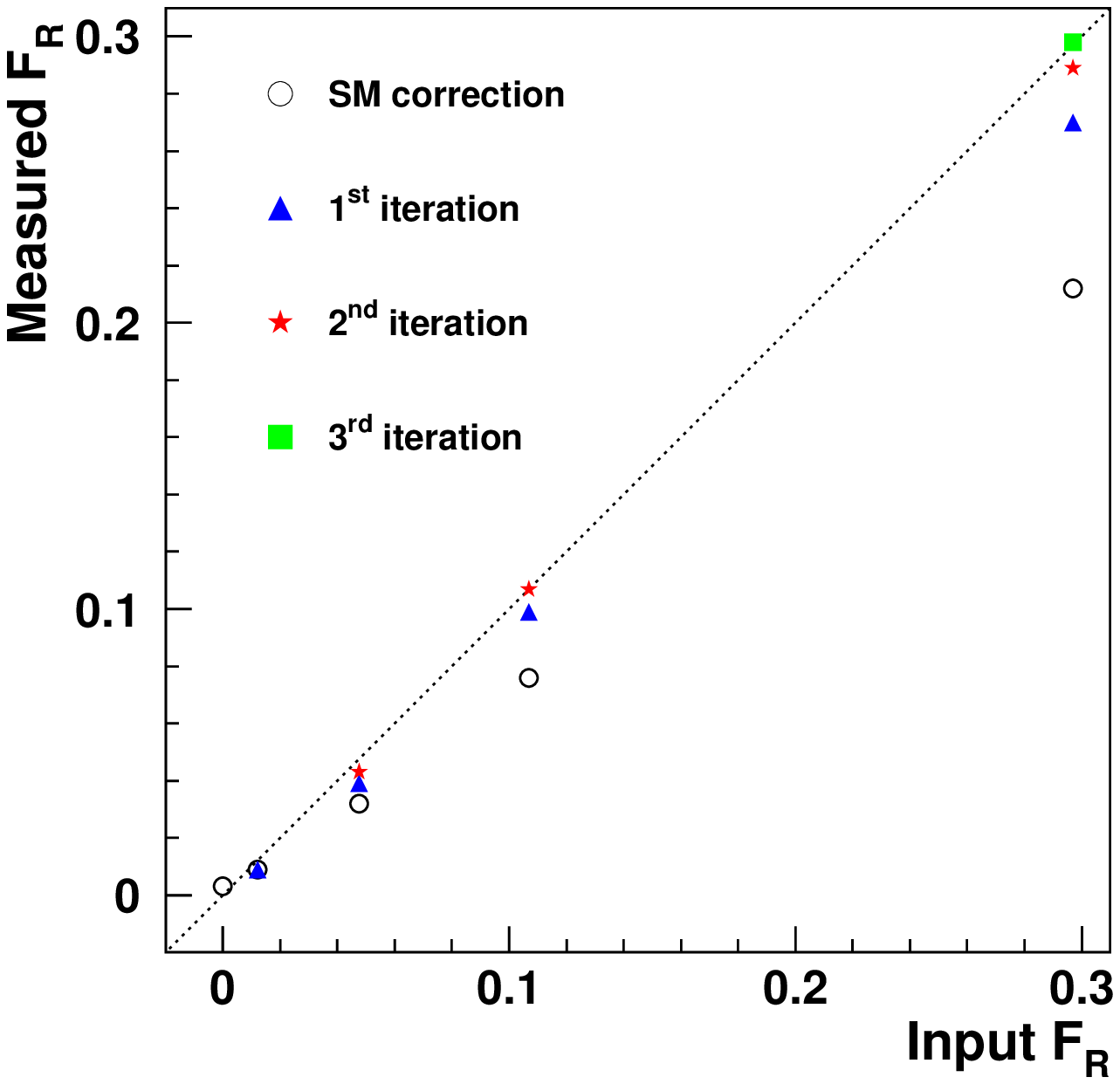, width=\linewidth,height=7cm,bbllx=10pt,bblly=20pt,bburx=390pt,bbury=375pt} 
    \caption{\it $F_R$ extracted from the fit as a function of $F_R$ input values.
    The open circles correspond to the measurement using the SM correction.
    Full triangles, stars and squares correspond to the first, second and third iteration, 
    respectively (see text). The dotted line is $y=x$.}
    \label{fig:iteration}
    \end{center}
  \end{minipage}
\end{figure}

The corrections functions of Figure~\ref{fig:correction} are extracted with a Standard 
Model scenario, assuming a pure V-A top decay vertex.
In case of deviation from the SM, the kinematic 
distributions, such as lepton p$_T$ or angles can be affected.
This is for example the case if a V+A component is present. In such a scenario, the fraction of 
longitudinal $W$~bosons will be unchanged, but a right component $F_R$, whose lepton spectrum is harder 
(see section~\ref{sec:wpola_observables}) will appear.
As a consequence, the correction function will be changed. This is illustrated in 
Figure~\ref{fig:bias} for different $F_R$ input values. In each case, the statistics of one 
LHC year of semileptonic events has been generated with AlpGen. Applying the SM correction function to these samples 
will therefore not correct completely for the bias induced by the selection cuts.
Figure~\ref{fig:iteration} shows the fraction $F_R$ extracted from the 
fit\footnote{In this case, $F_0$ is fixed to its SM value and the only fitted parameter is $F_R$.} 
as a function of the $F_R$ input value after applying the SM correction function (open circles).
The measurement is clearly biased. To overcome this problem we proceed iteratively. 
The SM correction function is first used. Then, in case of deviation of $F_R$ from zero, a new 
correction function is calculated with this new $F_R$ component, and applied. 
The process converges after a few iterations, as shown in Figure~\ref{fig:iteration}.

\subsection{Systematic uncertainties}
\label{sec:wpola_syst}

This section presents a detailed study of the systematic uncertainties related to
the $W$ polarization measurement. 

\subsubsection {Systematic uncertainties at generation level}
\label{sec:wpola_syst_mc}

In this subsection, five main sources of systematic uncertainties are considered:

\begin{itemize}

\item \underline{$Q$-scale}:
The uncertainty related to the $Q$-scale at parton generation is estimated by comparing samples 
generated with TopReX and AlpGen using default $Q^2$-scale: $p_T(t)^2 + \mathrm{M}_t^2$ and $\mathrm{M}_t^2$. 
In both cases, the same hadronization scheme (PYTHIA) is used.

\item \underline{Structure function}: The impact of the structure function 
is estimated by the maximum difference between the measurements obtained
with the standard parton density function (CTEQ5L) and three other ones, CTEQ6L~\cite{CTEQ6L}, MRST2002~\cite{MRST2002}
and GRV98~\cite{GRV98}.
It is conservatively estimated to be independent of the above $Q$-scale uncertainty.

\item \underline{ISR, FSR}: The presence of initial state radiations (ISR)
from incoming partons and especially final state radiations (FSR) 
can affect the $\Psi$ angle reconstruction since it impacts the top quark reconstruction. 
To estimate the effect due to ISR, the difference between the measurements
obtained with ISR switched on (usual data set) and off is computed. 
The same approach is used for FSR. The level of knowledge of ISR and FSR is around~10\%. 
Therefore, as a more conservative estimate, the systematics uncertainties 
have been taken to be 20\% of the corresponding differences.
It should be noted that more sophisticated methods exist to make this evaluation~\cite{TOP_ISRFSR}.

\item \underline{$b$-fragmentation}: The $b$-quark fragmentation 
is performed according to the Peterson 
para\-metrization\footnote{It has to be noticed that recent measurements can not be well fitted 
with the Peterson parametrization~\cite{benhaim}.}, with one free parameter $\epsilon_b$.
The default value is set to $\epsilon_b=-0.006$.
It has been changed to a more recent LEP value ($\epsilon_b=-0.0035$~\cite{abbaneo}), and
the differences on the results are taken as systematic uncertainties.

\item \underline{Hadronization scheme}: The angular distributions of jets and leptons
may be influenced by the hadronization scheme. 
Generating partons with AcerMC then processing the hadronization with PYTHIA or HERWIG~\cite{HERWIG} 
leads to different $W$ polarization measurements.
The related systematic uncertainty is extracted from this difference.

\end{itemize}

\subsubsection {Systematic uncertainties at reconstruction level}
\label{sec:wpola_syst_ana}

In this subsection, three main sources of systematic uncertainties are considered:

\begin{itemize}

\item \underline{$b$-tagging}: The impact of the $b$-tagging efficiency
is studied by increasing it from 50\% to 70\% by steps of 5\%, 
according to a parametrization coming from full simulation~\cite{B-TAGGING}.
Increasing the $b$-tagging efficiency degrades the $c$-jets and light jets rejection factors.
As an example, going from 55\% to 60\% decreases them respectively by 30\% and 80\%.
Figure~\ref{fig:wpola_syst}~(a) shows $F_L$, $F_0$ and $F_R$ measurements as a function of the 
$b$-tagging efficiency. Small and smooth dependences are observed.
The related error is computed with a realistic $\pm5$\% uncertainty on the $b$-tagging efficiency.

\item \underline{$b$-jet miscalibration}: The impact of the knowledge of the absolute $b$-jet 
energy scale is estimated by miscalibrating the reconstructed $b$-jet energy. 
Results are shown in Figure~\ref{fig:wpola_syst}~(b), for a miscalibration between $\pm5$\% by steps of 2\%.
The behaviors can be easily understood: a positive miscalibration overestimates 
the invariant mass of the lepton and the $b$-quark, M$_{lb}$ and therefore $\cos \Psi$, Equation~(\ref{eq:mlb}).
This bias the polarization toward higher values 
($F_L$ decreases and $F_R$ increases). The corresponding systematic is computed 
with a realistic $\pm3$\% uncertainty on the $b$-jet energy scale. 

\begin{figure}[htbp]
\begin{center}
\vspace*{-2cm}
\hspace*{1cm}\rotatebox{0}{\includegraphics[height=5.cm,width=.95\linewidth]{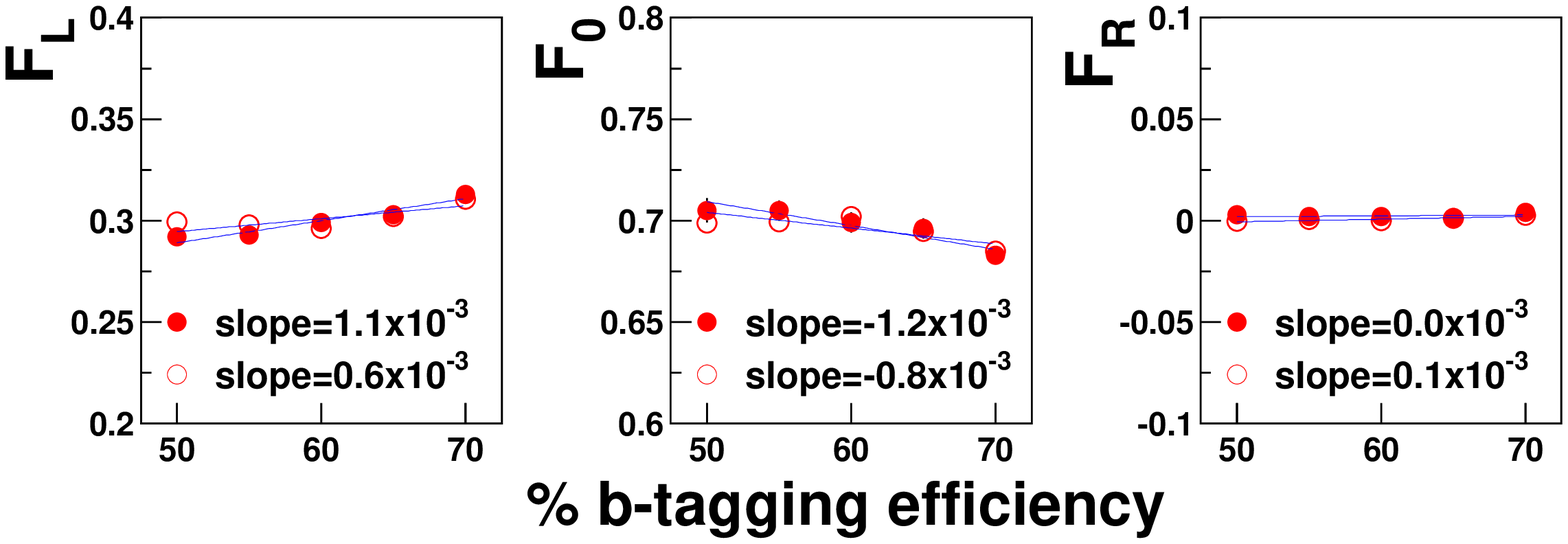}}
\vspace*{-.2cm}
\hspace*{1cm}\rotatebox{0}{\includegraphics[height=5.cm,width=.95\linewidth]{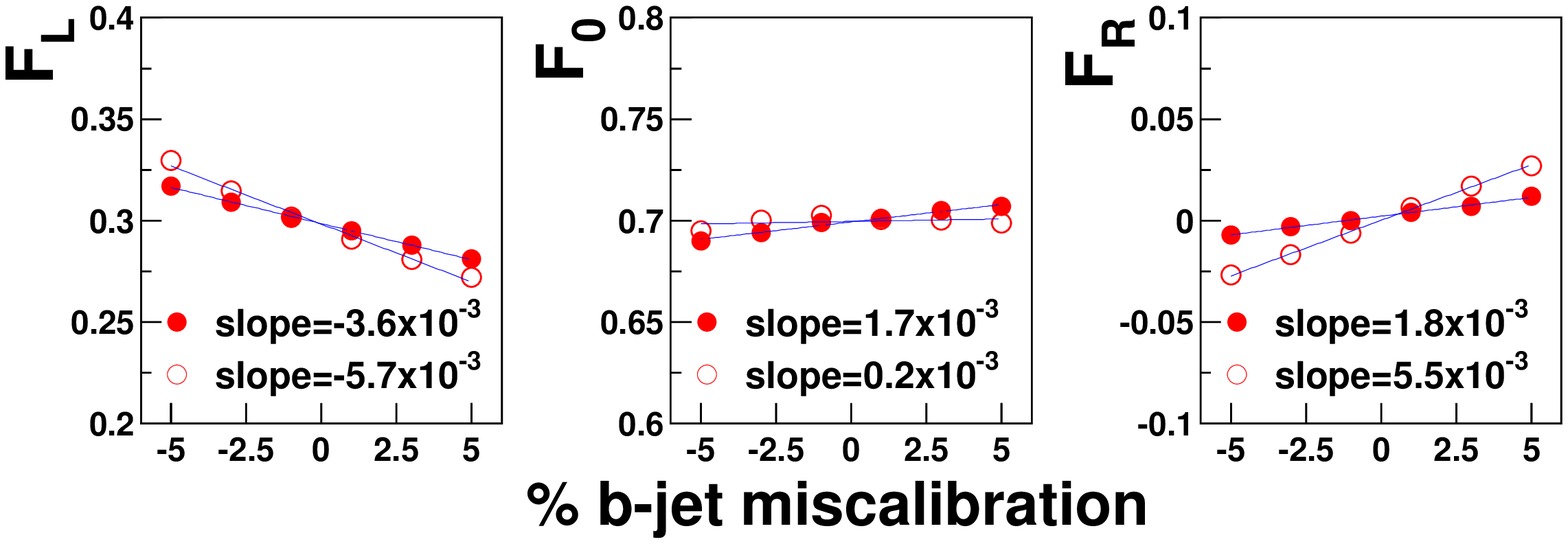}}
\vspace*{-.2cm}
\hspace*{1cm}\rotatebox{0}{\includegraphics[height=5.cm,width=.95\linewidth]{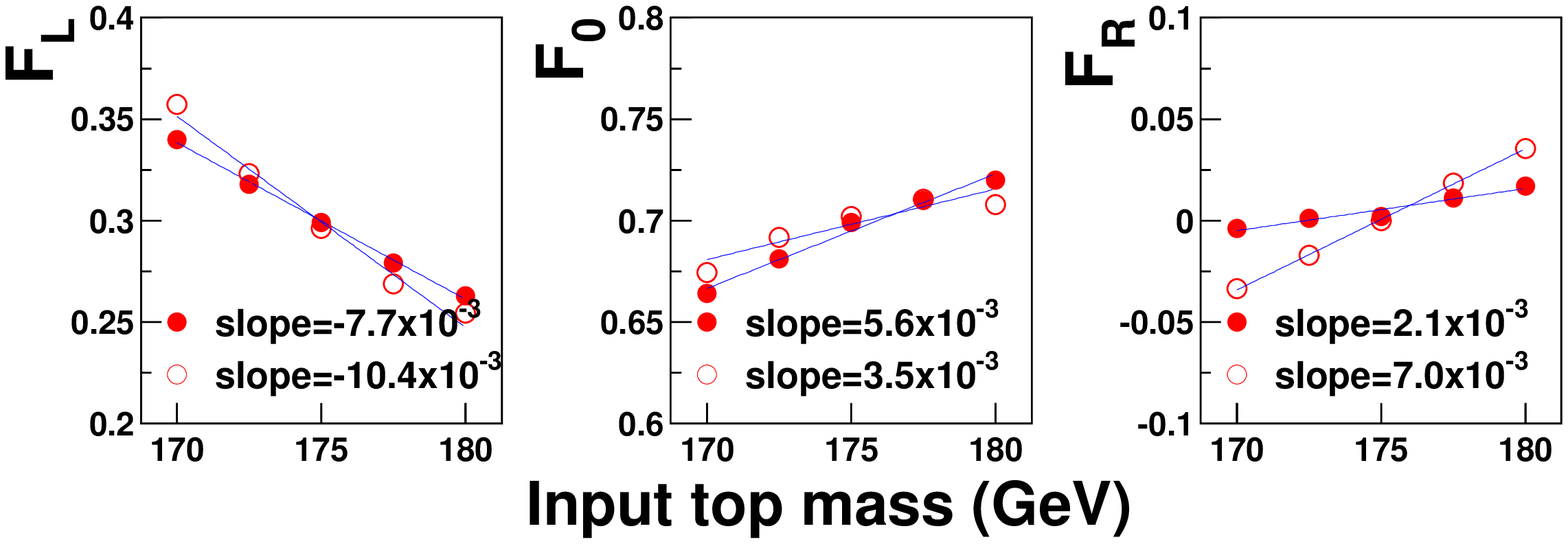}}
\vspace*{-.2cm}
\hspace*{1cm}\rotatebox{0}{\includegraphics[height=5.cm,width=.95\linewidth]{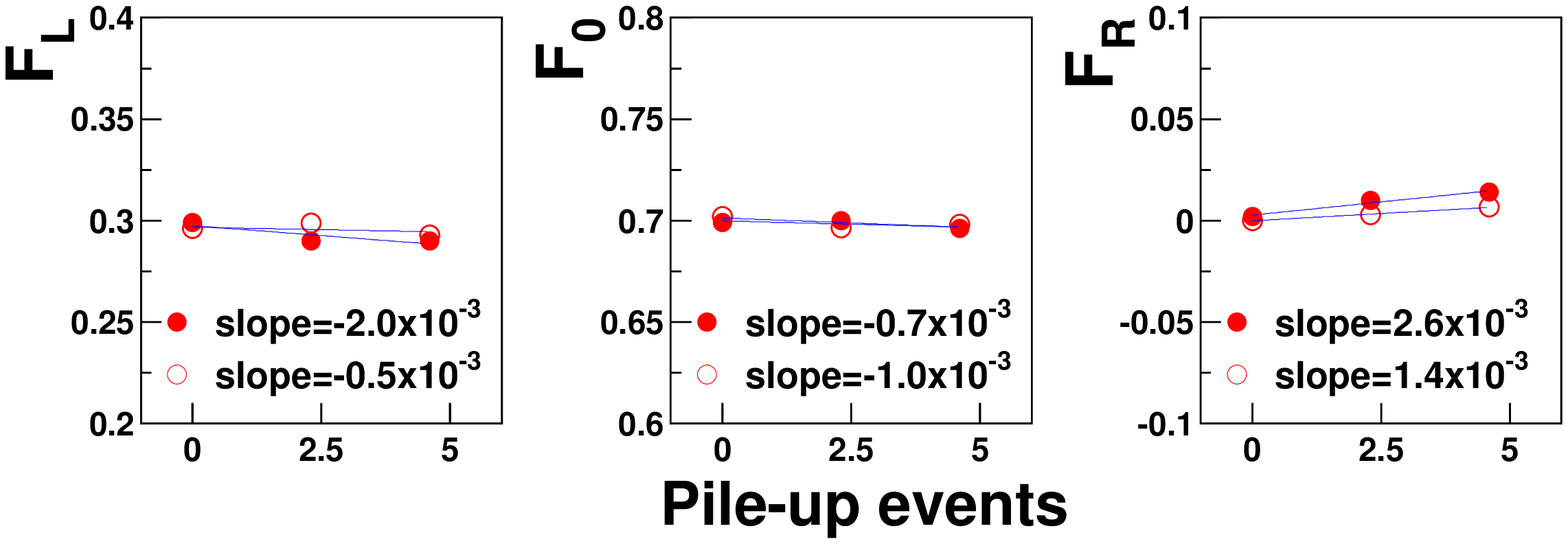}}
\end{center}
\begin{picture}(100,80)
\put(10,570){(a)}
\put(10,430){(b)}
\put(10,290){(c)}
\put(10,150){(d)}
\end{picture}
\vspace*{-3cm}
\caption{\it Measured $F_L$ (left), $F_0$ (middle) and $F_R$ (right) in the semileptonic (black circles) and dileptonic (open circles) 
$t\bar{t}$ channels
as a function of different parameters, see text for more details. 
Linear fits are superimposed in each case, and the corresponding slope is indicated.}
\label{fig:wpola_syst}
\end{figure}

\item \underline{Input top mass}: The SM $W$ polarization has a small dependence on
the top mass, with an increase (decrease) of $F_0$ ($F_L$) by 0.002 per GeV, Equation~(\ref{eq:fractions}).
Moreover, a 175~GeV top mass value is assumed in the event reconstruction, which
can impact the $\cos\Psi$ measurement if the real top mass is different.
Therefore, different samples of events were generated with a top mass between 170 and 180~GeV
by steps of 2.5~GeV. Results are shown in Figure~\ref{fig:wpola_syst}~(c). 
As for the case of a positive $b$-jet miscalibration, a high top mass increases M$_{lb}$, 
and therefore bias the polarization toward higher values.
The related systematics are -0.008 (-0.010) per GeV on $F_L$, 0.006 (0.003) on $F_0$ and 0.002 
(0.007) on $F_R$ in the semileptonic (dileptonic) channel.
The final uncertainty is computed assuming $\Delta$M$_{\mathrm{top}}$=2~GeV, which should be 
reached at Tevatron run~II~\cite{tev:mass}.

\end{itemize}
\vspace{1cm}

\subsubsection {Other sources of systematic uncertainties}
\label{sec:wpola_syst_other}

\begin{itemize}

\item \underline{Background}: As shown in section~\ref{sec:evt_rec}, the only
sizable background comes from $t\bar{t} \rightarrow \tau+X$ events, and is well under control.
A large variation of this background level by $\pm10$\% has a negligeable impact on the results
when considering the semileptonic channel because of the high signal over background ratio. In 
the dileptonic channel, this results in an uncertainty of 0.004 on $F_L$, 0.003 on $F_0$ and 0.001 on $F_R$.

\item \underline{Pile-up}: Pile-up events may influence the reconstruction and therefore impact 
on $W$ polarization measurement. 
They are generated with PYTHIA setting MSTP(131)=1 and using the default process definition MSTP(132)=4. 
Different samples were generated adding~2.3 or~4.6 pile-up events 
according to the Poisson law. These numbers are expected for a luminosity of 
10$^{33}$cm$^{-2}$s$^{-1}$ and 2$\times$10$^{33}$cm$^{-2}$s$^{-1}$,
respectively. Results are shown in Figure~\ref{fig:wpola_syst}~(d). 
It can be noticed that these results are conservative, as no jet recalibration is applied.
\end{itemize}

\subsubsection{Systematics summary}
\label{sec:wpola_syst_res}

\begin{figure}[htbp]
\begin{center}
\begin{tabular}{lccr}
\epsfig{file=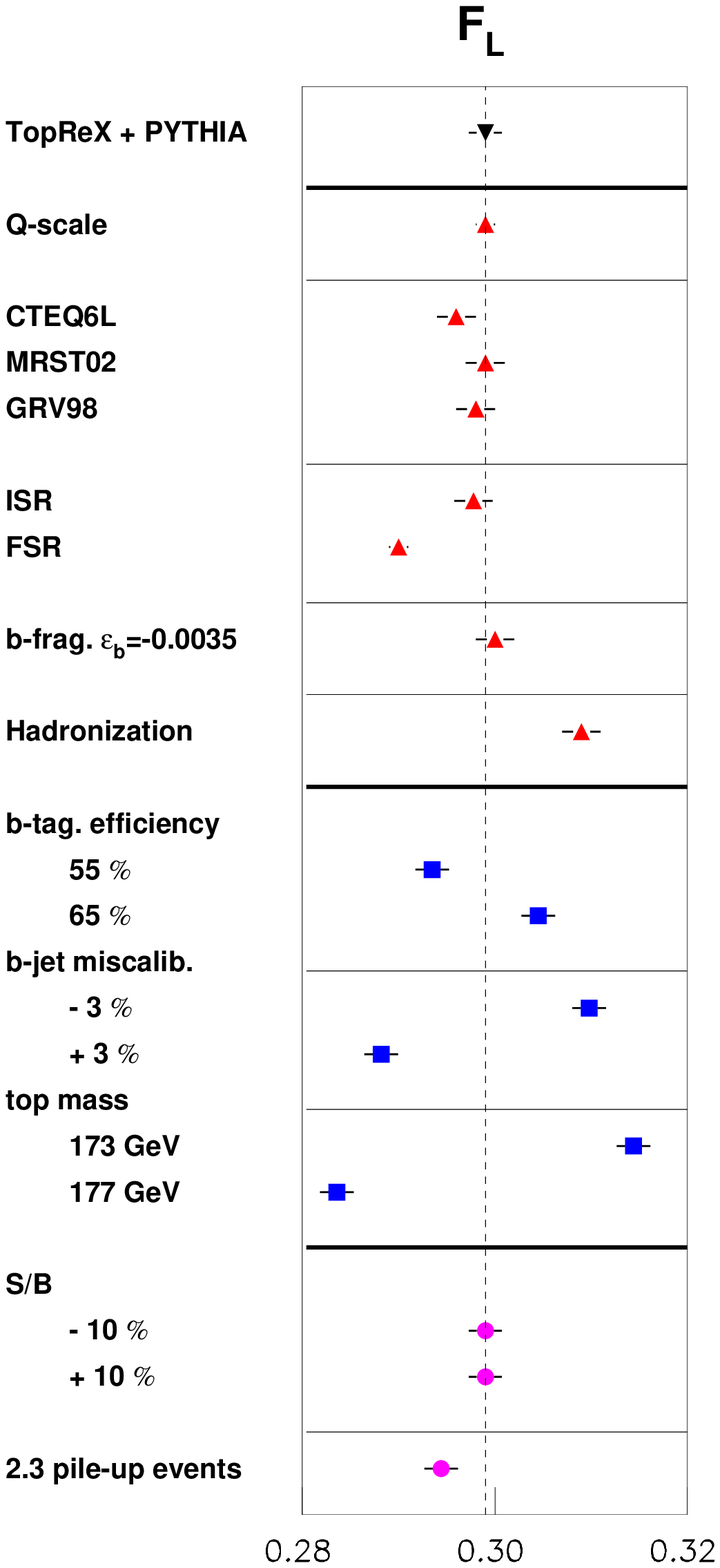,height=18cm,width=5.cm,bbllx=20pt,bblly=0pt,bburx=300pt,bbury=709pt} &
\epsfig{file=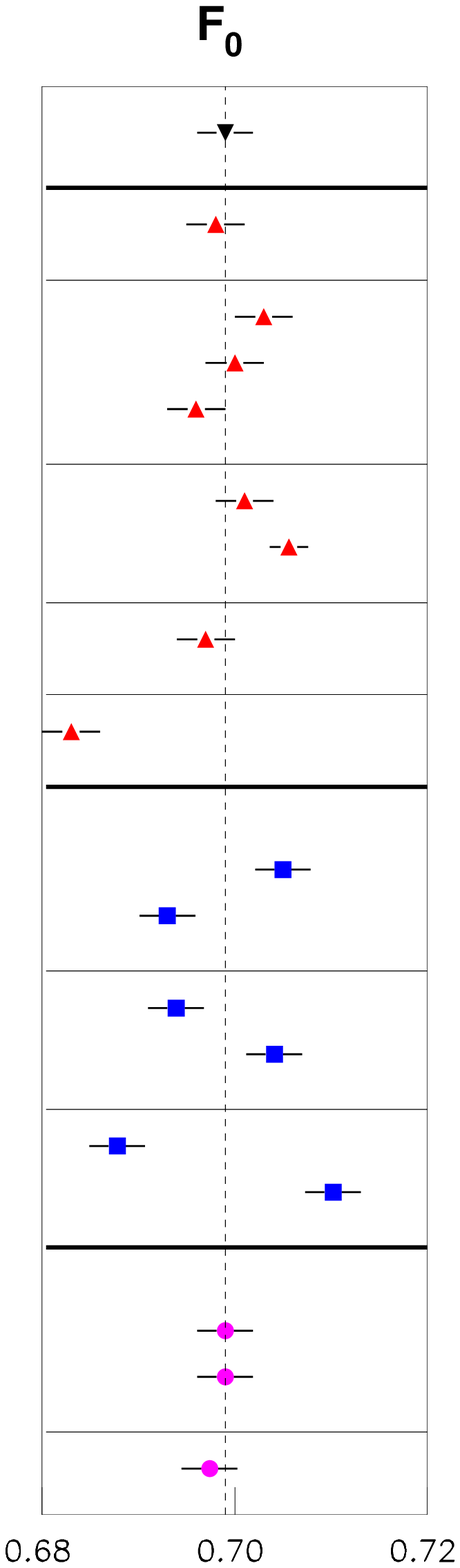,height=18cm,width=3.2cm,bbllx=119pt,bblly=0pt,bburx=300pt,bbury=709pt} &
\epsfig{file=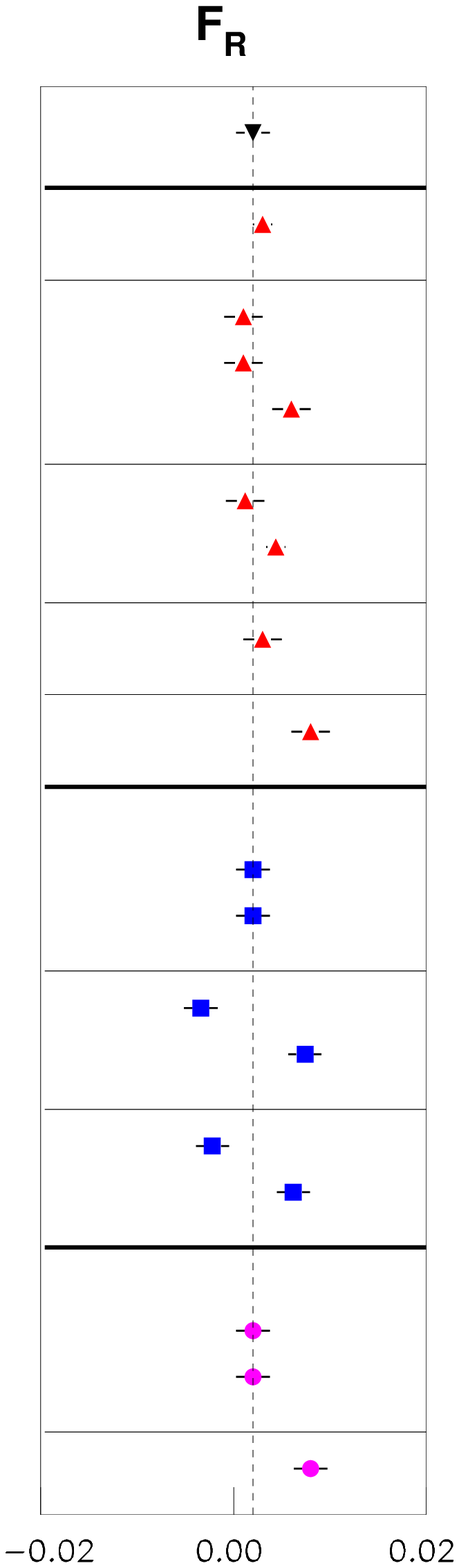,height=18cm,width=3.2cm,bbllx=119pt,bblly=0pt,bburx=300pt,bbury=709pt} &
\end{tabular}
\end{center}
\vspace*{-1cm}
\caption{\it Systematic uncertainties on $F_L$, $F_0$ and $F_R$ in the semileptonic $t\bar{t}$ channel.}
\label{fig:wpola_syst_all_semilep}
\end{figure}

\begin{figure}[htbp]
\begin{center}
\begin{tabular}{lccr}
\epsfig{file=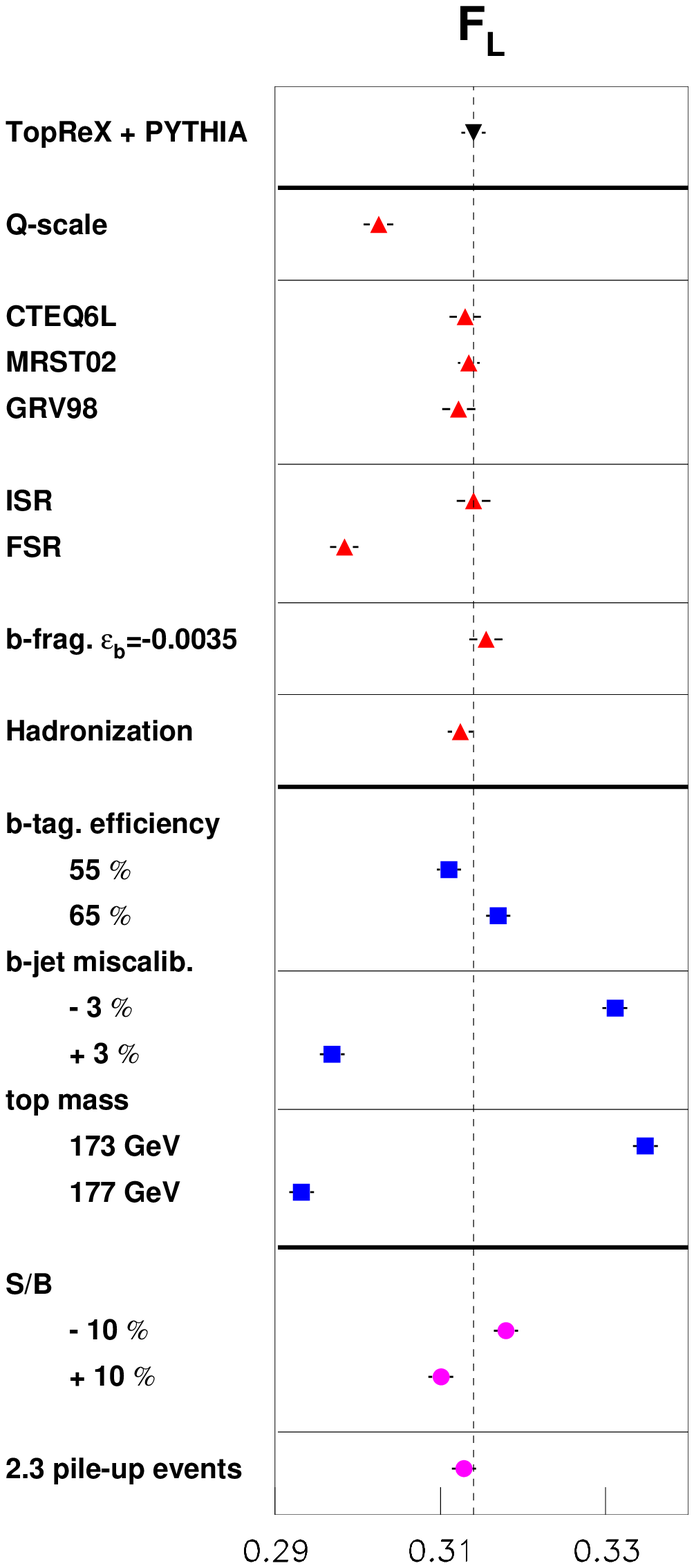,height=18cm,width=5.cm,bbllx=20pt,bblly=0pt,bburx=300pt,bbury=709pt} &
\epsfig{file=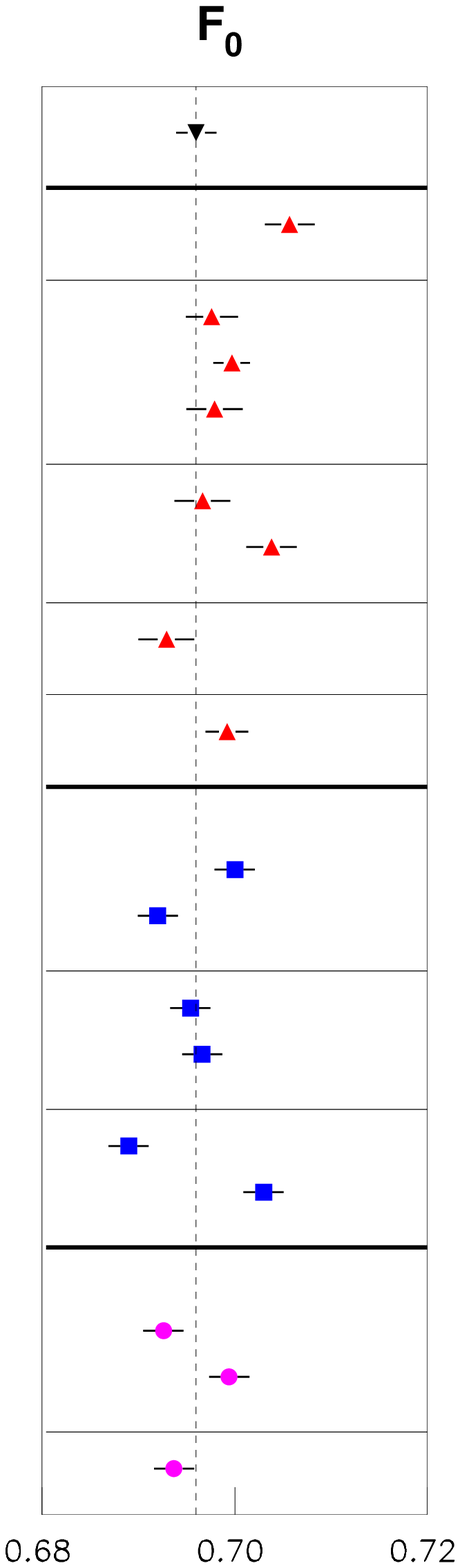,height=18cm,width=3.2cm,bbllx=119pt,bblly=0pt,bburx=300pt,bbury=709pt} &
\epsfig{file=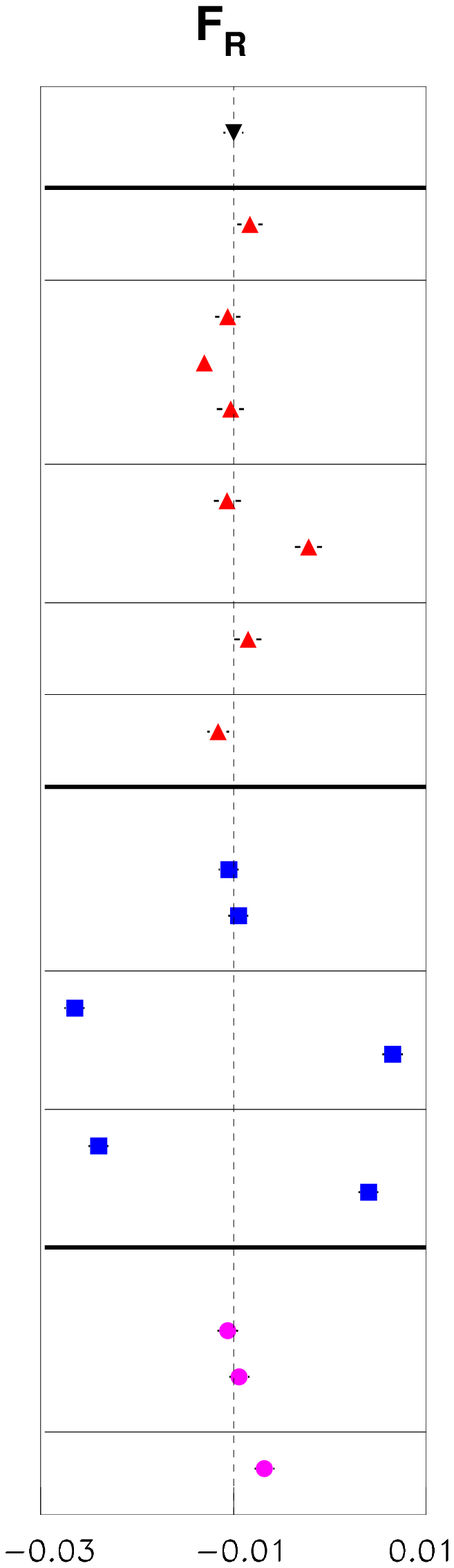,height=18cm,width=3.2cm,bbllx=119pt,bblly=0pt,bburx=300pt,bbury=709pt} &
\end{tabular}
\end{center}
\vspace*{-1cm}
\caption{\it Systematic uncertainties on $F_L$, $F_0$ and $F_R$ in the dileptonic $t\bar{t}$ channel.}
\label{fig:wpola_syst_all_dilep}
\end{figure}

All systematic uncertainties are listed 
in Table~\ref{tab:wpola_syst} and illustrated in Figures~\ref{fig:wpola_syst_all_semilep} 
and~\ref{fig:wpola_syst_all_dilep}
for the semileptonic and dileptonic channels, respectively.
Generation and reconstruction sources contribute roughly in the same proportion to the total error.
The dominant generation contributions come from the FSR knowledge, the hadronization scheme and the $Q$-scale.
while the reconstruction systematics are dominated by the $b$-jet miscalibration and the top mass uncertainty.
\vspace*{-.5cm}
\begin{table}[htbp]
\begin{center}
\begin{tabular}{|l||c|c|c||c|c|c|}
\hline
\hspace*{.5cm} Source of uncertainty            &  \multicolumn{3}{c||}{Semileptonic channel} &
   \multicolumn{3}{c|}{Dileptonic channel} \\
\cline{2-7}
                                & $F_L$ & $F_0$ & $F_R$ & $F_L$ & $F_0$ & $F_R$ \\
\hline
\hline
\hspace*{.5cm} {\bf Generation}&       &       &       &       &       &       \\
$Q$-scale                       & 0.000 & 0.001 & 0.001 & 0.011 & 0.010 & 0.002 \\
Structure function              & 0.003 & 0.003 & 0.004 & 0.002 & 0.004 & 0.003 \\
ISR                             & 0.001 & 0.002 & 0.001 & 0.000 & 0.001 & 0.001 \\
FSR                             & 0.009 & 0.007 & 0.002 & 0.016 & 0.008 & 0.008 \\
$b$-fragmentation               & 0.001 & 0.002 & 0.001 & 0.002 & 0.003 & 0.002 \\
Hadronization scheme            & 0.010 & 0.016 & 0.006 & 0.002 & 0.003 & 0.002 \\
\hline
\hspace*{.5cm} {\bf Reconstruction} &   &       &    	&       &       &       \\
$b$-tagging (5\%)               & 0.006 & 0.006 & 0.000 & 0.003 & 0.003 & 0.001 \\
$b$-jet miscalibration (3\%)    & 0.011 & 0.005 & 0.005 & 0.017 & 0.001 & 0.017 \\
Input top mass (2~GeV)          & 0.015 & 0.011 & 0.004 & 0.021 & 0.007 & 0.014 \\
\hline
\hspace*{.5cm} {\bf Others}     &       &       &       &       &       &       \\
S/B scale (10\%)  	 	& 0.000 & 0.000 & 0.000 & 0.004 & 0.003 & 0.001 \\
Pile-up  (2.3 events)   	& 0.005 & 0.002 & 0.006 & 0.001 & 0.002 & 0.003 \\
\hline
\hline
\hspace*{.5cm} {\bf TOTAL}      & 0.024 & 0.023 & 0.012 & 0.034 & 0.016 & 0.024 \\
\hline
\end{tabular}
\vspace*{.2cm}
\caption{\it Summary of systematics on $F_L$, $F_0$ and $F_R$ in the semileptonic and dileptonic $t\bar{t}$ channels.}
\label{tab:wpola_syst}
\end{center}
\end{table}

\subsection{Results}
\label{sec:wpola_results}

Table~\ref{tab:wpola_results} presents the expected Standard Model results for the $W$ polarization 
measurement in $t\overline{t}$ semileptonic and dileptonic channels
after one LHC year at low luminosity (10$^{33}$cm$^{-2}$s$^{-1}$, 10~fb$^{-1}$).
The sensitivity is driven by the systematic uncertainties, which largely dominates the statistical ones.
$F_L$ and $F_R$ parameters are more precisely measured in the semileptonic channel, while the accuracy on 
$F_0$ is slightly better in the dileptonic one. 
Combining the results of both channel studies, assuming a pessimistic 100\% correlation 
of systematic errors, lead to the results shown in Table~\ref{tab:wpola_results}, rightmost column. 
The only improvement of the combination concerns $F_0$ on which the absolute error is estimated to be 0.016.
It is worth to notice that $F_R$, which is expected to be zero in the SM, is the most precisely measured
with an accuracy of 0.012.

\begin{table}[htbp]
\begin{center}
\begin{tabular}{|l||c|c|c|}
\hline
		& Semileptonic ($\pm$stat$\pm$syst)& Dileptonic ($\pm$stat$\pm$syst) &  Semilep+Dilep \\
\hline
\hline
$F_L$	& 0.299 $\pm$ 0.003 $\pm$ 0.024  & 0.314 $\pm$ 0.007 $\pm$ 0.034 & 0.303 $\pm$ 0.003 $\pm$ 0.024 \\
$F_0$	& 0.699 $\pm$ 0.005 $\pm$ 0.023  & 0.696 $\pm$ 0.010 $\pm$ 0.016 & 0.697 $\pm$ 0.004 $\pm$ 0.015 \\
$F_R$	& 0.002 $\pm$ 0.003 $\pm$ 0.012  &-0.010 $\pm$ 0.005 $\pm$ 0.024 & 0.000 $\pm$ 0.003 $\pm$ 0.012 \\
\hline
\end{tabular}
\vspace*{.2cm}
\caption{\it Standard Model results for $W$ polarization components after one LHC year of data taking
(10$^{33}$cm$^{-2}$s$^{-1}$, 10~fb$^{-1}$) in semileptonic and dileptonic $t\bar{t}$ channels. A 
combination of both results is presented in the last column.} 
\label{tab:wpola_results}
\end{center}
\end{table}

This result is~3 times better than the statistical error foreseen with single top events ($\sim0.03$ on 
$F_R$ with 10~fb$^{-1}$)~\cite{ATLAS_SINGLE_TOP}.
It is also roughly 3 to 5 times better than the Tevatron run~II statistical expectations with 2~fb$^{-1}$
($\sim0.03$ on $F_R$~\cite{kilminster} and $\sim0.09$ on $F_0$~\cite{vickey}).
CDF and D0 published first measurements of the $W$ polarization in $t\bar{t}$ pairs based on Run~I 
data~\cite{CDF_WPOLA_FR,CDF_WPOLA_F0, D0_WPOLA}. 
For example, CDF results are $F_0=0.91\pm0.37$(stat)$\pm0.13$(syst) and $F_R=0.11\pm0.15$.
They are largely limited by statistical errors\footnote{A few tens of both 
dileptonic and semileptonic $t\bar{t}$ events, with an integrated luminosity
of 109~pb$^{-1}$ (125~pb$^{-1}$) for CDF (D0).}.
Preliminary studies started at Run~II~\cite{TEV_RUNII}, but statistical errors will remain 
large, even with the total integrated luminosity of 2~fb$^{-1}$~\cite{Chakra2}.\\

The results of Table~\ref{tab:wpola_results} were obtained assuming realistic uncertainties of 3\% on the $b$-jet energy scale and 2~GeV on the top mass.
More pessimistic assumptions (5\% and 3~GeV) lead to an increase of the total systematic errors on
$F_L$, $F_0$ and $F_R$ to 0.031, 0.018 and 0.013.
On the contrary, more optimistic assumptions (1\% and 1~GeV) lead to 0.018, 0.014 and 0.009.
In all cases, the absolute error on $F_R$ remains in the range 0.009-0.013 and that on $F_0$ in the range 0.015-0.018.
The assumptions on systematic uncertainties have therefore a small impact, assessing the robustness of the results.\\

All above results were obtained with a $10^{33}$cm$^{-2}$s$^{-1}$ luminosity.
The luminosity may be $2 \cdot 10^{33}$cm$^{-2}$s$^{-1}$
at the LHC start. In this case, two scenarios are considered in ATLAS for the single electron trigger:  
increase of the single electron p$_T$ cut at the trigger level from 20 to 25~GeV (scenario 1), or even to 
30~GeV (scenario 2). The complete study has been redone in the semileptonic channel for both scenarios, 
assuming the same hypothesis for each source of systematic uncertainty presented in 
Figure~\ref{fig:wpola_syst_all_semilep}.
The number of events will be multiplied by 1.8 (1.6) for scenario~1 (scenario~2), 
while systematic errors remain almost unchanged.
Consequently, the same precision will be achieved on the $W$ polarization measurement.
At high luminosity, $10^{34}$cm$^{-2}$s$^{-1}$, a possible improvement can be 
to consider leptonic final states with
J/$\psi$, in a similar way as what is performed for the top mass measurement~\cite{TT_MASS}.\\

The forward-backward asymmetry, $A_{FB}$, based on the angle between the charged lepton 
and the $b$-jet in the $W$ rest frame, is often discussed in literature~\cite{Aguila}.
It can be expressed in terms of $F_L$ and $F_R$~\cite{Groote}:
\begin{equation}
A_{FB}=\frac{3}{4} \left( F_L-F_R \right)
\label{eq:afb}
\end{equation}
Taking the correlation between $F_L$ and $F_R$ into account, the following measurement on $A_{FB}$
can be extracted from the previous results:
\begin{equation}
A_{FB}=0.227 \pm 0.003 \mathrm{(stat)} \pm 0.016 \mathrm{(syst)}
\label{eq:afbmeas}
\end{equation}
Nevertheless, it does not provide any more information than the separate measurements of the ratios $F_L$ and $F_R$.

\subsection{Sensitivity to new physics}
\label{sec:wpola_wtb}

As already stated in the introduction, the search for anomalous (i.e. non Standard Model) 
interactions is one of the main motivations for top quark physics.
The measurement of the $W$ polarization provides a direct test of our understanding 
of the $tWb$ vertex, responsible for practically all top quark decays in the Standard Model (SM).
The deviations from the SM expectations induced by new physics contributions have been 
calculated in the framework of a few models~\cite{SUSY_WPOLA,RPARITY_WPOLA,TC2_WPOLA,COMU_WPOLA}.
However, because of the great diversity of models beyond the SM (Supersymmetry, dynamical electroweak symmetry breaking models, 
extra dimensions, \dots), 
it is also useful to study these possible new interactions in a model independent approach~\cite{Larios,MALKAWI,PECCEI,BERN1}.
The unknown dynamics can be parametrized with couplings representing the strength of effective interactions, 
through the following Lagrangian~\cite{Kane}:
\vspace*{.3cm}
\begin{equation}
{\mathcal{L}} = \frac{g}{\sqrt{2}} W_{\mu}^- \bar{b} \gamma^{\mu} (f_{1}^{L} P_L + f_{1}^{R} P_R) t 
-\frac{g}{\sqrt{2}\Lambda} \partial_{\nu} W_{\mu}^- \bar{b} \sigma^{\mu\nu}(f_{2}^{L} P_L + f_{2}^{R} P_R) t  + h.c.
\label{eq:lagrangian}
\end{equation}
where $P_{R/L} = \frac{1}{2} (1 \pm \gamma_5)$,  
$\sigma^{\mu\nu} = \frac{i}{2} [\gamma^{\mu},\gamma^{\nu}]$, 
$g$ is the electroweak coupling constant and 
$\Lambda$ is the energy scale to which the new physics becomes apparent 
(in the following, $\Lambda=$M$_W$ is set to keep the notation used in the literature).
$f_{1}^{L}$ and $f_{1}^{R}$ are vector-like couplings, whereas $f_{2}^{L}$ and $f_{2}^{R}$ 
are tensor-like couplings. This is the most general CP-conserving Lagrangian
keeping only the leading (mass dimension~4, first term) and the next-to-leading 
(mass dimension~5, second term) effective operators in the low energy expansion.
In the SM, the values of the couplings at tree level are $f_{1}^{L} = V_{tb} =1$, $f_{1}^{R} = f_{2}^{L} = f_{2}^{R} 
=0$.\\

It will first be shown how the measurement of the $W$ polarization in top decay can probe these 
anomalous couplings (section~\ref{sec:wpola_wtbcoupling}) and then a review of their existing 
direct and indirect experimental limits will be given (section~\ref{sec:wpola_limit}).

\subsubsection{Probe of $tWb$ anomalous couplings}
\label{sec:wpola_wtbcoupling}

The contributions of $f_{1}^{L}$, $f_{1}^{R}$, $f_{2}^{L}$ and $f_{2}^{R}$ anomalous 
couplings to each fraction of helicity state, $F_L$, $F_0$ and $F_R$, have been calculated at 
LO~\cite{Kane,chen} and at NLO~\cite{Aguila}, NLO effects being small.
A deviation of $f_{1}^{L}$ from~1 has not been considered in the following, as the $W$ helicity is not sensitive to it.
This can be easily understood, as $f_{1}^{L}$ is proportional to $V_{tb}$, 
which can not be directly measured with $t\overline{t}$ pairs but only with single top quarks.\\

\begin{figure}[htbp]
\begin{center}
\rotatebox{0}{\includegraphics*[width=\linewidth,bbllx=0pt,bblly=0pt,bburx=655pt,bbury=280pt]{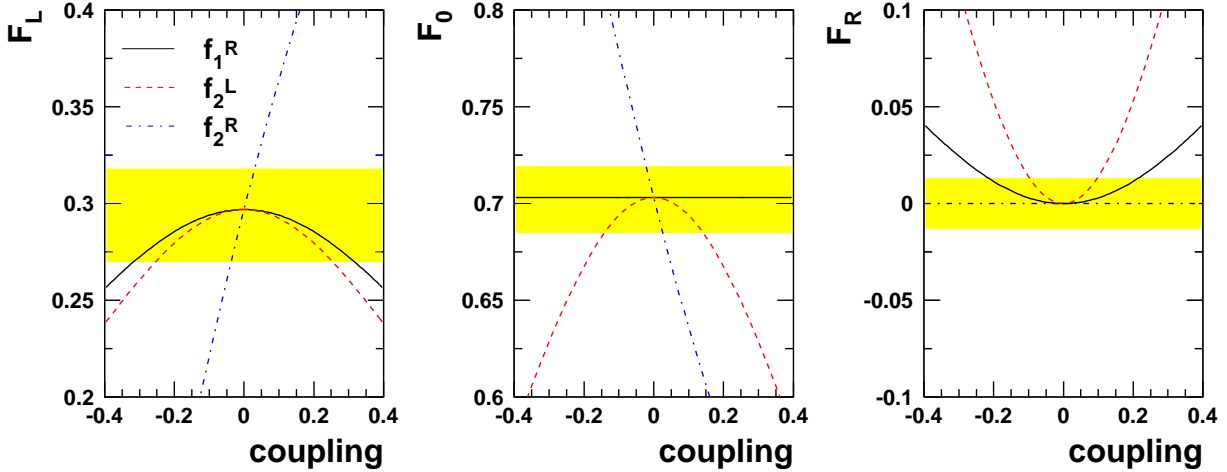}}
\end{center}
\caption{\it $F_0$, $F_L$ and $F_R$ dependence on the anomalous 
couplings $f_{1}^{R}$ (full lines), $f_{2}^{L}$ (dashed lines) and $f_{2}^{R}$ (dash-dotted lines). 
$f_1^L=1$ is assumed. The expected 1$\sigma$ uncertainties on $F_0$, $F_L$ and $F_R$ measured 
in $t\overline{t}$ pairs
after one LHC year at low luminosity (Table~\ref{tab:wpola_results}) are indicated with grey bands.}
\label{fig:wpola_couplings}
\end{figure}

In the following, an independent deviation of each anomalous coupling, 
$f_{1}^{R}$, $f_{2}^{L}$ and $f_2^R$, is assumed.
Figure~\ref{fig:wpola_couplings} shows the variation of $F_L$, $F_0$ and $F_R$ with these couplings.
$F_L$ and $F_R$ depend quadratically on $f_{1}^{R}$, whereas $F_0$ remains unchanged (full lines).
Similarly, the three fractions of helicity states are sensitive to $f_{2}^{L}$ in a quadratic way (dashed lines).
In these two cases, the sign of the coupling can not be determined and the sensitivity will be lowered
by the quadratic behavior. The last case is the most interesting (dash-dotted lines):
$F_0$ and $F_L$ depend almost linearly on $f_{2}^{R}$ with a slope=0.7~\cite{CPPM_WPOLA}, 
while $F_R$ is unchanged. Thus the sign of $f_2^R$ can be determined: 
$F_0>F_0^{SM}$ and $F_L<F_L^{SM}$ ($F_0<F_0^{SM}$ and $F_L>F_L^{SM}$) signs the presence of 
negative (positive) anomalous coupling $f_2^R$. The sensitivity is also higher than for $f_1^R$ and $f_2^L$.
The precision to which $F_L$, $F_0$ and $F_R$ can be measured (Table~\ref{tab:wpola_results})
sets the sensitivity to each anomalous coupling.
It is represented by grey bands in Figure~\ref{fig:wpola_couplings}.
$F_R$ is the most sensitive observable to probe $f_{1}^{R}$ and $f_{2}^{L}$ , whereas
$F_0$ is better for $f_{2}^{R}$.\\

\vspace{2cm}

\begin{figure}[htbp]
\begin{center}
\rotatebox{0}{\includegraphics*[width=\linewidth,bbllx=0pt,bblly=0pt,bburx=655pt,bbury=280pt]{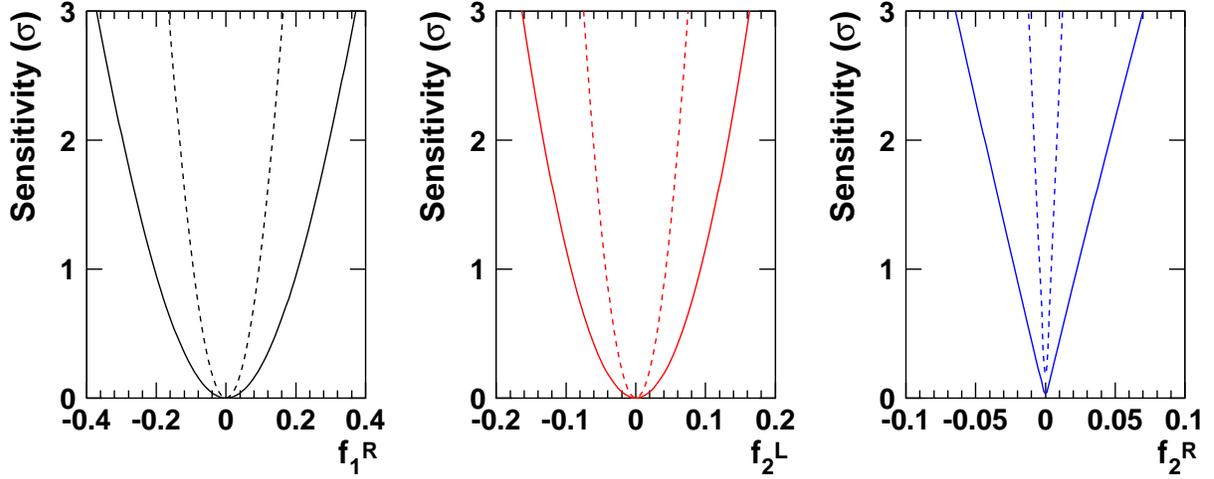}}
\end{center}
\vspace*{-1.0cm}
\caption{\it Sensitivity in $\sigma$ to the anomalous couplings $f_{1}^{R}$, $f_{2}^{L}$ and $f_{2}^{R}$
extracted from the $W$~polarization measurement in $t\overline{t}$ pairs after one LHC year 
at low luminosity (10~fb$^{-1}$).
Full lines indicate the overall (statistics+systematics) sensitivity, while dashed lines 
represent the statistical sensitivity only.}
\label{fig:couplings}
\end{figure}

Figure~\ref{fig:couplings} (full lines) shows the overall sensitivity (statistics+systematics) 
to each anomalous coupling that can be expected from the $W$ polarization measurement with $t\bar{t}$ pairs at LHC.
Dashed lines represent the statistical sensitivity only\footnote{A recent 
study at NLO on forward-backward asymmetry $A_{FB}$ in $t\overline{t}$ pairs~\cite{Aguila} 
indicates a $3\sigma$ statistical sensitivity 
with 100~fb$^{-1}$ on $f_{1}^{R} \sim 0.06$, $f_{2}^{L} \sim 0.03$ and $f_{2}^{R} \sim 0.003$.
This is in good agreement with our $3\sigma$ statistical sensitivity  
on $f_{1}^{R} \sim 0.17$, $f_{2}^{L} \sim 0.08$ and $f_{2}^{R} \sim 0.012$
obtained with 10~fb$^{-1}$ only.}.
The corresponding $2\sigma$ limits (statistics+systematics) are given in Table~\ref{tab:sensitivity}.
The best sensitivity is obtained on $f_{2}^{R}$ due to the presence of the large linear dependence.
It is of the order of the deviations expected by models like the Minimal Supersymmetric Standard Model (MSSM) 
or the Topcolor assisted Technicolor model (TC2)~\cite{chen}.

\subsubsection{Comparison with existing limits}
\label{sec:wpola_limit}

The only existing direct limits on the $tWb$ anomalous couplings 
can be placed from Tevatron $W$ polarization measurements 
in $t\overline{t}$ pairs, which are limited by the low statistics. 
As an example, the run~I result $F_R<0.18$ at 95\% C.L.~\cite{CDF_WPOLA_FR} translates in $f_1^R<0.8$.
From run~II expectations, statistical sensitivities to $f_1^R \sim 0.5$ and 
$f_2^{L,R} \sim 0.3$~\cite{chen} at 95\% C.L can be achieved.
The single top, which has not been experimentally observed so far, can provide further 
constraints on the $tWb$ anomalous couplings from its production rate and kinematic distributions.
At LHC, the expected $2\sigma$ limits are 
$-0.052 < f_2^L < 0.097$  and $-0.12 < f_2^R < 0.13$~\cite{Boos}
assuming a 5\% systematic uncertainty, and a statistical sensitivity to $f_1^R \sim 0.06$~\cite{Espriu}
with 100~fb$^{-1}$, one LHC year at high luminosity.
However, all these studies do not include any detector effect and detailed evaluation of 
systematic uncertainties. The related limits are summarized in Table~\ref{tab:sensitivity} for each coupling.\\

Indirect limits on the $tWb$ anomalous couplings have already been derived from precision measurements.
The $b \rightarrow s \gamma$ and $b \rightarrow s l^+l^-$ decays proceed via an electroweak radiative penguin process~\cite{bsgamma}.
As they include a $tWb$ vertex, an anomalous coupling will result in a change of the branching ratios.
The related limits on anomalous couplings are stringent: as an example, $f_1^R$ has to be less than 0.004~\cite{Larios2} at 95\% C.L.
As the $tWb$ coupling appears also in loop in $Z$ decays, electroweak measurements from LEP/SLC give 
other indirect limits, mainly competitive on $f_2^R$.
All these limits are presented for each coupling in the last two lines of Table~\ref{tab:sensitivity}.
However, they are indirect, SM-dependent, and scenarios can be envisaged where other contributions 
lead to cancellations that invalidate these bounds.\\

To conclude, it is worth to notice that our expected sensitivity to the right-handed tensor-like coupling $f_2^R$
is a factor 2-3 better than the best limit.
In any case, the $W$ polarization measurement in $t\overline{t}$ pairs 
and the single top studies at LHC will be complementary to determine the structure of 
the $tWb$ vertex as precisely as possible.\\

\begin{table}[htbp]
\begin{center}
\begin{tabular}{|c||c|c|c|}
\hline
		& \hspace*{.5cm}$f_1^R$ \hspace*{.5cm} & \hspace*{.5cm}$f_2^L$\hspace*{.5cm}  
                & \hspace*{.5cm}$f_2^R$ \hspace*{.5cm}    \\
\hline
$t\bar{t}$, LHC (10 fb$^{-1}$)	&  0.30 & 0.13 & 0.04 \\
(Stat.+ Syst.)	&  &  &  \\
\hline
\hline
$t\bar{t}$, Tevatron (2 fb$^{-1}$)	&  0.5 & 0.3 & 0.3 \\
(Stat. only)	&  &  &  \\
\hline
single top, LHC (100 fb$^{-1}$)	&  0.06 & 0.07 & 0.13 \\
(Stat.+ 5\% Syst.)	&  &  &  \\
\hline
\hline
$b \rightarrow s \gamma, sl^+l^-$, B-factories 	&  0.004 & 0.005 & 0.4 \\
(indirect)	&  &  &  \\
\hline
$Z$ decay, LEP    & - & - & 0.1 \\
(indirect)	&  &  &  \\
\hline
\end{tabular}
\vspace*{.5cm}
\caption{\it $2\sigma$ limits on anomalous couplings $f_1^R$, $f_1^L$ and $f_2^R$. 
At LO, in the SM, these couplings are equal to zero.
The first line presents our results
extracted from the $W$ polarization measurement in $t\overline{t}$ pairs after one LHC year 
at low luminosity (10~fb$^{-1}$). Expected limits at the Tevatron 
and with single top at LHC are shown in the next two lines. Current indirect limits from B-factories and LEP data
are presented in the last two lines.} 
\label{tab:sensitivity}
\end{center}
\end{table}

\newpage
\section{Sensitivity to top quark polarization in $t\bar{t}$ events}
\label{sec:cd}

As demonstrated in the previous section, the $W$ polarization measurement 
provides a direct probe of the top decay mechanism. Using the same events, it is also possible 
to test the $t\bar{t}$ production by measuring the top spin 
asymmetries, $A$ and $A_D$. Similarly as for $W$ polarization,
we will explain the method used to extract these asymmetries (section~\ref{sec:cd_method}),
present a complete study of systematic uncertainties in both semileptonic and dileptonic
$t\bar{t}$ channels (section~\ref{sec:cd_syst}), 
give the results combining both channels (section~\ref{sec:cd_results}), 
and finally discuss the related sensitivity to physics beyond the Standard Model (section~\ref{sec:cd_bsm}).

\subsection{Measurement method}
\label{sec:cd_method}

\begin{figure}[htbp]
\begin{center}
\vspace*{-1cm}
\rotatebox{0}{\includegraphics*[width=.8\linewidth,height=8.5cm,bbllx=0pt,bblly=20pt,bburx=655pt,bbury=680pt]{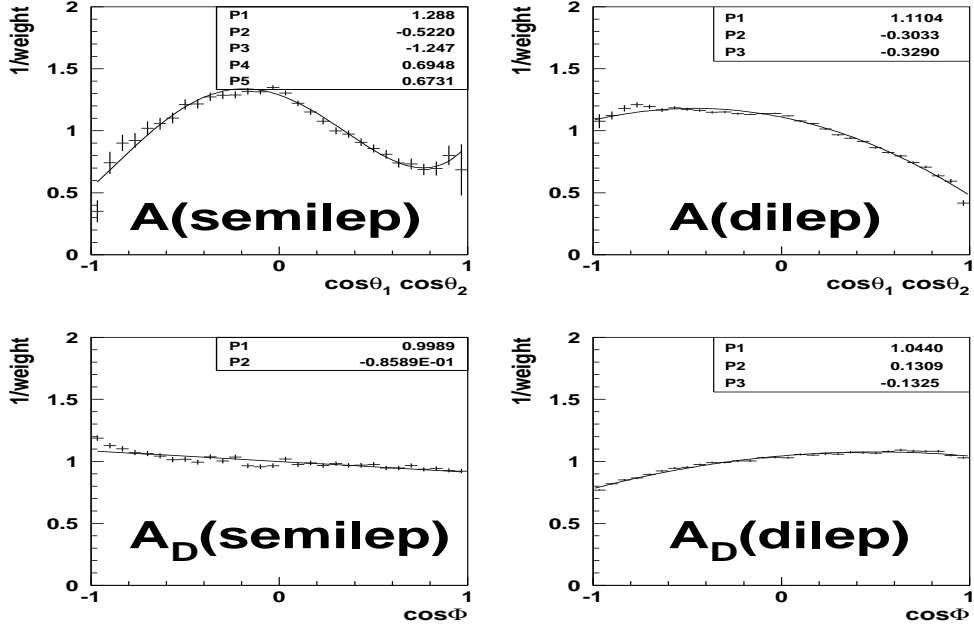}}
\end{center}
\vspace*{-.5cm}
\caption{\it Ratio between the distributions of $\cos \theta_1 \cos \theta_2$ (top) and $\cos \Phi$ (bottom)
after selection cuts and at parton level for semileptonic (left) and dileptonic (right) $t\bar{t}$ events.
The full lines are the results of polynomial fits.}
\label{fig:cd_weights}
\end{figure}

Similarly to the $W$ polarization analysis (section~\ref{sec:wpola}), 
selection cuts distort the parton level angular distributions.
Therefore, expressions given in Equation~(\ref{eq:estimators}) are no longer unbiased estimators of the spin correlation observables.
To correct for this bias, a weight is applied on an event by event basis, allowing
to get back, as much as possible, the original asymmetry.
One weight is applied per spin correlation observable ($A$ and $A_D$) and per channel (semileptonic and dileptonic). 
The weighting functions 
are computed by fitting the selection efficiency in $\cos \theta_1 \cos \theta_2$ ($\cos \Phi$) bins for $A$ ($A_D$).
This is shown in Figure~\ref{fig:cd_weights}, with a mean weight set to~1, which is the equivalent 
of Figure~\ref{fig:correction} in the $W$ polarization analysis.
The four ratios are fitted by a polynomial function to extract smooth corrections. 
The choice of polynomial order has been tuned to get the best fit quality.
The $A$ corrections have a wider range with respect to the $A_D$ ones, proving
that $A_D$ is less affected by selection cuts. 
The correction functions, computed on an independent data sample, are then applied event by event on
the analysis samples.\\

\begin{figure}[htbp]
\begin{center}
\rotatebox{0}{\includegraphics*[width=.48\linewidth,height=5.3cm,bbllx=50pt,bblly=140pt,bburx=540pt,bbury=630pt]{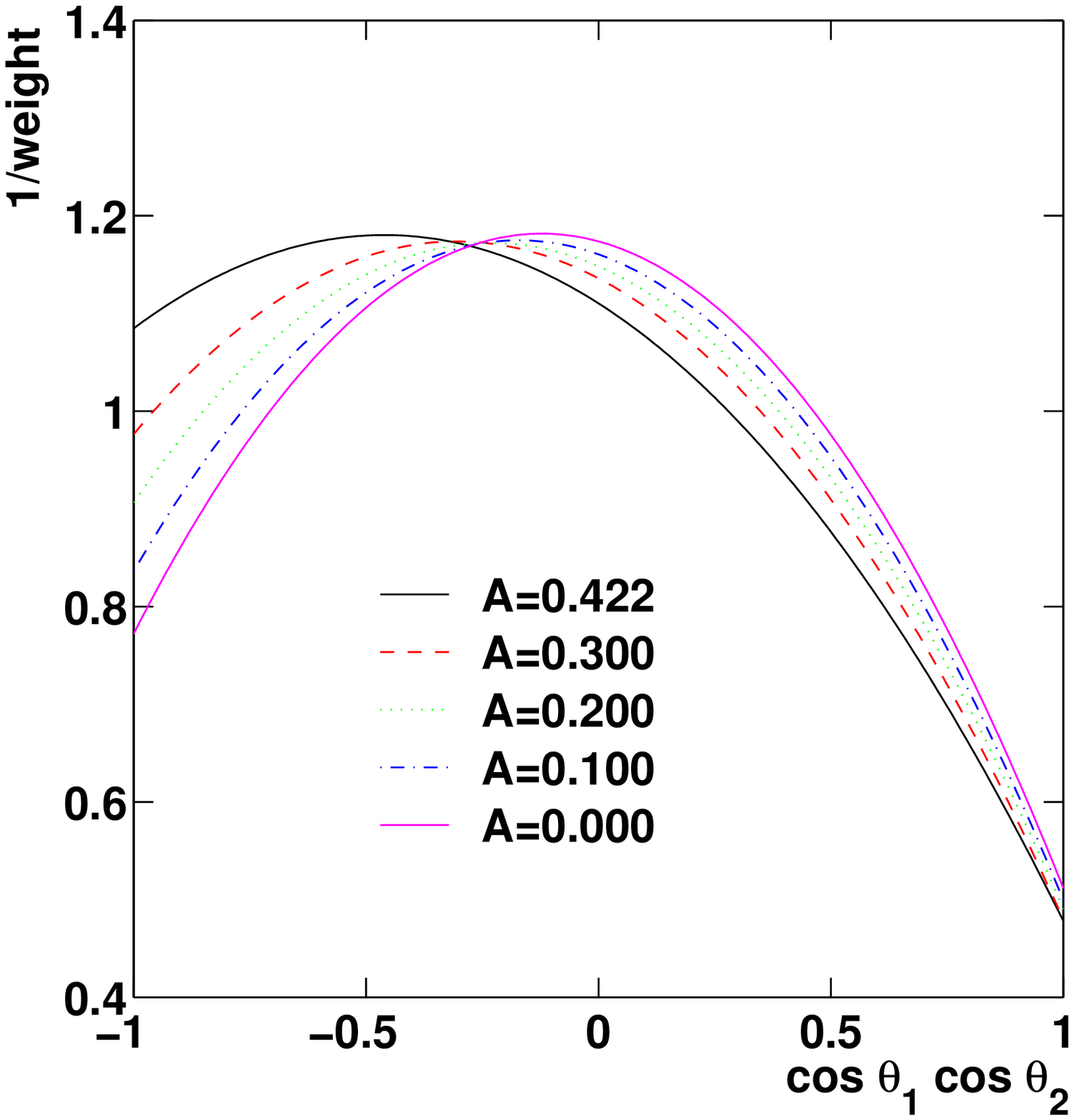}
\includegraphics*[width=.48\linewidth,height=5.3cm,bbllx=50pt,bblly=140pt,bburx=520pt,bbury=630pt]{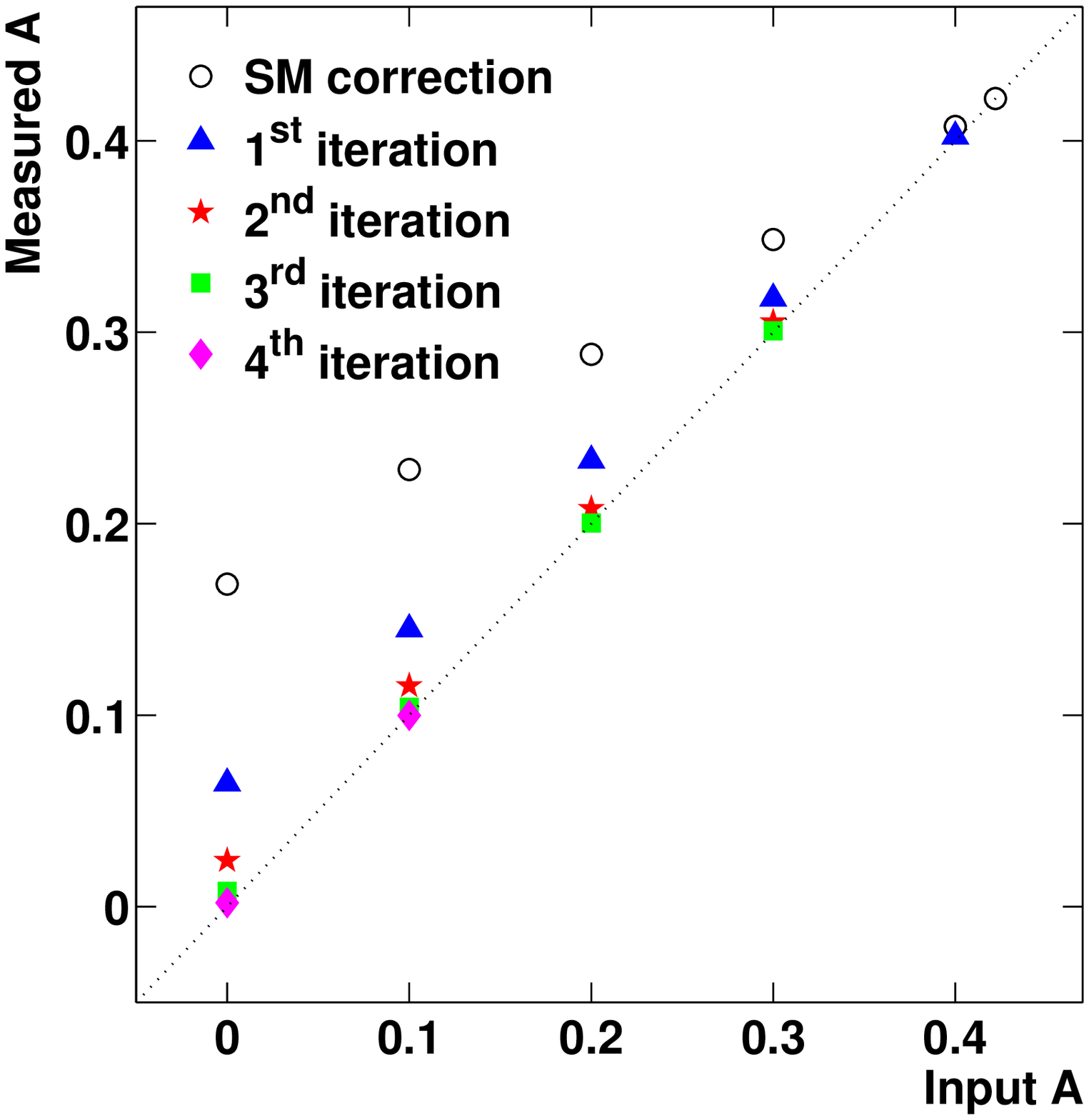}}
\rotatebox{0}{\includegraphics*[width=.48\linewidth,height=5.3cm,bbllx=50pt,bblly=140pt,bburx=540pt,bbury=630pt]{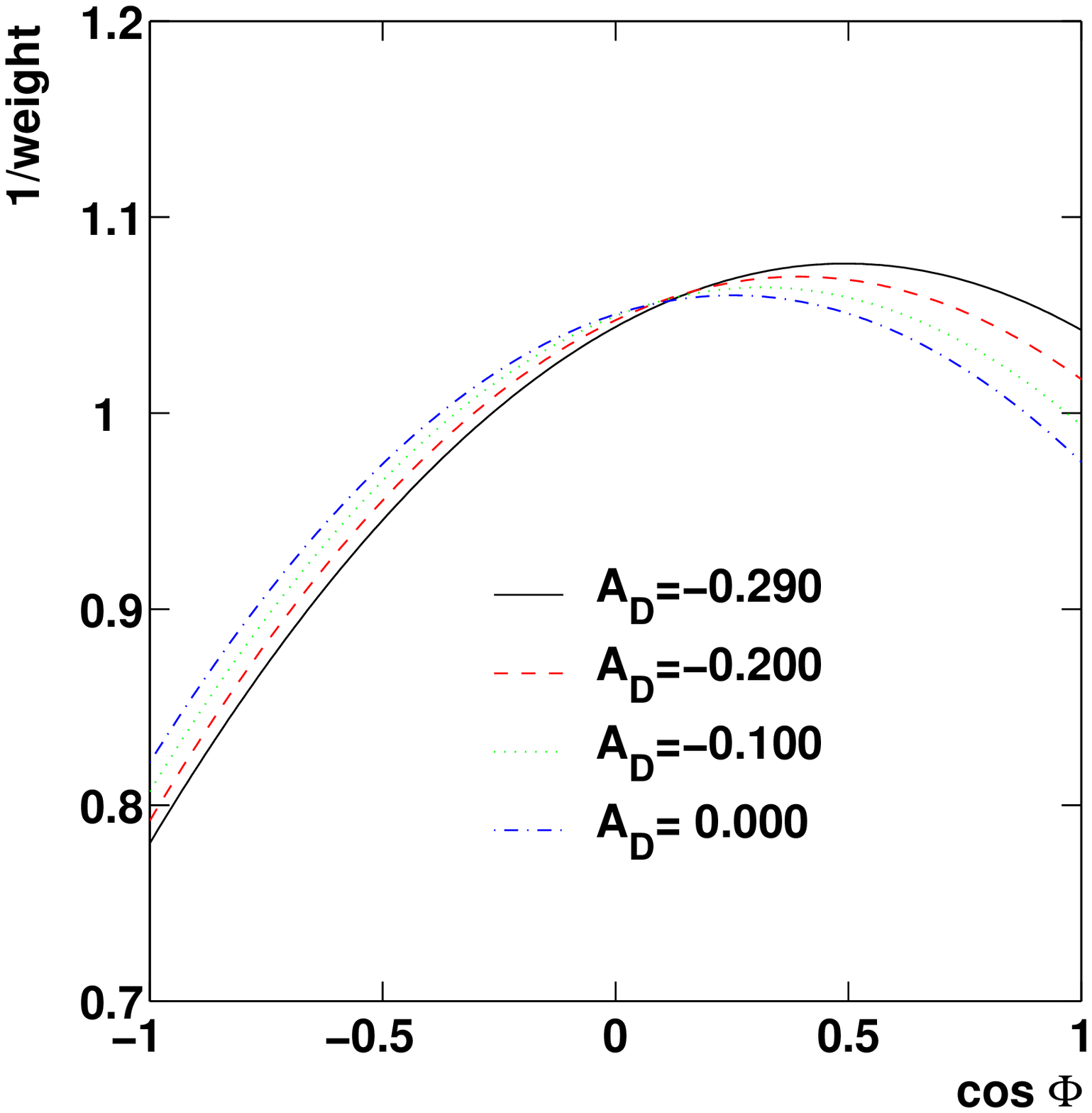}
\includegraphics*[width=.48\linewidth,height=5.3cm,bbllx=50pt,bblly=140pt,bburx=520pt,bbury=630pt]{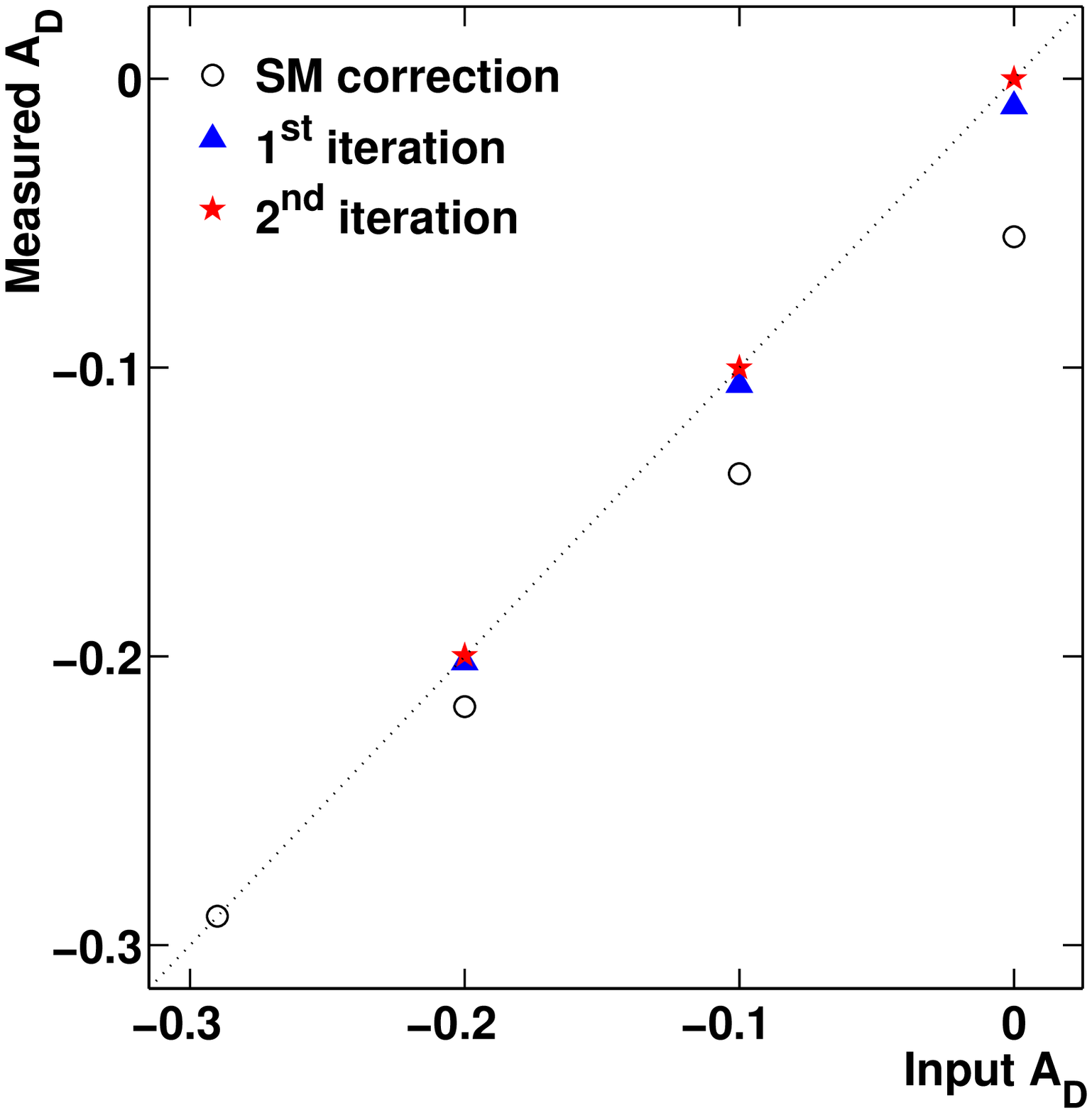}}
\end{center}
\vspace*{-.5cm}
\caption{\it Left: Correction functions for different $A$ (top) and $A_D$ (bottom) input values in the dileptonic channel. 
    The case $A=0.422$ and $A_D=-0.290$ corresponds to the SM functions of Figure~\ref{fig:cd_weights}.
    Right: $A$ (top) and $A_D$ (bottom) after selection cuts and correction as a function of their input values.
    The open circles correspond to the measurement using the SM correction.
    Full triangles, stars, squares and diamonds correspond to the first, second, third and fourth iteration, 
    respectively (see text). The dotted line is $y=x$.}
\label{fig:cin_vs_cout}
\end{figure}

The correction functions are extracted with a Standard Model scenario.
In case of deviation from the SM, the kinematic distributions can be affected, and
the correction functions will be changed.
This is illustrated in Figure~\ref{fig:cin_vs_cout} (left plots) for different $A$ (top) and $A_D$ (bottom) input values
in the dileptonic channel\footnote{This channel has been chosen to illustrate the method, 
as the bias due to the event reconstruction is more pronounced than in the semileptonic channel.}.
For this purpose, different mixtures of events with/without spin correlation effects have been generated.
Applying the SM correction function to these samples 
will therefore not correct completely for the bias induced by the selection cuts.
Figure~\ref{fig:cin_vs_cout} (right plots) shows with open circles the measured asymmetries $A$ (top) and $A_D$ (bottom)  
as a function of their input values after applying the SM correction function.
The measurement is clearly biased. As for Figure~\ref{fig:iteration} in the $W$ polarization measurement,
we proceed iteratively to overcome this problem. 
The SM correction function is first used. Then, in case of deviation from SM expectations a new 
correction function is calculated with this new asymmetry, and applied. 
The process converges after a few iterations, as seen in Figure~\ref{fig:cin_vs_cout}.

\subsection{Systematic uncertainties}
\label{sec:cd_syst}

The same sources of systematic uncertainties as for the $W$ polarization study (section~\ref{sec:wpola_syst}) 
are considered: 
five related to the generation ($Q$-scale, structure function, ISR-FSR,
$b$-fragmentation and hadronization scheme), three to the reconstruction ($b$-tagging, 
$b$-jet miscalibration, input top mass), the background normalization and the pile-up influence.
In the semileptonic channel, the light jet miscalibration is also taken into account as the least energetic jet 
in the top rest frame is used as spin analyzer.
A particular attention was paid to the proportion of $gg$ and $q\bar{q}$ processes 
involved in the $t\bar{t}$ pair production, which directly impacts the spin correlation (see Figure~\ref{fig:tt_cross_sect}).
To study separately this effect, samples with different proportions of $gg$/$q\bar{q}$ 
have been generated from 82\%/18\% to 90\%/10\% by steps of 2\%. Small and smooth dependences 
are observed with a slope of 0.006 (0.004) per \% of $gg$/$q\bar{q}$ variation for $A$ ($A_D$).\\

\begin{figure}[htbp]
\begin{center}
\rotatebox{0}{\includegraphics[height=4.2cm,width=13.cm]{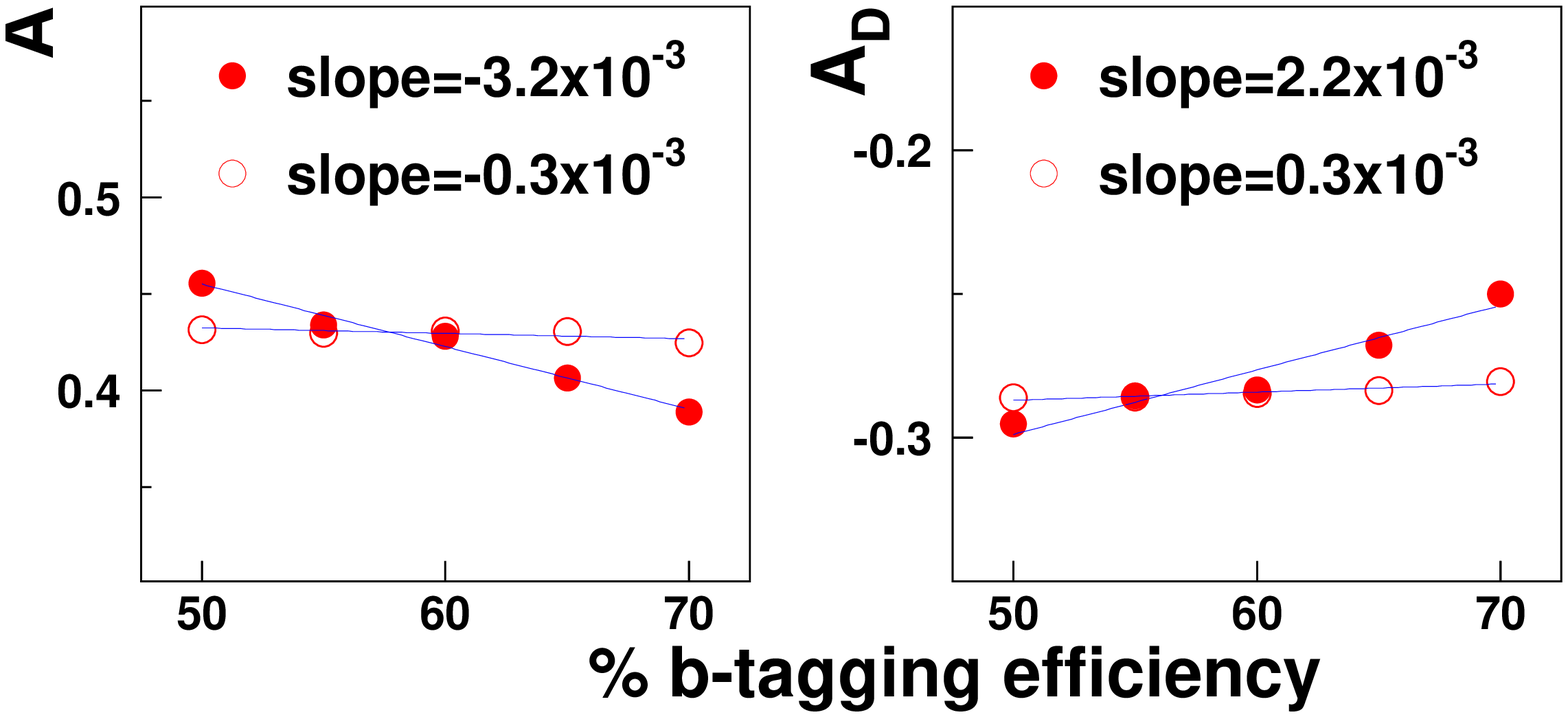}}
\vspace*{-.2cm}
\rotatebox{0}{\includegraphics[height=4.2cm,width=13.cm]{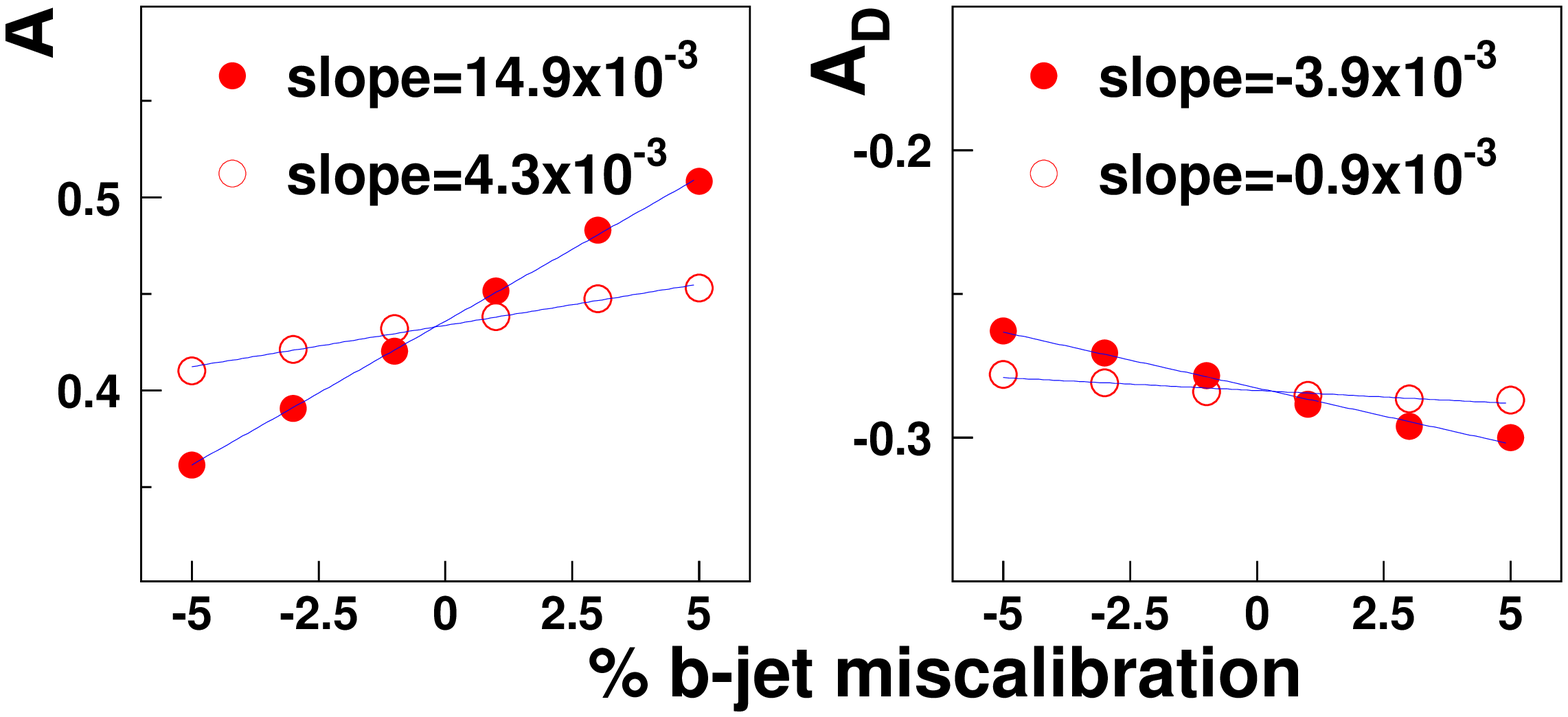}}
\vspace*{-.2cm}
\rotatebox{0}{\includegraphics[height=4.2cm,width=13.cm]{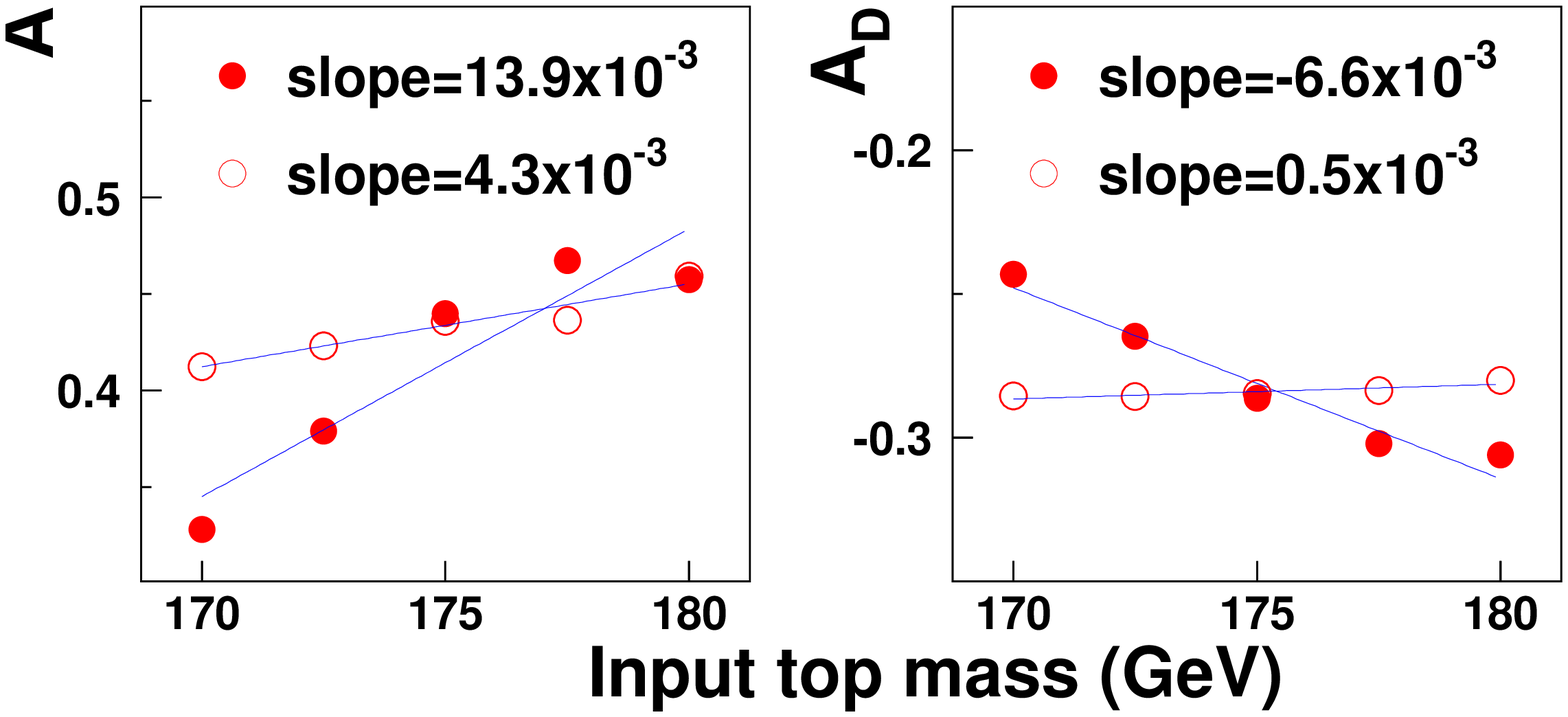}}
\vspace*{-.2cm}
\rotatebox{0}{\includegraphics[height=4.2cm,width=13.cm]{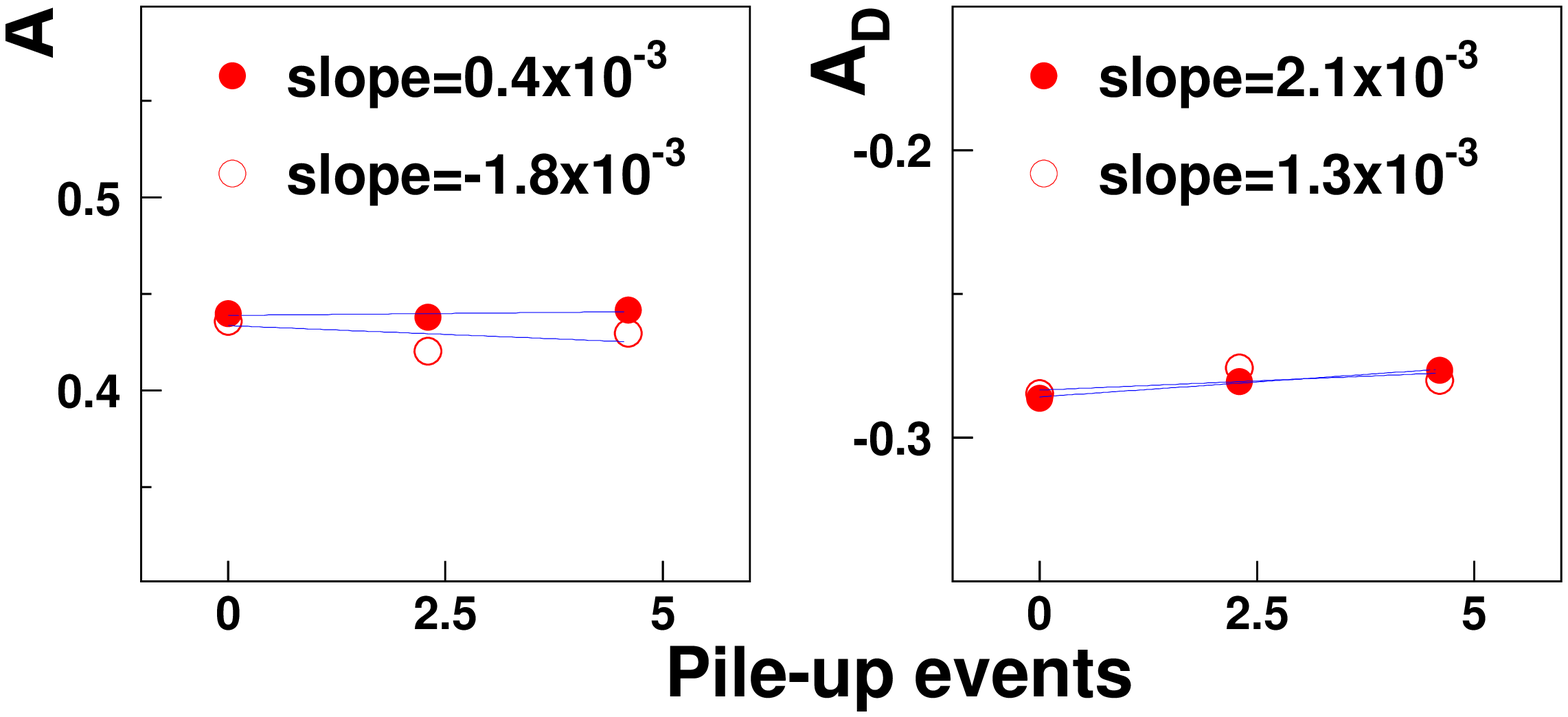}}
\end{center}
\begin{picture}(100,80)
\put(10,500){(a)}
\put(10,380){(b)}
\put(10,260){(c)}
\put(10,140){(d)}
\end{picture}
\vspace*{-3cm}
\caption{\it Measured $A$ (left) and $A_D$ (right) in the semileptonic (black circles) and dileptonic (open circles) $t\bar{t}$ channels
as a function of different parameters, see text for more details.
Linear fits are superimposed in each case, and the corresponding slope is indicated.}
\label{fig:cd_syst}
\end{figure}

\begin{figure}[htbp]
\begin{center}
\begin{tabular}{lccr}
\epsfig{file=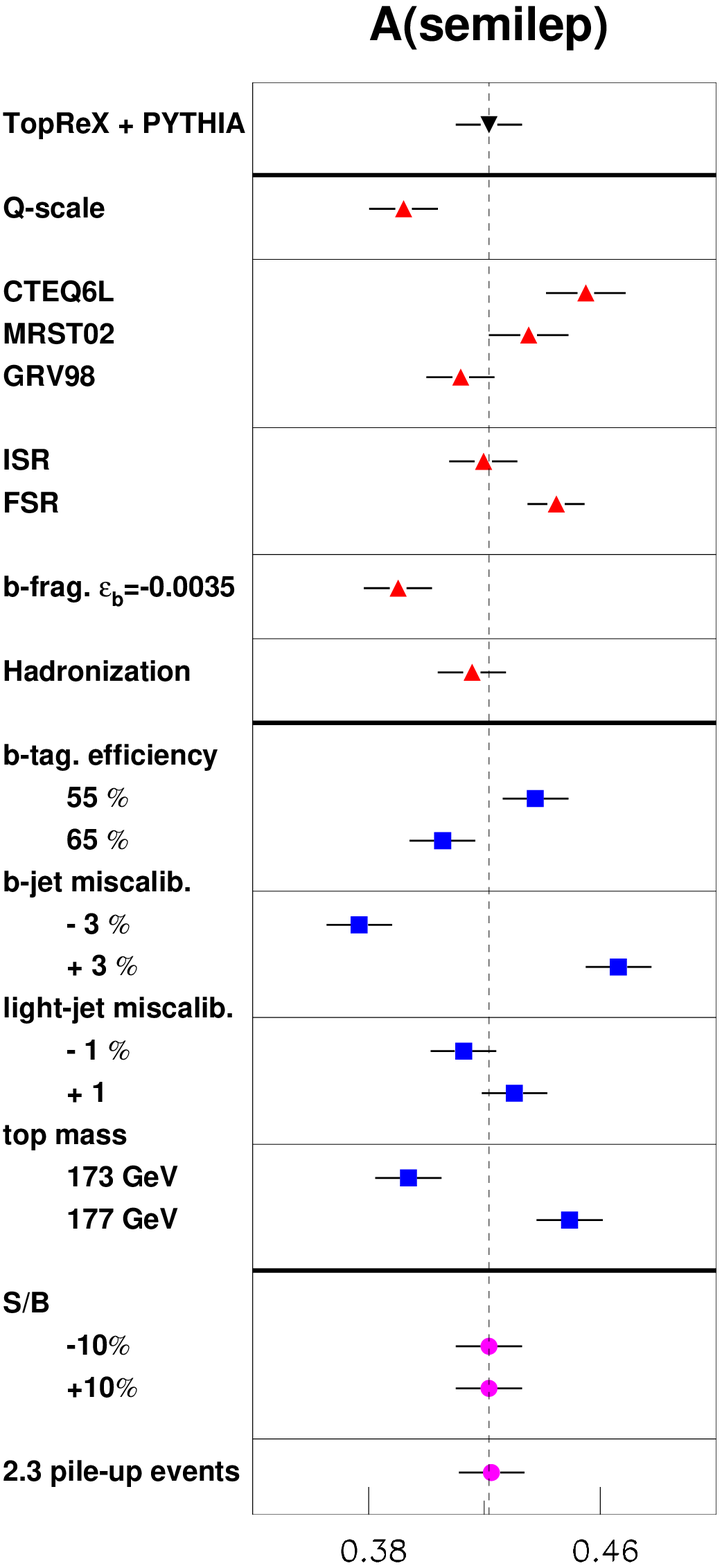,height=18cm,width=5.cm,bbllx=20pt,bblly=0pt,bburx=300pt,bbury=709pt} &
\epsfig{file=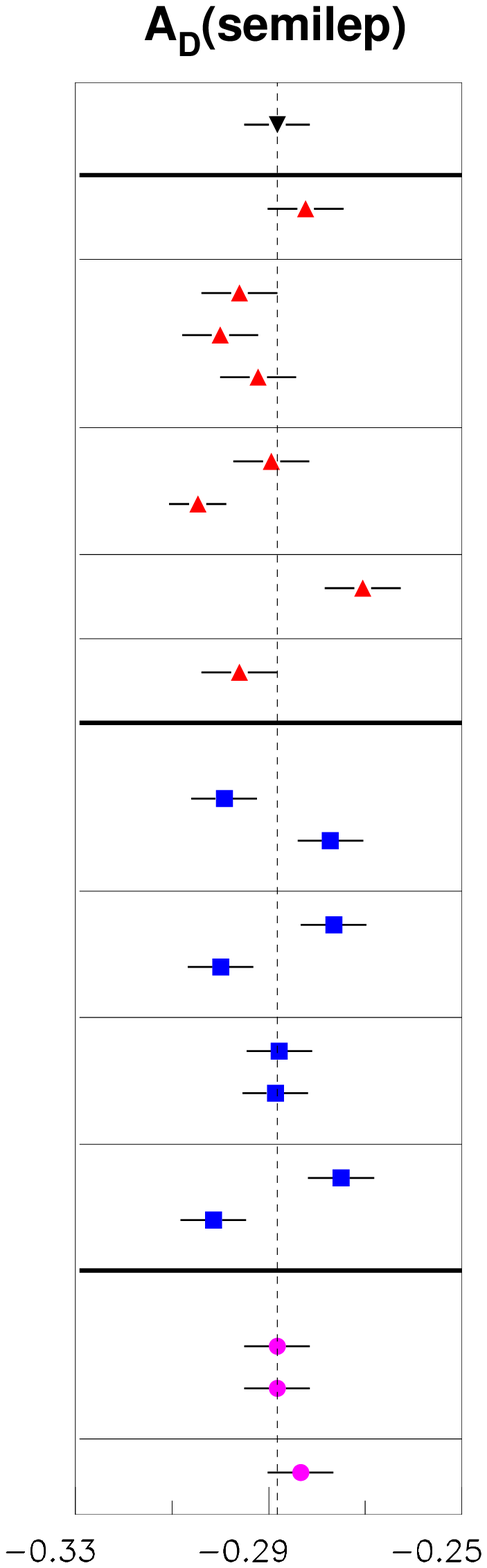,height=18cm,width=3.2cm,bbllx=119pt,bblly=0pt,bburx=300pt,bbury=709pt} &
\epsfig{file=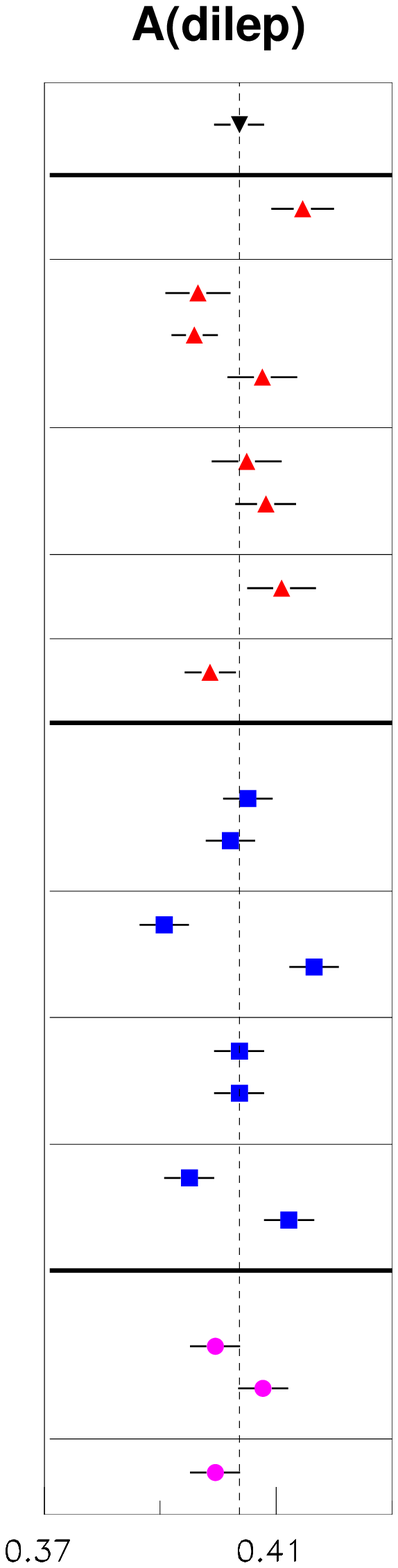,height=18cm,width=3.2cm,bbllx=119pt,bblly=0pt,bburx=300pt,bbury=709pt} &
\epsfig{file=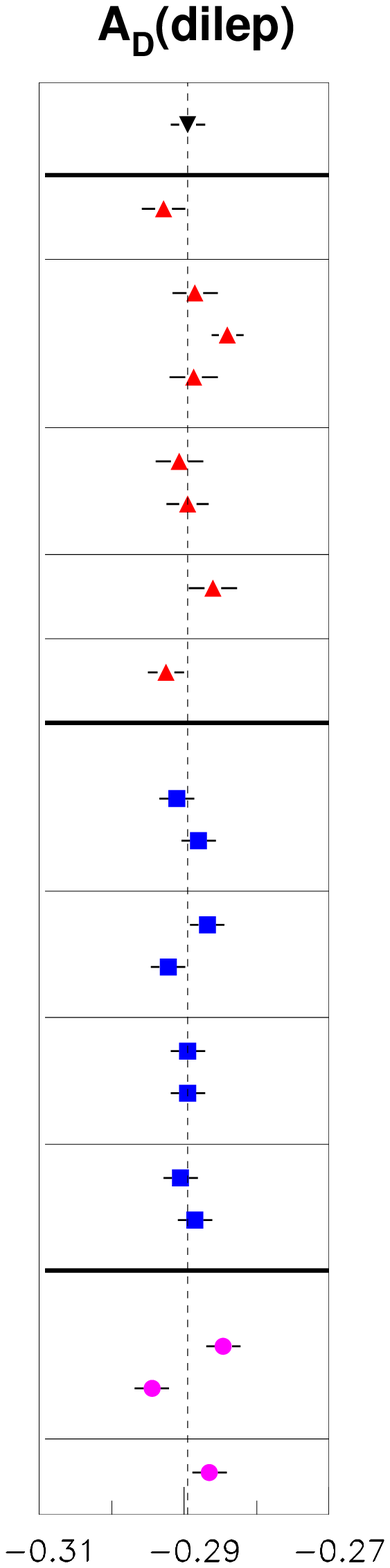,height=18cm,width=3.2cm,bbllx=150pt,bblly=0pt,bburx=300pt,bbury=709pt}
\end{tabular}
\end{center}
\vspace*{-1cm}
\caption{\it Systematic uncertainties on $A$ and $A_D$ in the semileptonic and dileptonic $t\bar{t}$ channels.}
\label{fig:cd_syst_all}
\end{figure}

The results obtained on $A$ and $A_D$ for different $b$-tagging efficiencies, $b$-jet miscalibrations, top masses and  
pile-up levels are detailed in Figure~\ref{fig:cd_syst}.
Linear behaviors are observed in both channels.
All systematic uncertainties are listed in Table~\ref{tab:cd_syst} and illustrated in Figure~\ref{fig:cd_syst_all}.
Generation and reconstruction sources contribute roughly in the same proportion to the total error.
The dominant generation contributions come from the $Q$-scale, the structure function and the $b$-fragmentation,
while the reconstruction systematics are dominated by the $b$-jet miscalibration and the top mass uncertainty.
The total systematic error for $A$ is 2.5 times higher than for $A_D$.
This is because the angles are computed in the $t\bar{t}$ rest frame, more difficult to reconstruct than
the top and anti-top rest frames separately.
The lower systematics in the dileptonic channel are explained by the choice of two ideal spin analyzers (charged leptons).
\vspace*{-.2cm}
\begin{table}[htbp]
\begin{center}
\begin{tabular}{|l||c|c|c|c|}
\hline
\hspace*{.5cm} Source of uncertainty  	&  \multicolumn{2}{c|}{Semileptonic channel} 
					&  \multicolumn{2}{c|}{Dileptonic channel} \\
\cline{2-5}
	                	& \ \ \  $A$ \ \ \  & $A_D$   & \ \ \ $A$ \ \ \ &  $A_D$   \\
\hline
\hline
\hspace*{.5cm} {\bf Generation}&       &        &       &	\\
$Q$-scale			& 0.029 & 0.006  & 0.011 & 0.003 \\
Structure function		& 0.033 & 0.012  & 0.008 & 0.005 \\
ISR				& 0.002 & 0.001  & 0.001 & 0.001 \\
FSR				& 0.023 & 0.016  & 0.005 & 0.000 \\
$b$-fragmentation		& 0.031 & 0.018  & 0.007 & 0.004 \\
Hadronization scheme    	& 0.006 & 0.008  & 0.005 & 0.003 \\
\hline
\hspace*{.5cm} {\bf Reconstruction}&    &        &       &       \\
$b$-tagging (5\%)		& 0.016 & 0.011  & 0.001 & 0.001 \\
$b$-jet miscalibration (3\%)	& 0.045 & 0.012  & 0.013 & 0.003 \\
light-jet miscalibration (1\%)	& 0.009 & 0.000  & 0.000 & 0.000 \\
Input top mass (2~GeV)		& 0.028 & 0.013  & 0.009 & 0.001 \\
\hline
\hspace*{.5cm} {\bf Others}	&	&        &       &       \\ 
S/B scale (10\%)		& 0.000 & 0.000  & 0.003 & 0.004 \\
Pile-up	(2.3 events)		& 0.001 & 0.005  & 0.004 & 0.003 \\
\hline
\hline
\hspace*{.5cm} 
{\bf TOTAL}			& 0.081 & 0.036   & 0.024 & 0.010  \\
\hline
\end{tabular}
\vspace*{.2cm}
\caption{\it Summary of systematics on $A$ and $A_D$ in the semileptonic and dileptonic $t\bar{t}$ channels.}
\label{tab:cd_syst}
\end{center}
\end{table}

\subsection{Results}
\label{sec:cd_results}

Table~\ref{tab:results_cd} presents the expected Standard Model results for $A$ and $A_D$ 
after one LHC year at low luminosity (10$^{33}$cm$^{-2}$s$^{-1}$, 10~fb$^{-1}$). 
In the semileptonic channel, the sensitivity is driven by the systematic uncertainties, which largely dominates 
the statistical ones, while both errors are comparable in the dileptonic channel. 
Combining the results of both channel studies, assuming a pessimistic 100\% correlation 
of systematic errors, lead to the results shown in Table~\ref{tab:results_cd}, rightmost column. 
They allow to observe and measure the Standard Model spin correlation with a 4\% precision.

\begin{table}[htbp]
\begin{center}
\begin{tabular}{|l||c|c|c|}
\hline
	& Semileptonic ($\pm$stat$\pm$syst)& Dileptonic ($\pm$stat$\pm$syst) &  Semilep+Dilep \\
\hline
\hline
$A$	&  0.422 $\pm$ 0.020 $\pm$ 0.081   &  0.404 $\pm$ 0.020 $\pm$ 0.024  &  0.406 $\pm$ 0.014 $\pm$ 0.023 \\
$A_D$	& -0.288 $\pm$ 0.012 $\pm$ 0.036   & -0.290 $\pm$ 0.011 $\pm$ 0.010  & -0.290 $\pm$ 0.008 $\pm$ 0.010 \\
\hline
\end{tabular}
\vspace*{.2cm}
\caption{\it Standard Model results for spin correlation observables after one LHC year of data taking
(10$^{33}$cm$^{-2}$s$^{-1}$, 10~fb$^{-1}$) in semileptonic and dileptonic $t\bar{t}$ channels. 
A combination of both results is presented in the last column.} 
\label{tab:results_cd}
\end{center}
\end{table}

This result can be compared with the 40\% precision expected from Tevatron run~II with 2~fb$^{-1}$, neglecting the systematics.
Experimentally, the $t\bar{t}$ spin correlation has never been observed.
The D0 experiment sets a lower limit on $A$ with 6 dilepton 
events from run~I (110~pb$^{-1}$)~\cite{D0}. This limit, $A > -0.25$ at 68\% confidence level,
can not be compared to the LHC values because the dominant production process at Tevatron is
$q \bar{q} \rightarrow t \bar{t}$, and the Standard Model prediction is $A=0.88$.\\

The results of Table~\ref{tab:results_cd} were obtained assuming realistic uncertainties of 3\% on the $b$-jet energy scale and 2~GeV on the top mass.
More pessimistic assumptions (5\% and 3~GeV) lead to an increase of the total systematic errors on
$A$ and $A_D$ to 0.030 and 0.011.
On the contrary, more optimistic assumptions (1\% and 1~GeV) lead to  0.015 and 0.009.
In all cases, $A_D$ remains in the range 4\%-5\%.
The assumptions on systematic uncertainties have therefore a small impact, assessing the robustness of this result.\\

All above results were obtained with a $10^{33}$cm$^{-2}$s$^{-1}$ luminosity.
As already discussed in section~\ref{sec:wpola_results}, the luminosity may be $2 \cdot 10^{33}$cm$^{-2}$s$^{-1}$
at the LHC start and modify the $p_T$ electron cut at the trigger level. 
The complete study has been redone in the semileptonic channel for the two foreseen scenarios, 
assuming the same hypothesis for each source of systematic uncertainty presented in Figure~\ref{fig:cd_syst_all}.
The number of events will be multiplied by 1.9 (1.8) for scenario~1 (scenario~2), while systematic errors
remain unchanged. Consequently, the same sensitivity will be achieved for $A$ and $A_D$ measurements.

\vspace{1cm}

\subsection{Sensitivity to new physics}
\label{sec:cd_bsm}

As already stated in the introduction,
a $t\bar{t}$ spin correlation observation would check that the top quark decays
indeed as a quasi-free quark, i.e. in particular before hadronization can take place
which could dilute the spin information.
A measurement of the expected Standard Model spin correlation would test
the top properties, with a left-handed coupling and a 1/2 spin.
On one hand, this would allow to set an upper limit on its lifetime, directly linked to 
Cabibbo-Kobayashi-Maskawa matrix elements~\cite{THEORY_21}.
On the other hand, this would offer a unique opportunity to study a ``bare'' quark, free from long distance effects
of QCD such as hadronization and confinement.\\

Therefore, a possible deviation of $t\bar{t}$ spin correlation from the SM prediction will be a hint of new physics. 
Its measurement can be used to probe the presence of new interactions.
For example, $gt\bar{t}$ anomalous couplings, 
linked to chromoelectric~\cite{CPVIOL} and chromomagnetic~\cite{chromomag} dipole moments
which naturally arise in dynamical electroweak symmetry breaking models such as technicolor or 
topcolor,
can affect the resultant $t\bar{t}$ spin correlation~\cite{Cheung}.
This is also the case in the presence of either a new heavy resonance in the $t\bar{t}$ production, such as a 
spin~0 neutral Higgs boson~\cite{HIGGS_RESO} (~$gg \rightarrow H \rightarrow t\bar{t}$~),
or spin~2 Kaluza-Klein (KK) gravitons~\cite{EXTRA_DIM}.
As an example, in theories with large extra dimensions~\cite{ADD}, the $s$-channel mediated by graviton 
KK modes gives rise to characteristic spin configurations and angular distributions for outgoing 
particles, which reflect the spin-2 nature of the intermediate KK gravitons.
With the sensitivity quoted in the previous section,  
a 5$\sigma$ deviation from the SM $t\bar{t}$ spin correlation 
can be observed if the fundamental scale of the extra dimensional theory is below 1.5~TeV.\\

New interactions in the decay can also affect the $t\bar{t}$ spin correlation. 
As an example~\cite{THEORY_43},
if a sufficiently light charged Higgs boson exists, such as in supersymmetric models, 
the decay $t \rightarrow H^+ b$ can compete with the
SM decay mode $t \rightarrow W^+ b$. 
As the charged Higgs decay to electrons and muons is largely suppressed, the deviation
on the $W$-polarization measurement can be small.
Contrarily, for m$_{H^+}<$150~GeV and at small $\tan\beta$ ($<2$), the decay in two jets is favored,
affecting the spin correlation in the semileptonic channel.
As a result, with m$_{H^+} \sim $80~GeV, a 5$\sigma$ deviation from the SM $t\bar{t}$ spin correlation 
can be observed if the branching ratio for top into charged Higgs plus $b$-quark is larger than 25\%.


\section{Conclusions}
\label{sec:conclu}

Because of its high mass, close to the electroweak symmetry breaking scale,
the top quark is an ideal place to search for physics beyond the Standard Model.
$W$ polarization in top decay and top spin observables reflect in detail the interactions involved
in top quark production and decay. Moreover, they can directly be inferred from the angular distributions of 
their respective decay products.
Therefore, they give a good opportunity for precise tests of these interactions 
and are sensitive probes of new physics.
Their precise measurements will be possible at the start of the LHC data taking, thanks to
the very large sample of top events that will be accumulated.
They will be complementary to $V_{tb}$ and cross section measurements, as NLO QCD corrections and 
theoretical errors are much smaller, of the order of 1\%.\\

The ATLAS capability to measure the $W$ polarization components $F_0$, $F_L$ and $F_R$ (for longitudinal, 
left-handed and right-handed helicity fractions) and $t\bar{t}$ spin asymmetries ($A$, $A_D$) 
has been studied in the complementary semileptonic and dileptonic $t\bar{t}$ channels.
The results of both channel studies have been combined.
Leading-order Monte Carlo generators were used as well as a fast simulation of the detector. 
The clean signature of semileptonic $t\bar{t}$ events, a high statistics (around $100\,000$~signal events 
after selection and reconstruction in one year at low luminosity, 10~fb$^{-1}$) 
and a high signal over background ratio (more than 10) are the attractive features of this channel. 
In the dileptonic channel,
the event topology reconstruction is complicated by the presence of two neutrinos 
in the final state, but feasible, the correct solution being found in 65\% of the events. 
Even if the statistics and the signal over background ratio are lower than in the semileptonic channel,
it is an attractive channel for the top spin asymmetry measurement, 
because the two charged leptons of the final state are the most powerful top spin analyzers. \\

In both channels, selection cuts bias the measurements. 
A weighting method was set up to correct for it, and its robustness assessed.
The sensitivity of the measurements is driven by the systematic uncertainties, 
which already dominates the statistical ones after one year at low luminosity. 
The main contributions to the total uncertainty come from 
the $Q$-scale, the hadronization scheme, the FSR  knowledge, the $b$-jet energy scale and the top mass.
After one LHC year, the Standard Model parameter $F_0$ can be measured with a 2\% accuracy and $F_R$ with 
a 1\% precision, comparable to the expected precision on the top mass. 
Using the same selected events, the Standard Model top spin asymmetry can be measured 
with a precision around 4\% with 10~fb$^{-1}$. 
These results are robust against other hypothesis for systematic uncertainties and trigger scenarios.\\

The sensitivity to physics beyond the Standard Model can be deduced from the above results.
This has been studied in a model independent approach on the decay side by introducing 
three $tWb$ anomalous couplings, $f_1^R$, $f_2^L$ and $f_2^R$, which parametrize new physics.
The best sensitivity, a $2\sigma$ limit of 0.04, is obtained on $f_2^R$, which is better
than indirect limits and expectations from other measurements.
Finally, the sensitivity of the top spin measurement to new interactions such as a top decay 
to charged Higgs boson or new s-channels (heavy resonance, gravitons) in $t\bar{t}$ production have also been assessed.\\

\vspace*{1cm}

{\bf Acknowledgments}

This work has been performed within the ATLAS collaboration, and we thank collaboration members
for helpful discussions.
We are indebted to P.~Uwer (CERN-TH) for long and fruitful discussions.
We thank S.~Bentvelsen (NIKHEF) for providing us with the W+4~jets events using AlpGen
generator, and our colleague J.B.~de Vivie de Regie (CPPM) for his help.
Last but not least, we are grateful to W.~Bernreuther and A.~Brandenburg (RWTH) for their helpful collaboration

{}

\end{document}